\pdfoutput=1
\documentclass[
prd,
amsmath,amssymb, aps, prd, nofootinbib ]{revtex4}
\usepackage{amsmath, amssymb, graphics}
\usepackage{slashed}
\usepackage{hyperref}
\usepackage[mathlines]{lineno}
\usepackage{color}
\usepackage{xcolor}
\usepackage{mathrsfs}
\usepackage{graphicx}
\usepackage[skip=2pt,font=footnotesize]{caption}
\usepackage{subfigure}
\usepackage{cleveref}
\usepackage[section]{placeins}
\def\beqa{\begin{eqnarray}}
\def\eeqa{\end{eqnarray}}
\newcommand{\tabincell}[2]{\begin{tabular}{@{}#1@{}}#2\end{tabular}}
\newcommand{\comment}[1]{}

\begin{document}

\preprint{ACFI-T17-17}

\title{CP-Violation in the Two Higgs Doublet Model: from the LHC to EDMs}
\author{Chien-Yi Chen}
\email{cchen@perimeterinstitute.ca}
\affiliation{Department of Physics and Astronomy, University of Victoria, Victoria, BC V8P 5C2, Canada}
\affiliation{Perimeter Institute for Theoretical Physics, Waterloo, ON N2L 2Y5, Canada
}
\author{Hao-Lin Li}
\email{haolinli@physics.umass.edu}
\affiliation{
Amherst Center for Fundamental Interaction Department of Physics, University of Massachusetts-Amherst, \\
Amherst, Massachusetts, 01003 USA
}
\author{Michael Ramsey-Musolf}
\email{mjrm@physics.umass.edu}
\affiliation{Amherst Center for Fundamental Interaction Department of Physics, University of Massachusetts-Amherst, \\
Amherst, Massachusetts, 01003 USA
}
\affiliation{Kellogg Radiation Laboratory, California Institute of Technology, Pasadena, CA 91125 USA}
\date{\today}
\begin{abstract}
We study the prospective sensitivity to CP-violating Two Higgs Doublet Models from the 14 TeV LHC and future electric dipole moment (EDM) experiments. We concentrate on the search for a resonant heavy Higgs that decays to a $Z$ boson and a SM-like Higgs $h$, leading to the $Z(\ell\ell)h(b\bar{b})$ final state. The prospective LHC reach is analyzed using the Boosted Decision Tree method. We illustrate the complementarity between the LHC and low energy EDM measurements and study the dependence of the physics reach on the degree of deviation from the alignment limit. In all cases, we find that there exists a large part of parameter space that is sensitive to both EDMs and LHC searches.
\end{abstract}
\maketitle
\section{Introduction\label{sec:introduction}}
With the discovery of the Higgs-like boson at the LHC~\cite{Aad:2012tfa,Chatrchyan:2012xdj}, the remaining particle predicted by the Standard Model (SM) has been found. 
Up to now, the measured properties of this new resonance show no significant deviation from the SM predictions. Nevertheless, the new 
boson could reside in a larger structure with an extended scalar sector that incorporates the SM.  The possibilities for such extended scalar sectors abound. Among the most widely considered and theoretically well-motivated are Two Higgs Doublet Models (2HDMs). Even with the rather minimal introduction of a second SU(2)$_L$ scalar doublet, the possible phenomenological consequences of 2HDMs are rich and diverse. 
The possibility of new sources of CP-violation is one of the most interesting but, perhaps, less extensively studied. 

Explaining the cosmic matter and anti-matter asymmetry requires the existence of additional CP-violation (CPV) beyond that of the SM.
Electroweak Baryogenesis (EWBG) is one of the most compelling solutions to this problem~\cite{Trodden:1998ym,Cline:2006ts,Morrissey:2012db}.
EWBG fulfills the Sakharov conditions for successful baryogenesis \cite{Sakharov:1967dj} ($B$ violation, out-of-equilibrium dynamics, and both C and CP-violation) through $B+L$ violating sphaleron transitions, a strong first order electroweak phase transition that proceeds through bubble nucleation, and CPV interactions at the bubble wall. 
While the SM would in principle provide these ingredients, it is known that the CPV effects generated by the Cabibbo-Kobayashi-Maskawa matrix and QCD $\theta$ term are too feeble and that the SM-like Higgs scalar is too heavy for a strongly first order electroweak phase transition ~\cite{Kajantie:1996mn,Csikor:1998eu,Rummukainen:1998as}. 

The 2HDMs provide possible solutions to these shortcomings. The viability of a strong first order electroweak phase transition and the favored parameter space of the 2HDMs have been studied in Refs.~\cite{Dorsch:2013wja,Fromme:2006cm,Cline:1996mga}. In the CPV sector, the LHC has already excluded the new boson as a pure CP odd scalar at 99.98\% CL and 97.8\% CL in Ref.~\cite{Khachatryan:2014kca} and Ref.~\cite{ATLAS:2013nma} respectively. 
 
 If the boson is a part of the 2HDM, it could nevertheless receive a small CP-odd admixture from CP-violating terms in the scalar potential.  This possibility for  2HDM CP-violation is strongly bounded by the non-observation of permanent electric dipole moments (EDMs) of the neutron, electron, and diamagnetic atoms, including mercury and radium ~\cite{Baron:2013eja,Baker:2006ts,Griffith:2009zz,Kumar:2013qya}, as analyzed recently 
in Refs.~\cite{Inoue:2014nva,Shu:2013uua,Bian:2014zka,Yamanaka:2017mef,Jung:2013hka}. The authors of Refs.~\cite{Chen:2015gaa,Bian:2017jpt,Akeroyd:2016ymd} also pointed out that LHC searches for additional, heavy scalars can be complementary to EDM searches,  especially in regions of 2HDM parameter space where strong cancellations between Barr-Zee EDM diagrams occur. Nonetheless, there exists a window for sufficient CPV to generate the matter-antimatter asymmetry, as shown in Ref.~\cite{Dorsch:2016nrg}. 

In what follows, we analyze the prospects for future LHC probes of the CPV 2HDM, building on the previous studies in Ref.~\cite{Inoue:2014nva} and Ref.~\cite{Chen:2015gaa}, where EDMs constraints and 8 TeV LHC results in CPV 2HDMs are analyzed in detail. We adopt the framework of CPV 2HDMs with a softly-broken $Z_2$ symmetry to avoid a problematic tree level flavor changing neutral currents (FCNCs). We consider future LHC searches for a heavy Higgs of mixed CP (denoted $h_{i=2,3}$) which decays to a $Z$ boson and a SM-like Higgs ($h_{1}$), and obtain the prospective reach for Run II and the high luminosity phase (HL-LHC). We  concentrate on the $llb\bar{b}$ final state, where the $Z$ boson decays to a pair of leptons ($e \ or \ \mu$), and the SM-like Higgs decays to a pair of b quarks, because it is one of the most sensitive channels and because the final state particles allow for a relatively high reconstruction efficiency. We first follow the cut-based analysis procedure described in Ref.~\cite{Aad:2015wra} to reproduce the ATLAS 8 TeV results and validate our Monte Carlo signal and background generation, then use the Boosted Decision Tree (BDT)~\cite{adaboost} method to obtain the 95\% CLs exclusion limit for future 14 TeV experiments with integrated luminosities equal to 300 fb$^{-1}$, and 3000 fb$^{-1}$, respectively. We subsequently translate the prospective exclusion limits into constraints on the parameter space, and find that a large portion of parameter space can be tested with both future LHC and EDMs experiments. 

From the global fit of Higgs coupling measurements~\cite{ATLAS:2014kua,CMS:2016qbe}, one find that the current data favor the 2HDMs to be close to the alignment limit: $\beta-\alpha=\pi /2$ where $\alpha$ and $\beta$ are defined in Sec.~\ref{sec:IB} and Sec.~\ref{sec:IA} respectively. Therefore, we summarize our finding in the following two categories with the combined analysis of the future EDMs and LHC exclusion bounds shown in Figs.~\ref{fig:T1} and ~\ref{fig:T2}.

\begin{itemize}
\item 2HDMs in alignment limit:
With a discovery at the future LHC, the Type-I 2HDM would imply observation of non-zero radium and electron EDMs in the next generation searches, while the Type-II 2HDM would imply non-zero neutron and radium EDMs.
A null result at the future LHC will still allow for the  CPV 2HDMs if the CPV effect in the model is sufficiently small. Future EDM may still yield non-vanishing results in this case.

\item 2HDMs away from the alignment limit:
With a discovery at the future LHC, one may or may not expect non-zero EDM results depending on the level of deviation. This is due to the fact that the production of the mostly CP odd Higgs in the model ($h_3$ defined in Sec.~\ref{sec:IB}) is sensitive to the deviation from the alignment limit which is not suppressed by a small CPV effect. As a result, the discovery at LHC may indicate a relatively large deviation from the alignment limit instead of a large CPV effect.
A null result at future LHC may not exclude the CPV 2HDMs if the CPV effect and the deviation from the alignment limit are sufficiently small. For a relatively large deviation from the alignment limit, any non-zero EDM results would disfavor the CPV 2HDMs.
\end{itemize}
The above conclusions are based on the detailed analysis discussed in Sec.~\ref{result}.

We also point out that our analysis will break down in some regions of parameter space that have both small values of $\tan\beta$ (ratio of the vacuum expectation values of the two neutral scalars) and the CPV Higgs mixing angle $\alpha_b$, where the interference between the resonant amplitude ( $gg\to h_{2,3}\to Zh_1$) and non-resonant amplitude (box diagram $gg\to Zh_1$) for $Zh_1$ production may become significant. We do not perform a full analysis of this effect, but rather  give a qualitative estimate, 
as this region does not appear to significantly impact the prospective Run II exclusion reach.

The organization of our paper is the following. In Sec.~\ref{sec:model}, we describe our set-up for CP violation 2HDMs. In Sec.~\ref{prodec}, we show the analytical formulas used to derive constraints on the parameter space. In Sec.~\ref{simulation} we describe details of our simulation and analyses. In Sec.~\ref{result}, we exhibit future LHC constraints and discuss possible issues arising from the interference between the resonant and non-resonant diagrams. Finally, we conclude in Sec.~\ref{conclusion}.  The distributions of kinematic variables used in BDT analysis are listed in Appendix~\ref{APPA}. The formulas for two-body decay rates of heavy Higgses are given in Appendix~\ref{APPB}.

\section{CPV 2HDM Model Description\label{sec:model}}
In this section, we describe details of the CPV 2HDM framework that will be used in the following discussions.
\subsection{General 2HDM Scalar Potential\label{sec:IA}}
The most general 2HDM scalar potential containing two Higgs doublets $\phi_1$ and $\phi_2$ can be expressed in the following form:
\begin{eqnarray}                           \label{pot}
V(\phi_1, \phi_2)&=&-\frac{1}{2}\left[m_{11}^2(\phi_1^\dagger\phi_1)
+\left(m_{12}^2 (\phi_1^\dagger\phi_2)+{\rm h.c.}\right)
+m_{22}^2(\phi_2^\dagger\phi_2)\right] 
\nonumber \\
&&+ \frac{\lambda_1}{2}(\phi_1^\dagger\phi_1)^2
+\frac{\lambda_2}{2}(\phi_2^\dagger\phi_2)^2+\lambda_3(\phi_1^\dagger\phi_1) (\phi_2^\dagger\phi_2) 
+\lambda_4(\phi_1^\dagger\phi_2) (\phi_2^\dagger\phi_1) 
\nonumber \\
&&+\frac{1}{2}\left[\lambda_5(\phi_1^\dagger\phi_2)^2 + \lambda_6 (\phi_1^\dagger\phi_2) (\phi_1^\dagger\phi_1) 
+ \lambda_7 (\phi_1^\dagger\phi_2) (\phi_2^\dagger\phi_2) +{\rm h.c.}\right] \ . 
\label{pot_gen}
\end{eqnarray}
Two fields $\phi_1$ and $\phi_2$ can be expressed as
\begin{eqnarray}\label{decompose}
\phi_1=\begin{pmatrix}
H_1^+ \\
\frac{1}{\sqrt2} (v_1 + H_1^0 + i A_1^0)
\end{pmatrix}, \ \ 
\phi_2=\begin{pmatrix}
H_2^+ \\
\frac{1}{\sqrt2} (v_2 + H_2^0 + i A_2^0)
\end{pmatrix} \ 
\end{eqnarray}
with in general $v_1$ and $v_2$ complex and $v=\sqrt{|v_1|^2+|v_2|^2}=246$
 GeV. We also denote that $\tan\beta=|v_2|/|v_1|$. One can always perform a $SU(2)_L\times U(1)_Y$ gauge transformation to go into a basis where $v_1$ is real while $v_2=|v_2|e^{i\xi}$ is still complex.

To guarantee that there are no FCNCs at tree level, one can assign $Z_2$ charges to the two Higgs doublets as well as the fermion fields such that each fermion can only couple to one of the Higgs doublets. Depending on the transformation of the fermion fields under the $Z_2$ symmetry, there can be various types of 2HDMs that we will introduce in Sec. \ref{Yukawaint}. The $Z_2$ symmetry implies the potential parameters $m_{12}^2$ and $\lambda_{6,7}$ vanish, which in turn forbids the presence of CP phases in the potential. Therefore, we retain the $m_{12}^2$ term which only softly breaks the $Z_2$ symmetry. In general, this soft $Z_2$ symmetry breaking term together with quartic $Z_2$ conserving term would induce new quartic $Z_2$ breaking terms by renormalization, but they are at one-loop level and thus do not induce new FCNC at tree level.

Hermicity implies that there are only two complex parameters, $m_{12}^2$ and $\lambda_5$, in the potential. With the global phase redefinition of the fields $\phi_j \to e^{i\theta_j}\phi_j$, one may define two rephasing invariant phases as in Ref ~\cite{Inoue:2014nva},
\begin{eqnarray}
\nonumber
\delta_1 & = & \mathrm{Arg}\left[\lambda_5^\ast(m_{12}^2)^2\right] \ ,\\
\delta_2 & = & \mathrm{Arg}\left[\lambda_5^\ast(m_{12}^2) v_1 v_2^\ast\right]\ .
\end{eqnarray}
The minimization of the potential yields that:
\begin{eqnarray}\label{mini}
&&m_{11}^2 = \lambda_1 v^2 \cos^2\beta + (\lambda_3 + \lambda_4) v^2 \sin^2\beta - {\rm Re} (m_{12}^2 e^{i\xi}) \tan\beta + {\rm Re} (\lambda_5 e^{2i\xi}) v^2\sin^2\beta \ ,
\label{mini1}\\
&&m_{22}^2 = \lambda_2 v^2 \sin^2\beta + (\lambda_3 + \lambda_4) v^2 \cos^2\beta - {\rm Re} (m_{12}^2 e^{i\xi}) \cot\beta + {\rm Re} (\lambda_5 e^{2i\xi}) v^2\cos^2\beta \ ,
\label{mini2}\\
&&{\rm Im} (m_{12}^2 e^{i\xi})=v^2 \sin\beta\cos\beta {\rm Im} ( \lambda_5 e^{2i\xi} ) \ . \label{mini3} 
\end{eqnarray}
Eq.~\ref{mini3}  above indicates that the value of $\xi$ is determined by given $m_{12}^2$ and $\lambda_5$. Expressing this equation with rephasing invariant phases implies:
\begin{equation}
\label{eq:invar1}
|m_{12}^2| \sin(\delta_2-\delta_1) = |\lambda_5 v_1 v_2|\sin(2\delta_2-\delta_1)\ \ \ .
\end{equation}
In short, there is only one CP independent phase in the potential after electroweak symmetry breaking(EWSB). Using this rephasing freedom of the fields, we will work in a basis where $\xi=0$ and encode this invariant CPV phase into a CPV angle in the diagonalization matrix for the neutral Higgs sector.

\subsection{Higgs Mass Eigenstates\label{sec:IB}}
After EWSB, we can use the following relations to diagonalize the mass matrix for the charged Higgs sector, which separates the physical charged Higgs and would-be Goldstone bosons:
\beqa
\left(\begin{array}{c} H^+ \\ G^+ \end{array}\right)=
\left(
\begin{array}{cc}
-s_\beta &c_\beta \\
c_\beta &s_\beta 
\end{array}
\right)
\left(\begin{array}{c} {H_1}^+ \\ {H_2}^+ \end{array}\right)
\eeqa
This leads to a relationship between the mass of the charged Higgs and parameters in the scalar potential:
\begin{eqnarray}\label{MC}
m^2_{H^+} = \frac{1}{2} \left(2\nu - \lambda_4 - {\rm Re}\lambda_5 \right)  v^2, \ \ \ \nu \equiv\frac{{\rm Re}m_{12}^2 \csc\beta \sec\beta}{{2v^2}} \ .
\end{eqnarray}
where the parameter $\nu$ sets the hierarchy between the SM-like Higgs and charged Higgs.
The mass term in the Lagrangian is given by 
${\cal L}^\mathrm{mass}_\mathrm{neutral}=-(H_1^0,H_2^0,A^0){\cal M}^2(H_1^0,H_2^0,A^0)^T$ gives,
\begin{equation}\label{MM}
{\cal M}^2=v^2
\begin{pmatrix}
\lambda_1c_\beta^2+\nu s_\beta^2 & (\lambda_{345} - \nu)c_\beta s_\beta & -\frac{1}{2}{\rm Im}\lambda_5 \, s_\beta \\
(\lambda_{345} - \nu)c_\beta s_\beta & \lambda_2 s_\beta^2+\nu c_\beta^2 & -\frac{1}{2}{\rm Im}\lambda_5\, c_\beta\\
-\frac{1}{2}{\rm Im}\lambda_5\, s_\beta & -\frac{1}{2}{\rm Im}\lambda_5\, c_\beta & -{\rm Re}\lambda_5+\nu
\end{pmatrix} \ ,
\end{equation}
where $\lambda_{345}$ represents $\lambda_3+\lambda_4+{\rm Re}(\lambda_5)$. A rotation matrix $R$ defined below can be used to diagonalize the mass matrix:
\begin{eqnarray}\label{R}
R  =\begin{pmatrix}
-s_{\alpha}c_{\alpha_b} & c_{\alpha}c_{\alpha_b} & s_{\alpha_b} \\
s_{\alpha}s_{\alpha_b}s_{\alpha_c} - c_{\alpha}c_{\alpha_c} & -s_{\alpha}c_{\alpha_c} - c_{\alpha}s_{\alpha_b}s_{\alpha_c} & c_{\alpha_b}s_{\alpha_c} \\
s_{\alpha}s_{\alpha_b}c_{\alpha_c} + c_{\alpha}s_{\alpha_c} & s_{\alpha}s_{\alpha_c} - c_{\alpha}s_{\alpha_b}c_{\alpha_c} & c_{\alpha_b}c_{\alpha_c}
\end{pmatrix} \ ,
\end{eqnarray}
where $s_\alpha$ and $c_\alpha$ are short hands for $\sin\alpha$ and $\cos\alpha$. Under this rotation matrix, we have ${\cal M}^2=R^T{\rm diag}(m_{h_1}^2,m_{h_2}^2,m_{h_3}^2)R$, and $R(H_1^0,H_2^0,A^0)^T=(h_1,h_2,h_3)^T$. We demand that three rotation angles are in the following range:
\beqa
-\frac{\pi}{2}<\alpha,\alpha_b,\alpha_c <\frac{\pi}{2}
\eeqa
With this diagonalization procedure, one can obtain six linearly independent equations which can be solved for the parameters in the scalar potential in terms of the physical parameters, as shown below~\cite{Inoue:2014nva},
\begin{eqnarray}
\label{eq:lambda1}
\lambda_1 &=& \frac{m_{h_1}^2 \sin^2\alpha \cos^2\alpha_b + m_{h_2}^2 R_{21}^2 
+ m_{h_3}^2 R_{31}^2}{v^2 \cos\beta^2} - \nu \tan^2\beta \ , \\
\lambda_2 &=& \frac{m_{h_1}^2 \cos^2\alpha \cos^2\alpha_b + m_{h_2}^2 R_{22}^2 
+ m_{h_3}^2 R_{32}^2}{v^2 \sin\beta^2} - \nu \cot^2\beta \ , \\
{\rm Re}\lambda_5 &=& \nu - \frac{m_{h_1}^2 \sin^2\alpha_b + \cos^2\alpha_b (m_{h_2}^2 \sin^2\alpha_c + m_{h_3}^2 \cos^2\alpha_c)}{v^2} \ , \\
\lambda_3 &=& \nu - \frac{m_{h_1}^2 \sin\alpha \cos\alpha \cos^2\alpha_b - m_{h_2}^2R_{21}R_{22} - m_{h_3}^2R_{31}R_{32}}{v^2\sin\beta\cos\beta} - \lambda_4 - {\rm Re}\lambda_5 \ , \\  \nonumber
{\rm Im}\lambda_5 &=& \frac{2 \cos\alpha_b \left[ (m_{h_2}^2-m_{h_3}^2) \cos\alpha \sin\alpha_c \cos\alpha_c +  (m_{h_1}^2 - m_{h_2}^2 \sin^2\alpha_c-m_{h_3}^2\cos^2\alpha_c)^2 \sin\alpha \sin\alpha_b \right]}{v^2 \sin\beta} \label{eq:imlambda5}  \ , \\  \\
\tan\beta &=& \frac{(m_{h_2}^2 -m_{h_3}^2) \cos\alpha_c \sin\alpha_c + (m_{h_1}^2 -m_{h_2}^2 \sin^2\alpha_c-m_{h_3}^2 \cos^2\alpha_c) \tan\alpha \sin\alpha_b}
{(m_{h_2}^2 -m_{h_3}^2) \tan\alpha \cos\alpha_c \sin\alpha_c - (m_{h_1}^2 -m_{h_2}^2 \sin^2\alpha_c-m_{h_3}^2 \cos^2\alpha_c) \sin\alpha_b} \ .\label{ab}
\end{eqnarray}
The last equation relates the two CPV angles, $\alpha_c$ and $\alpha_b$, and indicates that there exists only one independent CPV phase in our model. Using Eq.~(\ref{MC}) and the minimization condition Eq.~(\ref{mini}) we obtain the full relationships between model parameters ($\lambda_1$, $\lambda_2$, $\lambda_3$, $\lambda_4$, ${\rm Re}\lambda_5$, ${\rm Im}\lambda_5$, $m_{11}^2$, $m_{22}^2$, ${\rm Re}m_{12}^2$, ${\rm Im}m_{12}^2$) and phenomenological parameters ($v$, {$\tan\beta$}, $\nu$, $\alpha$, $\alpha_b$, $\alpha_c$,  $m_{h_1}$, $m_{h_2}$, $m_{h_3}$, $m_{H^+}$). Through Eq.~(\ref{ab}), one can solve for the angle $\alpha_b$ in terms of $\alpha_c$,
\begin{eqnarray}\label{ac2ab}
\alpha_b =- \arcsin \left[\frac{(m_{h_2}^2-m_{h_3}^2) \sin2\alpha_c\cot(\beta+\alpha)}{2(m_{h_1}^2 - m_{h_2}^2 \sin^2\alpha_c-m_{h_3}^2 \cos^2\alpha_c)} \right] \ .
\end{eqnarray}
Conversely, one could obtain the formula for $\alpha_c$ in terms of $\alpha_b$. However, two solutions will be generated when solving the second order equation for $\tan\alpha_c$. Here we adopt the convention in Ref.~\cite{Chen:2015gaa},
\begin{eqnarray}          \label{ac}              
\alpha_c = \left\{\begin{array}{ll} 
\alpha_c^-, & \hspace{0.3cm} {\rm \alpha+\beta\leq0} \\
\alpha_c^+, & \hspace{0.3cm} {\rm \alpha+\beta>0} 
\end{array} \right., \hspace{0.6cm}
\tan\alpha_c^\pm\!=\!\frac{\mp| \sin\alpha_b^{\rm max}| \!\pm\! \sqrt{ \sin^2\alpha_b^{\rm max} - \sin^2\alpha_b }}{\sin\alpha_b}
\sqrt{\frac{m_{h_3}^2-m_{h_1}^2}{m_{h_2}^2-m_{h_1}^2}}\ . \hspace{-0.8cm}\nonumber \\
\end{eqnarray}
where $\sin \alpha_b^{{\rm max}}$ sets a theoretical bound on the CPV angle $\alpha_b$ which comes from the requirement of the existence of a real solution for $\tan\alpha_c$:
\begin{eqnarray}\label{prejudice}
\sin^2\alpha_b \leq \frac{(m_{h_3}^2-m_{h_2}^2)^2 \cot^2(\alpha+\beta)}{4 (m_{h_2}^2-m_{h_1}^2) (m_{h_3}^2-m_{h_1}^2)} \equiv \sin^2\alpha_b^{\rm max} \ .
\end{eqnarray}

\subsection{Interaction Terms}\label{Yukawaint}
To eliminate the tree level FCNCs, one can assign $Z_2$ charges to different fermion fields. In general, this would lead to four possible arrangements in the Yukawa sector, which are often dubbed Type-I, Type-II, Lepton-specific and Flipped 2HDMs ~\cite{Glashow:1976nt,Barger:1989fj,Branco:2011iw}. In this work, we only concentrate on the first two, since Type-I (Type-II) differs from Lepton-specific (Flipped) only in the lepton sector and they should behave similarly to the first two in our collider and EDMs experiments. Under the $Z_2$ symmetry fermion fields transform as
\beqa
Q_L\to Q_L\, \quad u_R\to u_R\, \quad d_R\to d_R, & \quad \mathrm{Type\ I}\ \ \ ,\\
Q_L\to Q_L\, \quad u_R\to u_R\, \quad d_R\to -d_R, & \quad \mathrm{Type\ II}\ \ \ .
\eeqa
The corresponding Yukawa interactions invariant under the $Z_2$ symmetry are:
\begin{eqnarray}\label{cff}
\mathcal{L}_{\rm I} = - Y_U \overline Q_L (i\tau_2) \phi_2^* u_R - Y_D \overline Q_L \phi_2 d_R + {\rm h.c.} \ , \\
\mathcal{L}_{\rm II} = - Y_U \overline Q_L (i\tau_2) \phi_2^* u_R - Y_D \overline Q_L \phi_1 d_R + {\rm h.c.} \ .
\end{eqnarray}
The interaction of the physical Higgs with fermions and with vector bosons can be parametrized as
\begin{eqnarray}\label{Lcpv}
\mathcal{L}_{int} = - \frac{m_f}{v} h_i  \left( c_{f,i} \bar f f+ \tilde c_{f,i} \bar f i\gamma_5 f  \right) + a_i h_i \left( \frac{2m_W^2}{v} W_\mu W^\mu + \frac{m_Z^2}{v} Z_\mu Z^\mu \right) \ ,
\end{eqnarray}
where $c_{f,i}(\tilde{c}_{f,i})$ represents the scalar (pseudo-scalar) component of the physical Higgs $h_i$ coupling to fermions while $a_i$ stands for the coefficient of $h_i$ coupling to the vector bosons.  Analytic expressions for these coefficients are given in terms of the phenomenological parameters in Table~\ref{Hcouplings}.
\begin{table}[h]
\centering{\begin{tabular}{ c c c c c c }
\hline
\hline
&  $c_{t,i}$ & $c_{b,i}=c_{\tau,i}$ & $\tilde c_{t,i}$ & $\tilde c_{b,i}=\tilde c_{\tau,i}$ & $a_i$ \\

Type I & $R_{i2}/\sin\beta$ & $R_{i2}/\sin\beta$ & $-R_{i3}\cot\beta$ & $R_{i3}\cot\beta$ & 
$R_{i2}\sin\beta+R_{i1}\cos\beta$ \\

Type II & $R_{i2}/\sin\beta$ & $R_{i1}/\cos\beta$ & $-R_{i3}\cot\beta$ & $-R_{i3}\tan\beta$ & 
$R_{i2}\sin\beta+R_{i1}\cos\beta$ \\
\hline
\end{tabular}} .\ \ \ 
\caption{Couplings to Higgs mass eigenstates.}\label{Hcouplings}
\end{table}
Higgs global fits to the CP conserving 2HDM from current LHC measurements indicate that the couplings are close to the alignment limit: $\beta-\alpha=\pi/2$~\cite{ATLAS:2014kua,CMS:2016qbe}. 

Hence, we concentrate on the region having only small deviations from this limit in our study. 
The interaction between the heavy Higgses, SM Higgs and $Z$ bosons can be parametrized in the following form:
\begin{eqnarray}\label{Lhizh}
\mathcal{L}_{h_i\to Zh_1} = g_{iz1}Z^\mu(\partial_\mu h_i  h_1-h_i  \partial_\mu h_1) \ ,
\end{eqnarray}
with the coefficient $g_{iz1}$ expressed as:
\beqa \label{giz1}
g_{iz1}=\frac{e}{\sin2\theta_W}\left[(-\sin\beta R_{11}+\cos\beta R_{12}) R_{i3} - (-\sin\beta R_{i1}+\cos\beta R_{i2}) R_{13} \right] . \
\eeqa
We parametrize the deviation from the alignment limit by a small variable $\theta$ where $\beta-\alpha=\pi /2+\theta$. Then we expand coupling $g_{iz1}$ in the limits of small $\alpha_b$ (CPV angle) and $\theta$ , which gives,
\beqa
g_{2z1}&\propto &-\alpha_b+O(\alpha_b\theta)\label{g2z1} \\
g_{3z1}&\propto &-\theta+O(\alpha_b^2)\label{g3z1}
\eeqa
Thus, near the alignment limit, the decay $h_2\to Z h_1$ could occur only if $\alpha_b\not=0$, assuming it is kinematically allowed. In contrast, the decay $h_3\to Z h_1$ could arise even in the $\alpha_b=0$ limit so long as there exists a departure from exact alignment.
Consequently, one may interpret null results of any search for a heavy scalar decaying to a $Z$-boson and a SM-like Higgs boson in terms of constraints on either $\alpha_b$ or $\theta$. In what follows we will, thus, consider the present and prospective constraints on $\alpha_b$ in two cases: $\theta = 0$ and $\theta\not=0$.

\section{Production and Decay of Heavy Higgs\label{prodec}}
\subsection{Production of Heavy Higgs\label{proh}}
At the LHC, the dominant production mode for a heavy Higgs is via gluon fusion. Therefore, we restrict our study on this specific production mode. The one loop gluon fusion production cross-section of a heavy Higgs is obtained by rescaling the value of the production cross-section for the SM-like Higgs:
\begin{eqnarray}\label{ggHXsec} \nonumber
\sigma (gg\to h_i)= \sigma (gg\to H_{\rm SM})
\frac{\left|c_{t,i} F_{1/2}^H(\tau^i_{t}) + c_{b,i} F_{1/2}^H(\tau^i_{b})\right|^2 + \left|\tilde c_{t,i} F_{1/2}^A(\tau^i_{t}) + \tilde c_{b,i} F_{1/2}^A(\tau^i_{b})\right|^2}{\left|F_{1/2}^H(\tau^i_{t}) + F_{1/2}^H(\tau^i_{b})\right|^2}, \\
\end{eqnarray}
with $\tau^i_f=m_{h_i}^2/(4m_f^2)$, the ratio of the mass squared of the heavy Higgs ($h_i$) to 4 times the mass squared of the fermion running in the loop. Here, $\sigma (gg\to H_{\rm SM})$ represents the gluon fusion production cross-section of a heavy Higgs with SM couplings. The functions $F_{1/2}^H$ and $F_{1/2}^A$ are defined in the following:
\begin{eqnarray}\label{formfactors}
\label{fh}
F_{1/2}^H(\tau) &=& 2\left(\tau +(\tau-1) f(\tau) \right) \tau^{-2} \ ,  \\
\label{fa}
F_{1/2}^A(\tau) &=& 2 f(\tau) \tau^{-1} \ ,  \\
\label{ftau}
f(\tau) &=& \left\{ \begin{array}{ll}
{\rm arcsin}^2\left( \sqrt{\tau} \right), & \hspace{0.5cm} \tau\leq 1 \\
\frac{1}{4} \left[ \log\left( \frac{1+\sqrt{1-\tau^{-1}}}{1-\sqrt{1-\tau^{-1}}} \right) -i\pi \right]^2, & \hspace{0.5cm} \tau> 1
\end{array}\right.\, .
\end{eqnarray}
As one can see from Eq.~(\ref{ggHXsec}), the numerator involves the sum of two contributions arising from the CP-odd and CP-even components of the physical Higgs boson, respectively. Denoting ${\cal M}^{gg\to h_i}_{\mathrm{CP-odd}}$ and ${\cal M}^{*gg\to h_i}_{\mathrm{CP-even}}$ as the CP-odd and CP-even parts of the gluon fusion matrix elements, we see that
the interference term ${\cal M}^{gg\to h_i}_{\mathrm{CP-odd}}{\cal M}^{*gg\to h_i}_{\mathrm{CP-even}}$ vanishes after integrating over final state phase space due to parity.
The heavy Higgs production cross-section in this form automatically takes into account the K-factor, if one uses the production $\sigma (gg\to H_{\rm SM})$ with higher order corrections. Here we obtain the values of $\sigma (gg\to H_{\rm SM})$ from the website~\cite{HXsecpage}. 

\subsection{Decay of Heavy Higgs\label{dech}}
The dominant two body decay modes of the heavy Higgses are taken into account with $\Gamma_\mathrm{tot}$ expressed in the following form:

\begin{eqnarray}
\Gamma_\mathrm{tot}(h_i) &=& \Gamma(h_i \to gg) + \Gamma(h_i \to Z h_1) + \Gamma(h_i \to W^+W^-) + \Gamma(h_i \to ZZ) + \Gamma(h_i \to t\bar t) \nonumber \\
&+&\Gamma(h_i \to b\bar b) + \Gamma(h_i \to \tau^+ \tau^-) +  \Gamma(h_i \to h_1 h_1) +\cdots \ ,
\end{eqnarray}
where the \lq\lq $+\cdots$ denote the tiny decay rates to a pair of light fermions and photons, and $Z$ boson and photon, which we have neglected. In addition, we ignore decay rate of a heavy Higgs to one SM-like Higgs and another heavy Higgs,  as well as a pair of heavy Higgses  because they are forbidden by kinematics due to the mass hierarchy we choose in our benchmark model. The analytical expression for each two-body decay rate can be found in the Appendix ~\ref{APPB}.

\section{Simulation detail\label{simulation}}
In this section, we will discuss details of our collider simulation. We first reproduce the result of 8 TeV ATLAS exclusion limit on $\sigma(gg\to h_i)\times \mathrm{Br}(h_i\to Zh_1)\times \mathrm{Br}(h_1\to b\bar{b})$ obtained by searching for a heavy Higgs $h_{i=2,3}$ decaying to $Z(\ell^+\ell^-)h_1(b\bar{b})$~\cite{Aad:2015wra} (As in Ref.~\cite{Aad:2015wra} we do not include a $\mathrm{Br}(Z\to\ell^+\ell^-)$ factor because it is assumed to have the SM value) . We then use a BDT method to perform events classification and derive the projected exclusion limit for a future 14 TeV search. Events are generated by \texttt{MadGraph 5 aMC@NLO}~\cite{Alwall:2014hca} and then passed through PYTHIA6~\cite{Sjostrand:2006za} for parton showering. Finally Delphes3~\cite{deFavereau:2013fsa} is used for fast detector simulation.

\subsection{8 TeV Result Reproduction}\label{8TEVANA}
We use the cuts described in Ref.~\cite{Aad:2015wra} as follow:
\begin{itemize}
\item The events must have 2 electrons or 2 opposite charged muons with $p^{e,\mu}_T>7$ GeV and $|\eta_e|(|\eta_\mu|)<2.5(2.7)$
\item The leptons must have $p^{e,\mu}_{T,lead}>25$ GeV, and if the leptons are $\mu^+\mu^-$ pairs, then one of the $\mu$ must satisfy $|\eta_\mu|<2.5$
\item The events must have exactly 2 tagged b-jets with $p^{\rm lead}_{b,T}>45$ GeV and $p^{\rm sub}_{b,T}>20$ GeV
\item The reconstructed invariant mass for dilepton and dijet systems should satisfy: $83<m_{\ell\ell}<99$ GeV and $95<m_{bb}<135$ GeV
\item $E^{\rm miss}_T/\sqrt{H_T}<3.5\ {\rm GeV}^{1/2}$ where $H_T$ is defined as the scalar sum of all jets and leptons in the events
\item $p^{\rm Z}_T>0.44\times M_{h{2,3}}^{\rm rec}-106$ GeV where $M_{h{2,3}}^{\rm rec}$ is the reconstructed mass of heavy Higgs.
\end{itemize}
For the detector simulation, we use the default Delphes ATLAS cards with following modifications. The values are modified to be consistent with those used in the ATLAS analysis~\cite{Aad:2015wra}:
\begin{itemize}
\item The isolation conditions for leptons: 

Change DeltaRMax from 0.5(default) to 0.2; Change PTMin from 0.5(default) to 0.4(1) for electron(muon); Change PTRatio from 0.1(default) to 0.15. These changes will increase the lepton identification in the boosted regime.
\item Change the ParameterR for jet-clustering(anit-kt) algorithm from 0.6   to 0.4.
\end{itemize}
For the signal process, we only take into account the gluon fusion production mode of the heavy Higgs. As for the background processes, we consider the two major backgrounds $Z bb$ and $t\bar{t}$ as well as to sub-leading backgrounds SM $Zh$ and diboson $ZZ$ backgrounds. For all the backgrounds, we generate events with one additional jet with jet matching. The numbers of events generated and the corresponding acceptance times efficiency are given in  Table~\ref{table:BGtable8}. The cross-sections are normalized to the values with higher order corrections. The K-factors for $Zbb$ , $t\bar{t}$, $Zh$, $ZZ$ are calculated based on the result in Ref.~\cite{Cordero:2009kv,Czakon:2013goa,ZhXsec8TeV,ATLAS:2013gma}. One can observe that the $Z(\ell\ell)bb$ background is a bit larger than the ATLAS result in Table~\ref{table:BGtable8}. This maybe due to the fact that ATLAS used a data-
driven method to estimate the number of $Z(\ell\ell)bb$ background events, which may include some effects that our fast detector simulation cannot fully replicate. However, one can also see that these kinds of effects are at a controllable level; our simulation result agrees with ATLAS results within at most 20\% uncertainty. Since we may also expect the same kind of effect in 14 TeV simulations,  our projected exclusion limit result will be conservative. 
\begin{table}[h]
\begin{tabular}{ c c c c c c }
\hline
\hline
\tabincell{c}{Backgrounds/\\Signal} & $\sigma$(pb) & \ $\sigma\times \int {\cal L}$ & \tabincell{c}{simulated \# of \\ events after cuts}  &\tabincell{c}{\# of expected \\ event in Ref.~\cite{Aad:2015wra}}& A $\times$ $\epsilon$  \\
\hline
$Z(\ell\ell)bb$ & 12.91 & 2.620$\times 10^5$ & 1,788 & 1443$\pm$60  & 6.825$\times 10^{-3}$  \\

$t(bl\nu)\bar{t}(bl\nu)$ & 18.12 & 3.678$\times 10^5$ & 359 & 317$\pm$28 & 9.761$\times 10^{-4}$  \\

${\rm SM} \ Z(\ell\ell)h(bb)$ & 0.02742 & 5.566$\times 10^2$ & 47 & 31$\pm$1.8 & 8.443$\times 10^{-2}$  \\

${\rm Diboson}(Z(\ell\ell)Z(bb))$ & 0.2122 & 4.308$\times 10^3$ & 28 & 30$\pm$5 & 6.679$\times 10^{-3}$  \\

${\rm Signal}(500\ GeV)$ & 0.03 & 4.06$\times 10^2$ & 54 & - & 1.332$\times 10^{-1}$  \\
\hline
\end{tabular}

\caption{Summary of the 8 TeV simulation. The second column gives the cross-section of each background process at 8 TeV LHC with generator level cuts. The signal distributions are normalized to 0.03 pb as suggested in Ref.~\cite{Aad:2015wra}. The third column is the total number of events produced at 8TeV LHC with the integrated luminosity equal to 20.3 fb$^{-1}$. The fourth column is the number of events left for each background after all the cuts with the integrated luminosity equal to 20.3 fb$^{-1}$. The fifth column gives the number of events left with the same cuts estimated by ALTAS in Ref.~\cite{Aad:2015wra}. The last column gives the acceptance times the efficiency after all the cuts obtained by our simulation.}
\label{table:BGtable8}
\end{table}

\begin{figure}
\centering     
\subfigure[ $\ M^{\rm rec}_{h_2}$ Distribution]{\label{fig:mllbb8}\includegraphics[width=80mm]{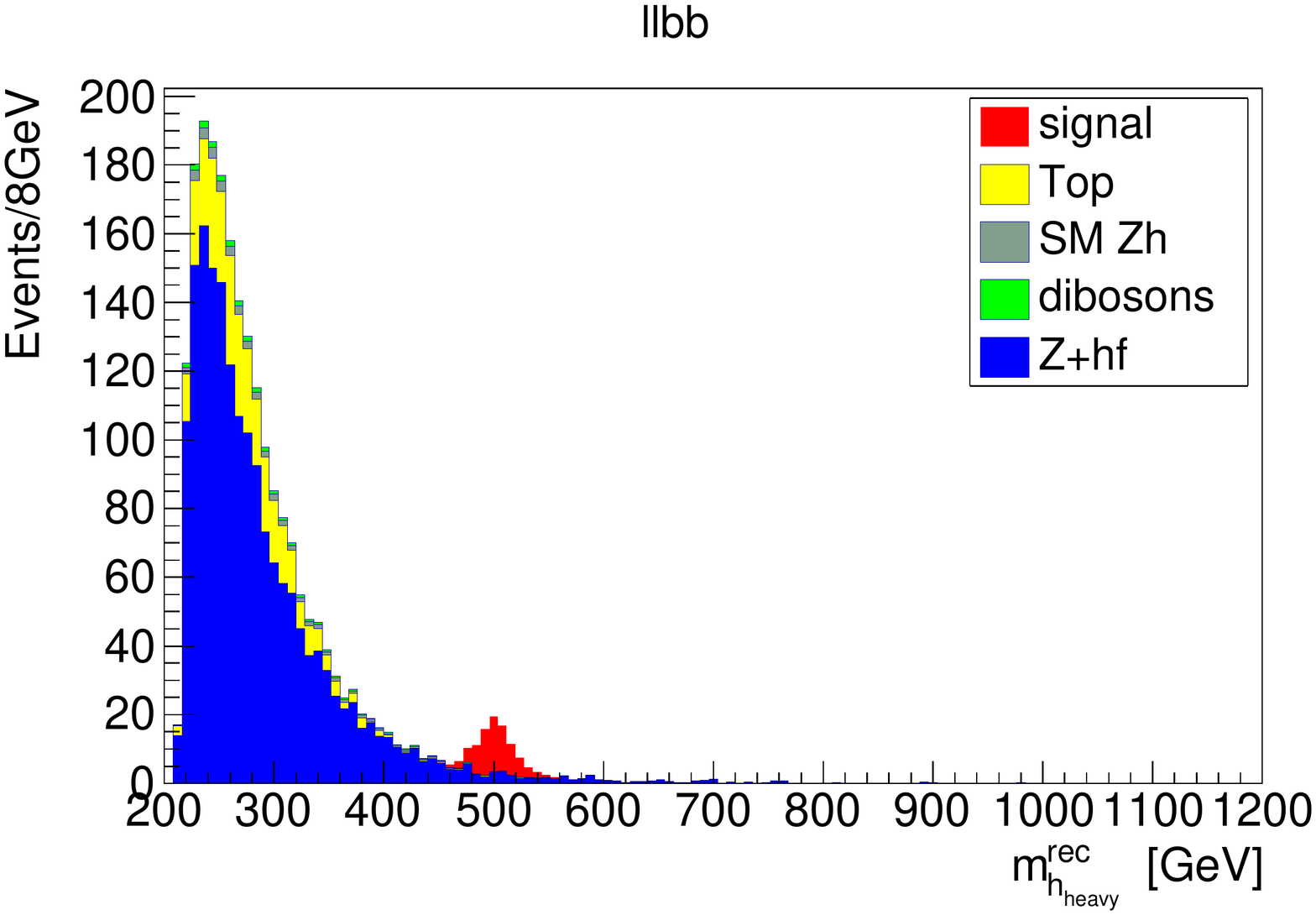}}
\subfigure[\ 95\% CLs exclusion limit]{\label{fig:exclusion8}\includegraphics[width=70mm]{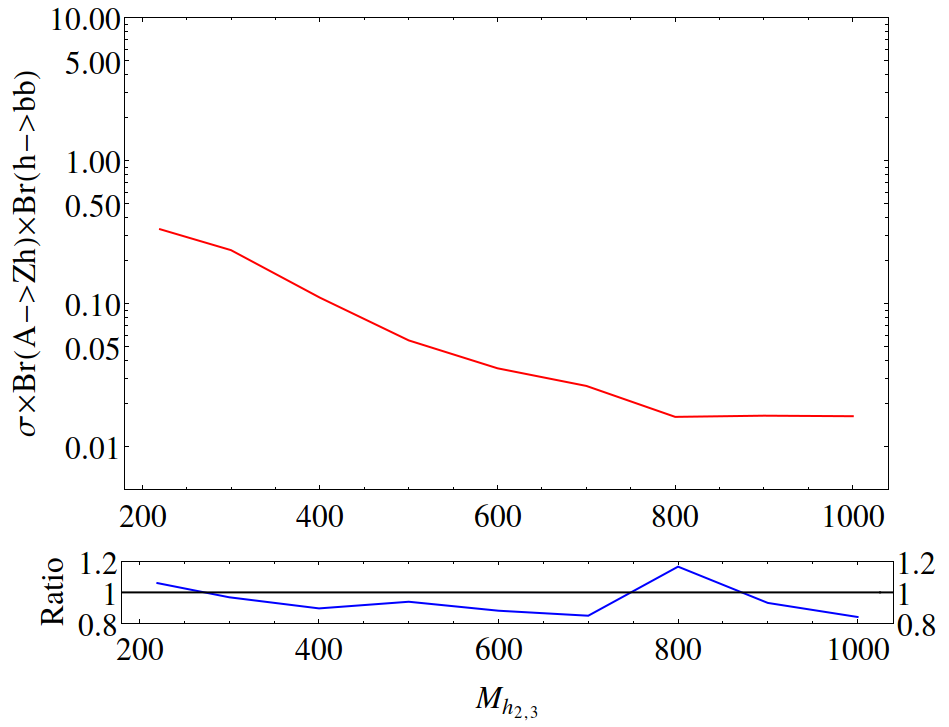}}
\caption{Fig.(a) shows the reconstructed invariant mass distributions with $\ell^+\ell^- b \bar{b}$ final state. The signal comes from a heavy Higgs of mass 500 GeV and production cross-section 0.03 pb with an integrated luminosity 20.3 fb$^{-1}$. Fig.(b) demonstrates the 95\% exclusion limit on the signal $\sigma(gg\to A)Br(A\to Zh)Br(h\to bb)$. The red curve is our result using the distribution in Fig.(a) with profile likelihood method while the blue curve is the ratio of the ATLAS result (in Fig.3(b) of Ref.~\cite{Aad:2015wra}) to our reproduced expected exclusion limit. }
\end{figure}

We present the reconstructed invariant mass of the heavy Higgs in Fig.~\ref{fig:mllbb8} which can be compared with the ATLAS result in Fig.3(b) in Ref. ~\cite{Aad:2015wra}. With this binned distribution we use a profile likelihood method as used in the ATLAS paper to reproduce the 95\% CLs exclusion limit.  A comparison with ATLAS result is given in Fig.~\ref{fig:exclusion8}, the red curve is our reproduced exclusion limit, and the blue curve is the ratio of the ATLAS results to our reproduced values. One can see that, the ratio is generally less than one which corresponds to the excess of $Z(\ell\ell)bb$ background in our simulation. The peak at 800 GeV is due to the lack of background statistics and downward fluctuation near $m_{hi}=$ 800 GeV, but for our benchmark model where $m_{h1}=550, m_{h2}=600$ GeV, the ratio seems reasonably close to one.

\subsection{14 TeV Prediction}\label{sec:14Pre}
We use the same Delphes card when generating events for the 14 TeV case. The preselection cuts we use are almost the same as those for the 8 TeV case. In order to get a sufficiently large sample for BDT analysis, we expand the mass window for $m_{bb}$ from $95\sim 135$ GeV to $60\sim 140$ GeV. Also, rather than implementing the $E^{miss}_T/\sqrt{H_T}$ and $p^Z_T$ cuts, we allow the BDT to optimize them. The numbers of events generated and the acceptance times efficiency after preselection for signal and backgrounds are given in Table~\ref{table:BGtable14}.
\begin{table}[h]
\centering{\begin{tabular}{ c c c c c c }
\hline
\hline
\tabincell{c}{Backgrounds/\\Signal} & $\sigma$(pb) & \# of events generate & \tabincell{c}{\# of events remaining \\after cuts} & Preselsction Efficiency  \\
\hline
$Z(\ell\ell)bb$ & 36.57 & 7.084$\times 10^6$ & 94,323 & 1.331$\times 10^{-2}$   \\

$t(bl\nu)\bar{t}(bl\nu)$ & 68.11 & 3.276$\times 10^7$ & 120,627 & 3.680$\times 10^{-3}$  \\

${\rm SM} \ Z(\ell\ell)h(bb)$ & 0.0502 & 1.429$\times 10^5$ & 14,380 & 1.006$\times 10^{-1}$  \\

${\rm Diboson}(Z(bb)Z(\ell\ell))$ & 0.3833 & 1.780$\times 10^6$ & 80,887 & 4.554$\times 10^{-3}$  \\

${\rm Signal}(550\ GeV)$ & 0.06 & 1.0$\times 10^5$ & 20,645 & 0.2065  \\

${\rm Signal}(600\ GeV)$ & 0.06 & 1.0$\times 10^5$ & 21,392 & 0.2139 \\
\hline
\end{tabular}} \ 
\caption{Summary of the 14 TeV simulation after preselection cuts. The second column gives the cross-sections for each background process after generator level cuts at the 14 TeV LHC. The signal distributions are normalized to 0.06 pb. The third column gives the number of events generated in our simulation. The fourth column shows the number of events left after the preselection cuts described in Sec.~\ref{sec:14Pre} before training the BDT. The last column gives the efficiency of the preselection cuts for each process.}\label{table:BGtable14}
\end{table}
After preselection, we use a built-in package in ROOT, Toolkit for Multivariate Data Analysis (TMVA)~\cite{Hocker:2007ht} and the BDT method for the classification of signal and background events. The variables used for the classification are listed below:
\beqa
{p_{T,\ell}^{\rm lead},p_{T,\ell}^{\rm sub},p_{T,b}^{\rm lead},p_{T,b}^{\rm sub},m_{\ell\ell},m_{bb},p^Z_T,p^h_T,E^{\rm miss}_T/\sqrt{H_T},\Delta R_{\ell\ell},\Delta R_{jj},\Delta R_{Zh},\Delta \phi_{Zh}},
\eeqa
where $p^{\rm lead,sub}_{T,(j,\ell)}$ represent the leading and subleading $p_T$ of leptons and jets; $m_{\ell\ell}$ and $m_{bb}$ are the invariant masses of dijet and dilepton systems, respectively; $p^{h,Z}_T$ stands for the reconstructed $p_T$ of the $Z$ boson and the SM Higgs; $E^{\rm miss}_T/\sqrt{H_T}$ is the ratio of the missing transverse energy to $\sqrt{H_T}$ defined in the previous subsection; $\Delta R_{\ell\ell,bb,Zh}$ are the angular separations of two leptons, two bjets and reconstructed $Zh$, respectively, with $\Delta R_{ab}=\sqrt{(\eta_a-\eta_b)^2+(\phi_a-\phi_b)^2}$. $\Delta \phi_{Zh}$ is the separation of the azimuthal angles between $Z$ and $h$. The distributions of these variables are shown in~\Cref{fig:MVA1,fig:MVA2,fig:MVA3,fig:MVA4} in Appendix~\ref{APPA}.

\begin{figure}
\centering     
\subfigure[ \tiny{\ BDT output for $M_{h2}=550$ GeV}]{\label{fig:BDT550}\includegraphics[width=75mm]{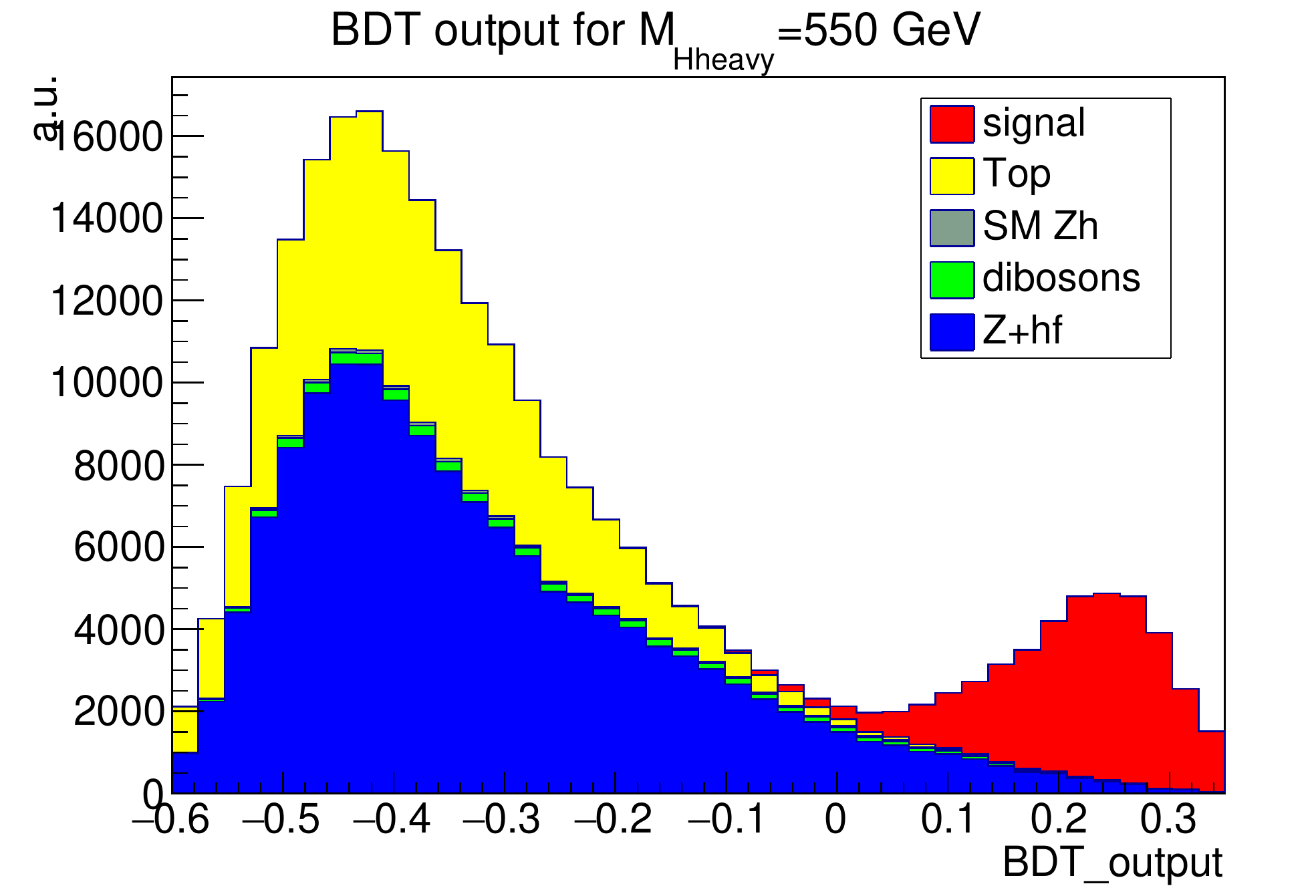}}
\subfigure[\tiny{\ BDT output for $M_{h3}=600$ GeV}]{\label{fig:BDT600}\includegraphics[width=75mm]{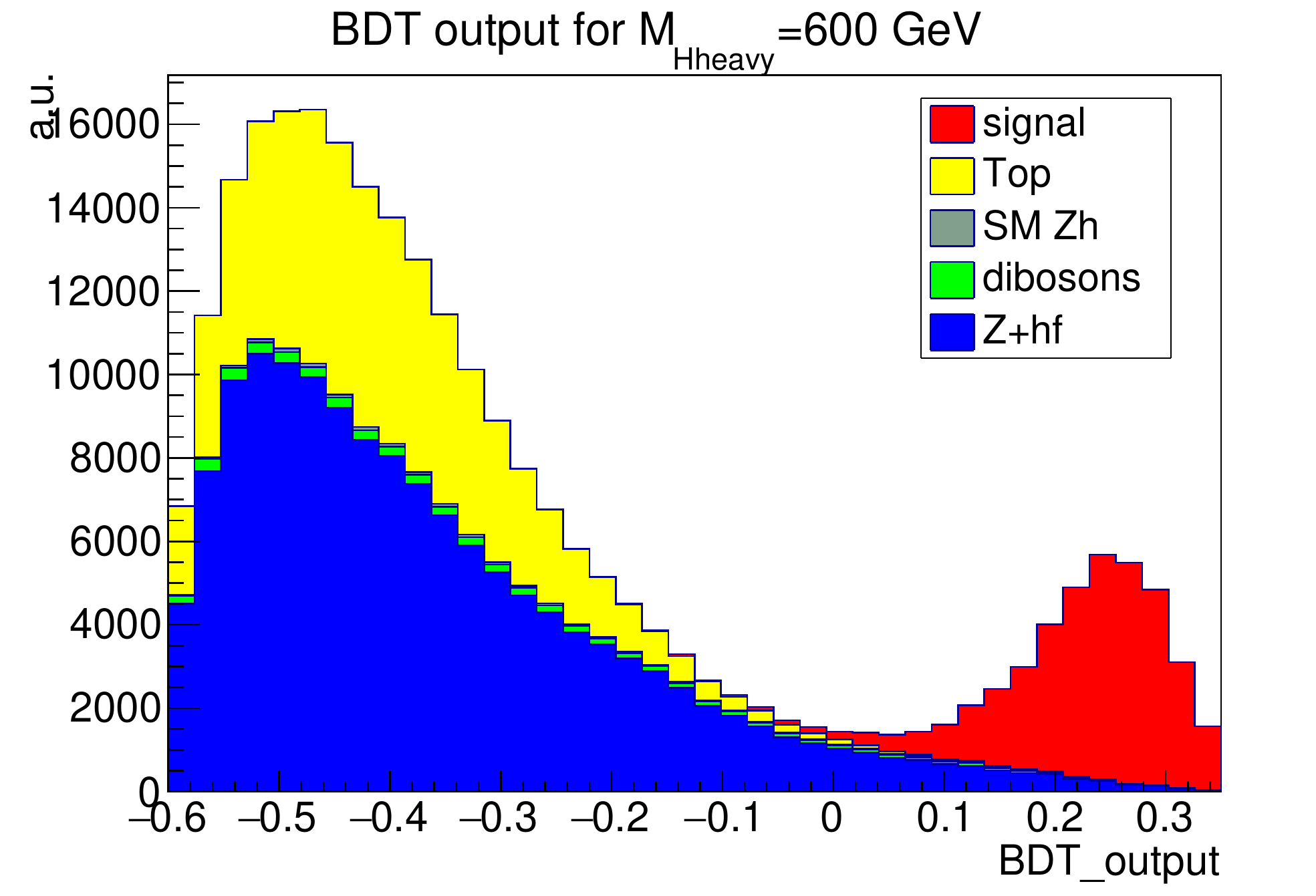}}
\caption{The BDT output distributions for both signal and backgrounds. The signals in Figs. (a) and (b) are for heavy Higgses with masses 550 and 600 GeV, respectively. The background distributions are normalized to the actual 14 TeV cross-sections in Ref.~\cite{HXsecpage}, while the signal distributions are normalized to 0.06 pb. }{\label{fig:BDTDis}}
\end{figure}

We select a representative point with $M_{h2}=550$ and $M_{h3}=600$ GeV as the signal to train the BDT. The BDT algorithm settings in TMVA are:
\beqa \nonumber
&&NTrees=850:MiniNodeSize=2.5\%:MaxDepth=3:BoostType=AdaBoost\\ \nonumber &&:AdaBoostBoostBeta=0.5:UseBaggedBoost:BaggedSampleFraction=0.5\\&&:SeparationType=GiniIndex:nCuts=20\ . \nonumber
\eeqa
The distributions of the BDT ouput for two heavy Higgs masses are shown in~\cref{fig:BDT550,fig:BDT600}. One could find that the discriminating power is a bit better for the heavier Higgs as we expected. 

The next step is to select a cut on the BDT output to obtain the most stringent 95\% exclusion limits. After applying the BDT cuts shown in Table.~\ref{table:BDT550limit} and ~\ref{table:BDT600limit}, we use the reconstructed heavy Higgs mass distribution of the remaining events to derive the 95\% exclusion limit on $\sigma(gg\to h_{2,3})Br(h_{2,3}\to Zh_1)Br(h_1\to b\bar{b})$.  We show the resulting prospective exclusion limits in \Cref{table:BDT550limit,table:BDT600limit}. We also perform a cut-based analysis with the same ATLAS cuts described in ~\ref{8TEVANA}, and the results are shown in the \lq\lq cut-based result" column in \Cref{table:BDT550limit,table:BDT600limit}. One can see that the exclusion limits of our BDT analysis are 30\% to 50\% better (lower) than the cut-based analysis results. 

\begin{table}
\begin{tabular}{ c c c c  }
 \hline
 \hline
 \multicolumn{4}{c}{$M_{h_2}=550$ GeV (14 TeV)} \\
 \hline
 Luminosity(${\rm fb}^{-1}$)& Best Cut  &exclusion limit(pb) &cut-based result\\

 100   & 0.22    &0.0299  & 0.0443\\
 300&   0.22  & 0.0167   &  0.0261\\
 3000 &0.22 & 0.00510    & 0.00782\\
 \hline
\end{tabular}\caption{Exclusion limits for $\sigma (gg\to h_{2,3})\times Br(h_{2,3}\to Zh_1)\times Br(h_1\to b\bar{b})$ and best cuts on BDT output of different luminosities for $M_{h_2}=550$ GeV. The column ``cut-based result"  gives the exclusion limit derived from the ATLAS cut-based analysis described in Section~\ref{8TEVANA}.}\label{table:BDT550limit}
\end{table}

\begin{table}
\begin{tabular}{  c c c c   }
 \hline
 \hline
  \multicolumn{4}{ c }{$M_{h_3}=600$ GeV (14 TeV)}\\
 \hline
  Luminosity(${\rm fb}^{-1}$) & Best Cut  &exclusion limit(pb)&cut-based result\\
 
   100  &   0.21  &   0.0248 &0.0340\\
     300   &0.22     &0.0138  & 0.0192\\
    3000    &  0.22    &0.00423 &0.00598\\
 \hline
\end{tabular}\caption{Exclusion limits and best cuts on BDT output of different luminosities for $M_{h_3}=600$ GeV. The column ``cut-based result" gives the exclusion limit derived from the ATLAS cut-based analysis described in ~\ref{8TEVANA}.}\label{table:BDT600limit}
\end{table}

\section{Results and Discussion\label{result}}
\begin{figure}
\centering     
\subfigure[\ Type-I Alignment limit constraint on $h_2$]{\label{fig:T1Aligh2}\includegraphics[width=70mm]{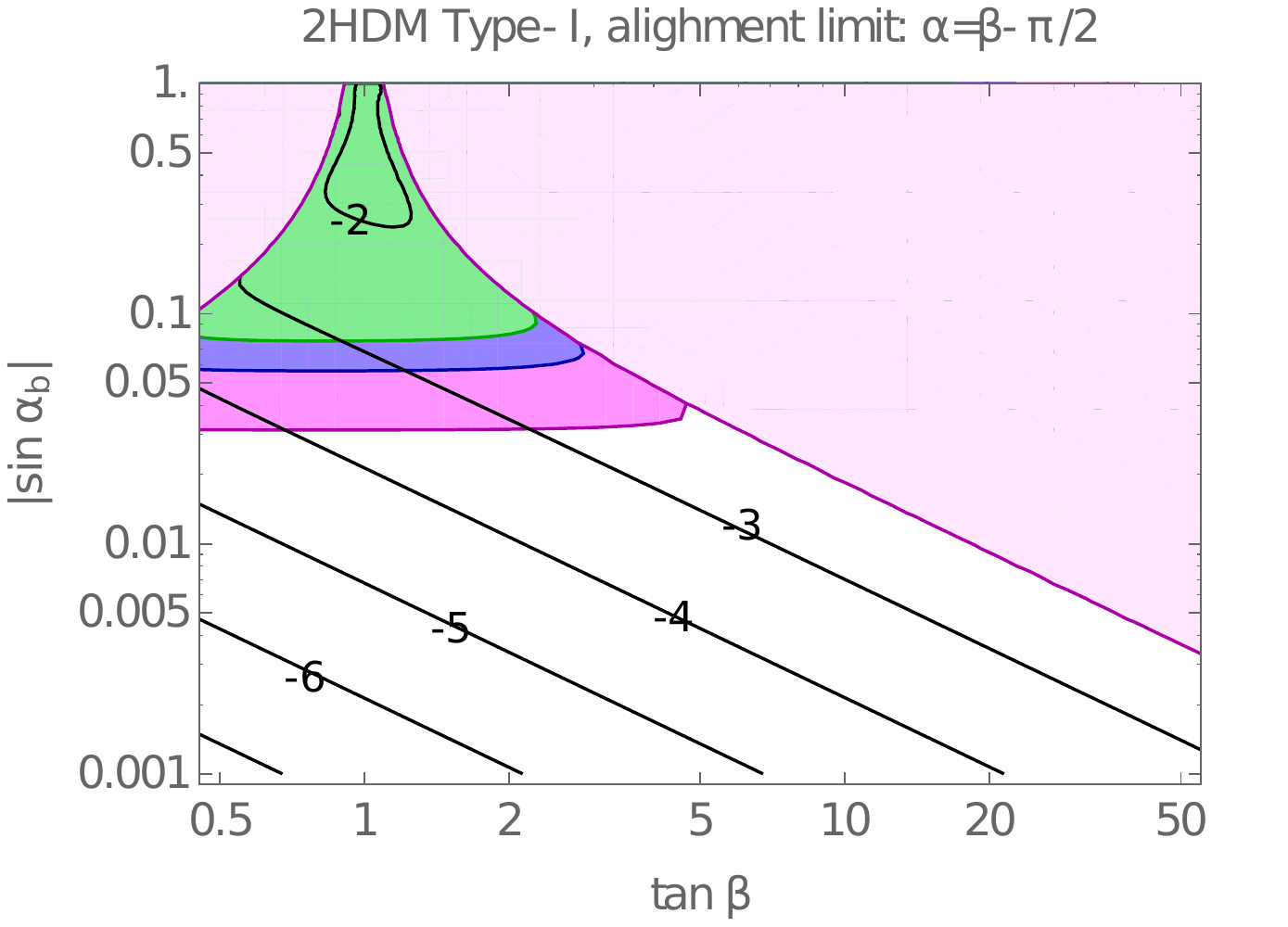}}
\subfigure[\ Type-I Alignment limit constraint on $h_3$]{\label{fig:T1Aligh3}\includegraphics[width=70mm]{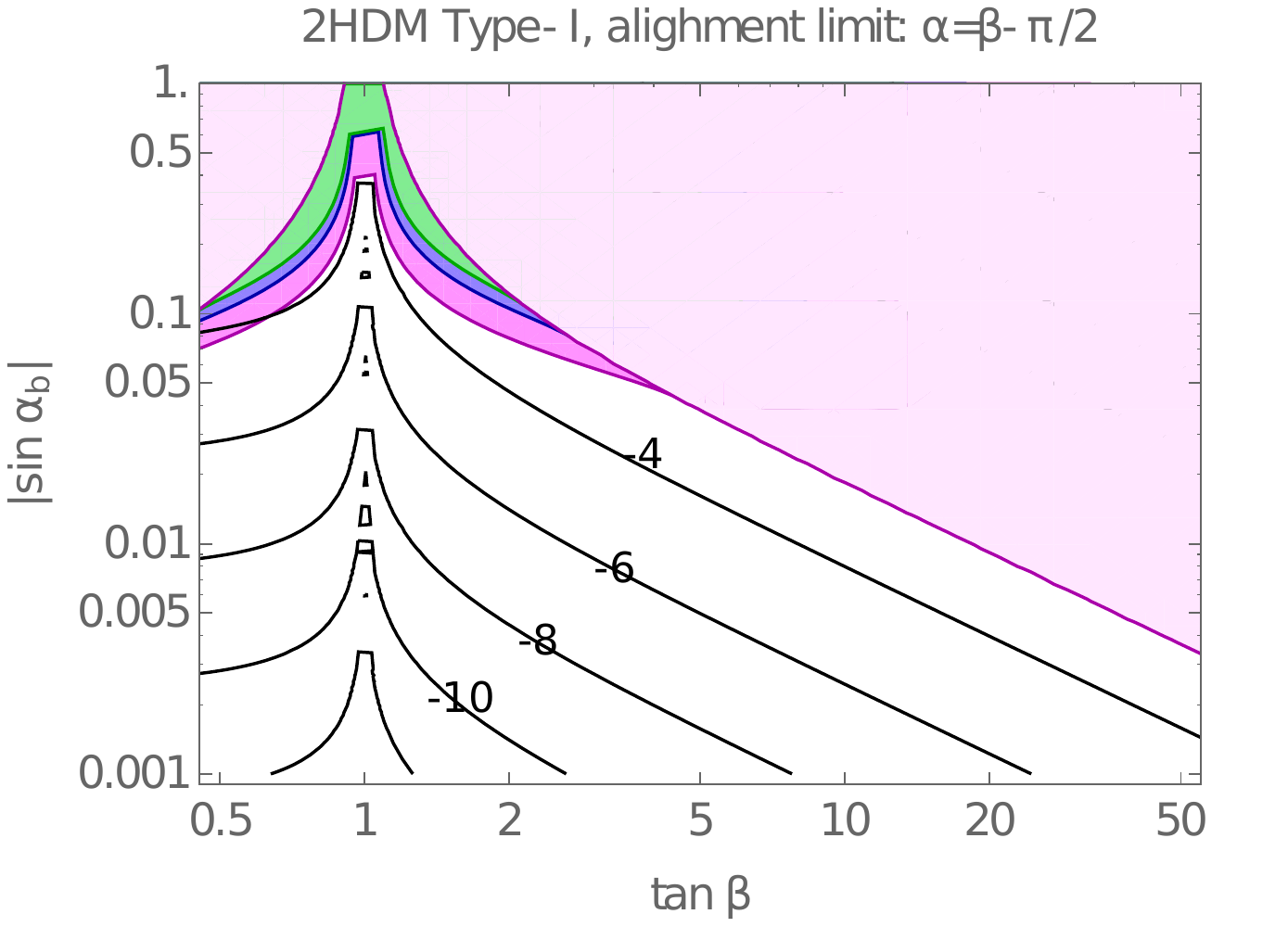}}\\
\subfigure[\ Type-II Alignment limit constraint on $h_2$]{\label{fig:T2Aligh2}\includegraphics[width=70mm]{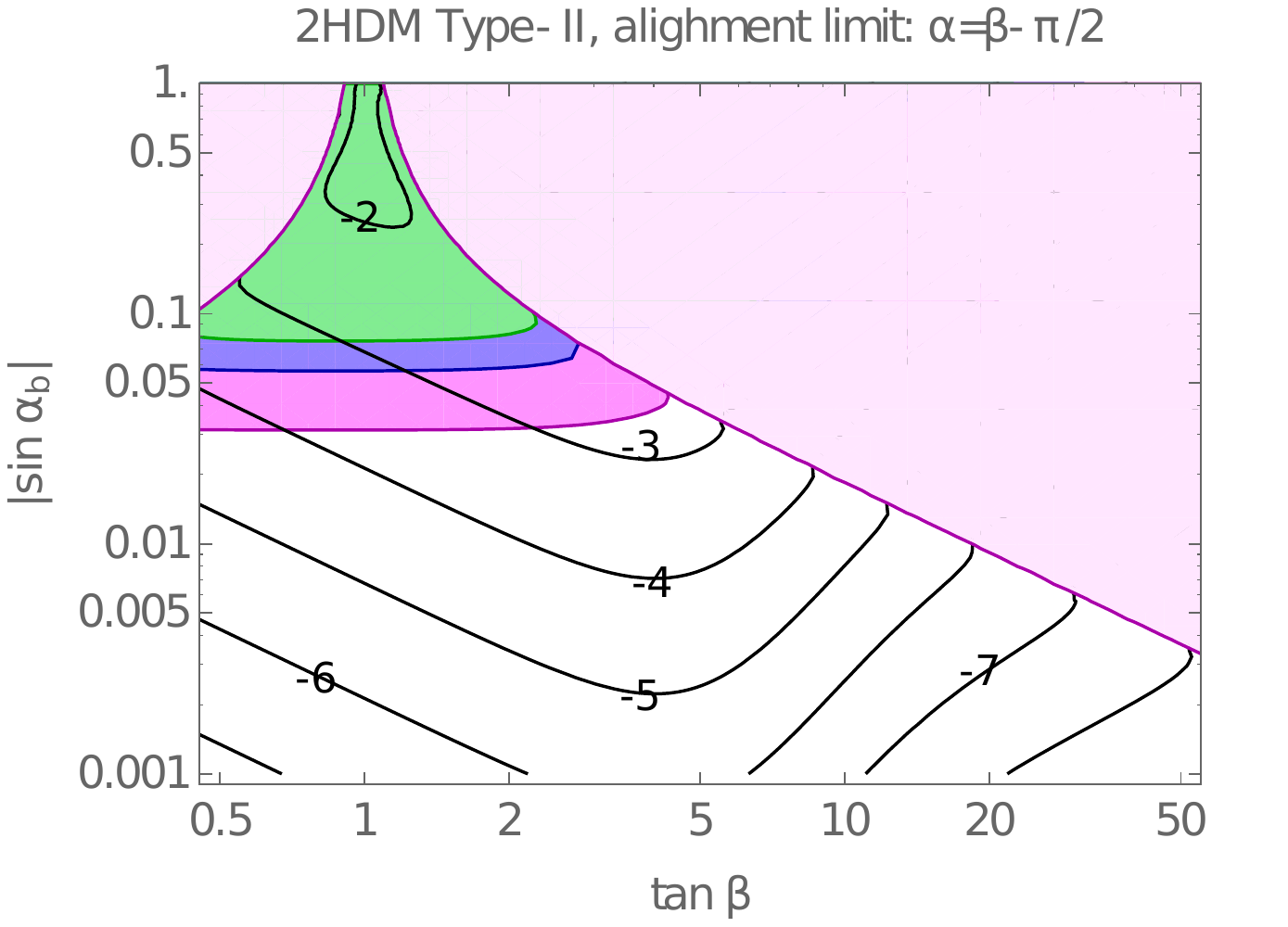}}
\subfigure[\ Type-II Alignment limit constraint on $h_3$]{\label{fig:T2Aligh3}\includegraphics[width=70mm]{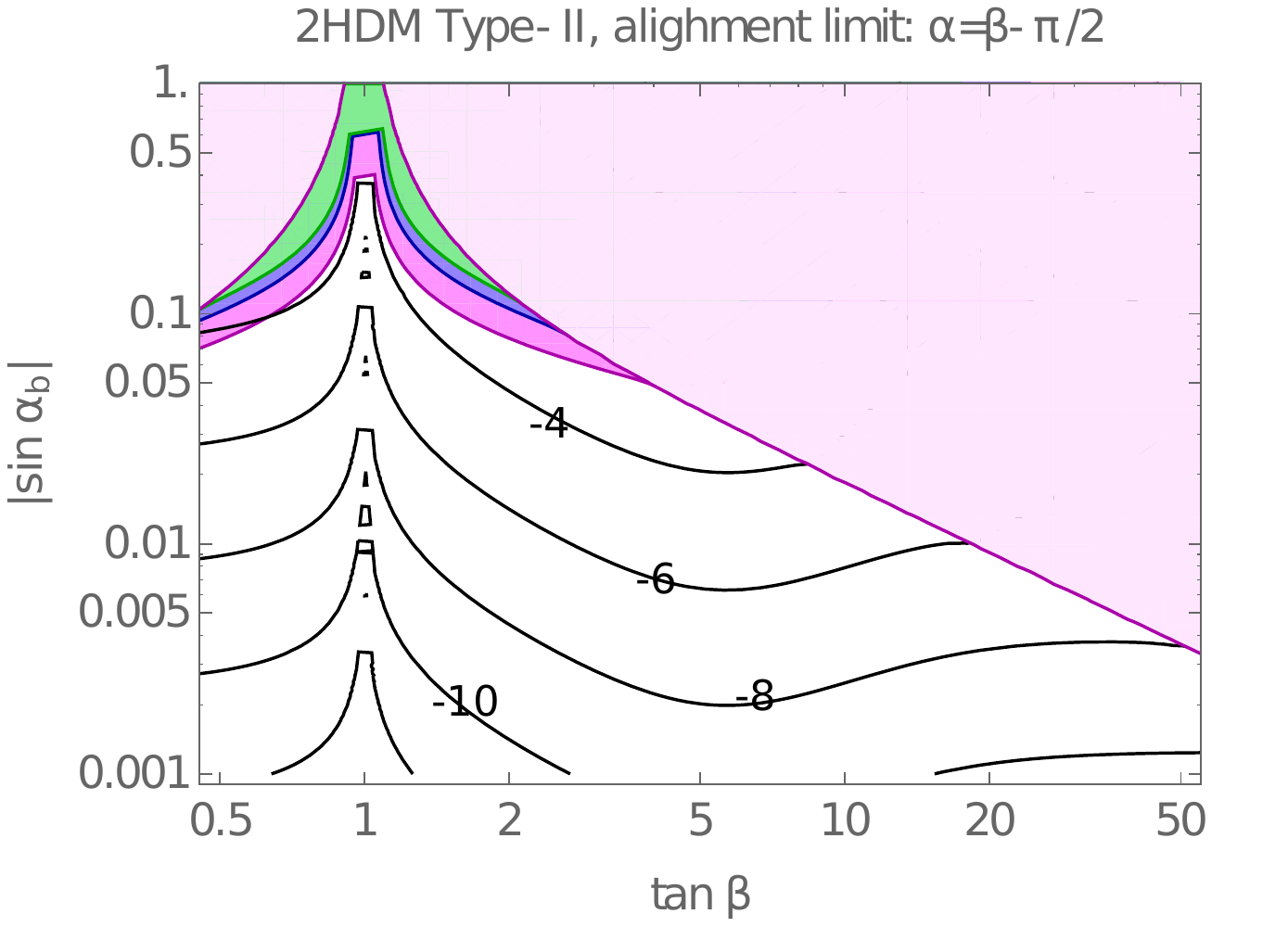}}
\caption{Exclusion limits for the heavy Higgs resonant productions in the alignment limit. The plots in the first and second row represent the constraints for the Type-I and Type-II models, respectively. The left (right) column shows the constraints from the resonant production of $h_2 (h_3)$. The pink region is theoretically inaccessible as described in the text. The green, blue, and magenta regions represent the exclusion limits for the LHC integrated luminosities equal to 100 fb$^{-1}$, 300 fb$^{-1}$ and 3000 fb$^{-1}$, respectively. The black contour represents the logarithm of $\log\overline{|M|^2_i}$ in Eq. \ref{amp2} with $s=m_{h_{2,3}}$ for $i=2,3$. }
\label{fig:Aligh2h3}
\end{figure}

\begin{figure}
\centering
\subfigure[\ Type-I constraint on $h_2$]{\label{fig:T1cbma002h2}\includegraphics[width=70mm]{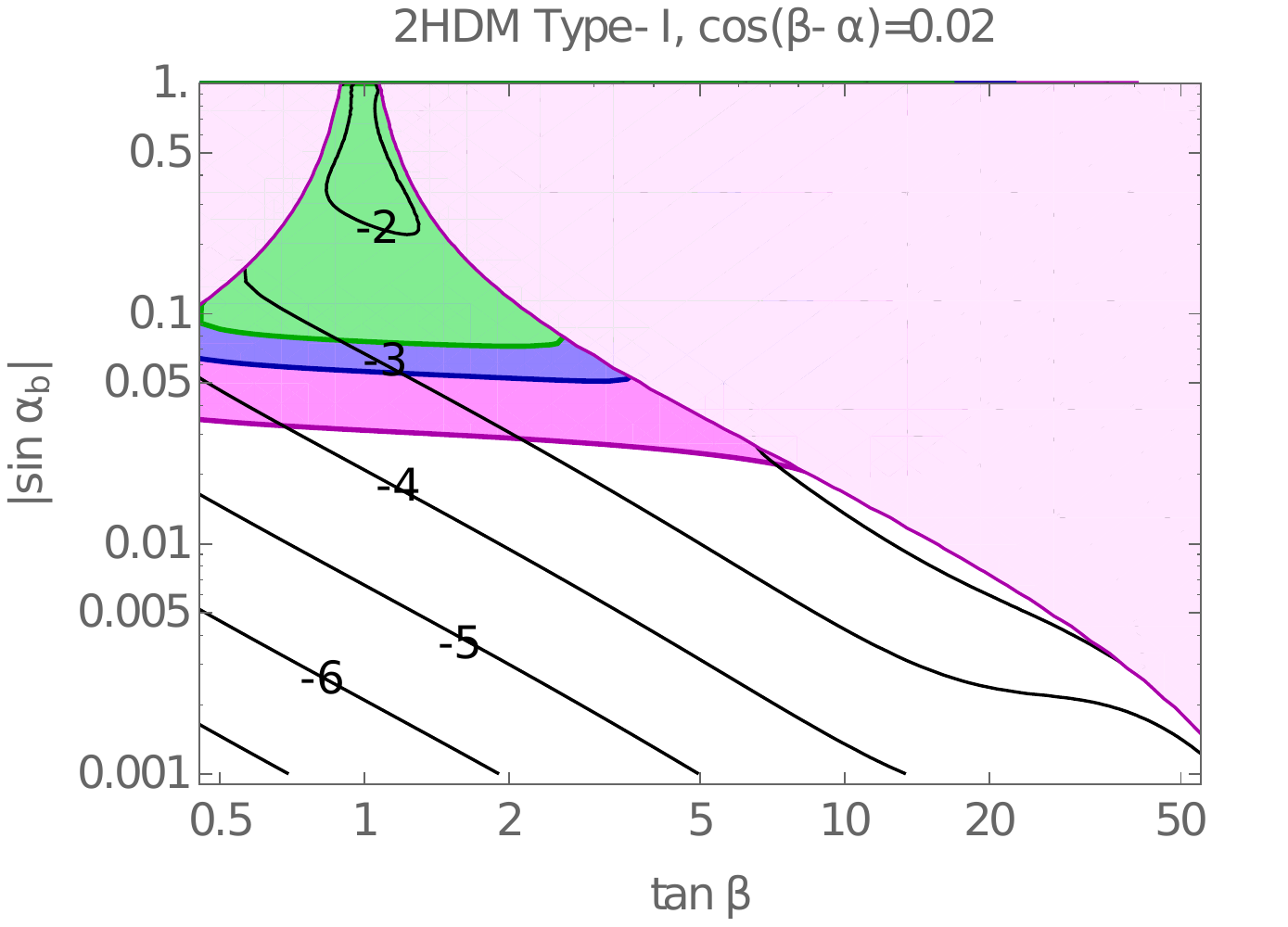}}
\subfigure[\ Type-I constraint on $h_3$]{\label{fig:T1cbma002h3}\includegraphics[width=70mm]{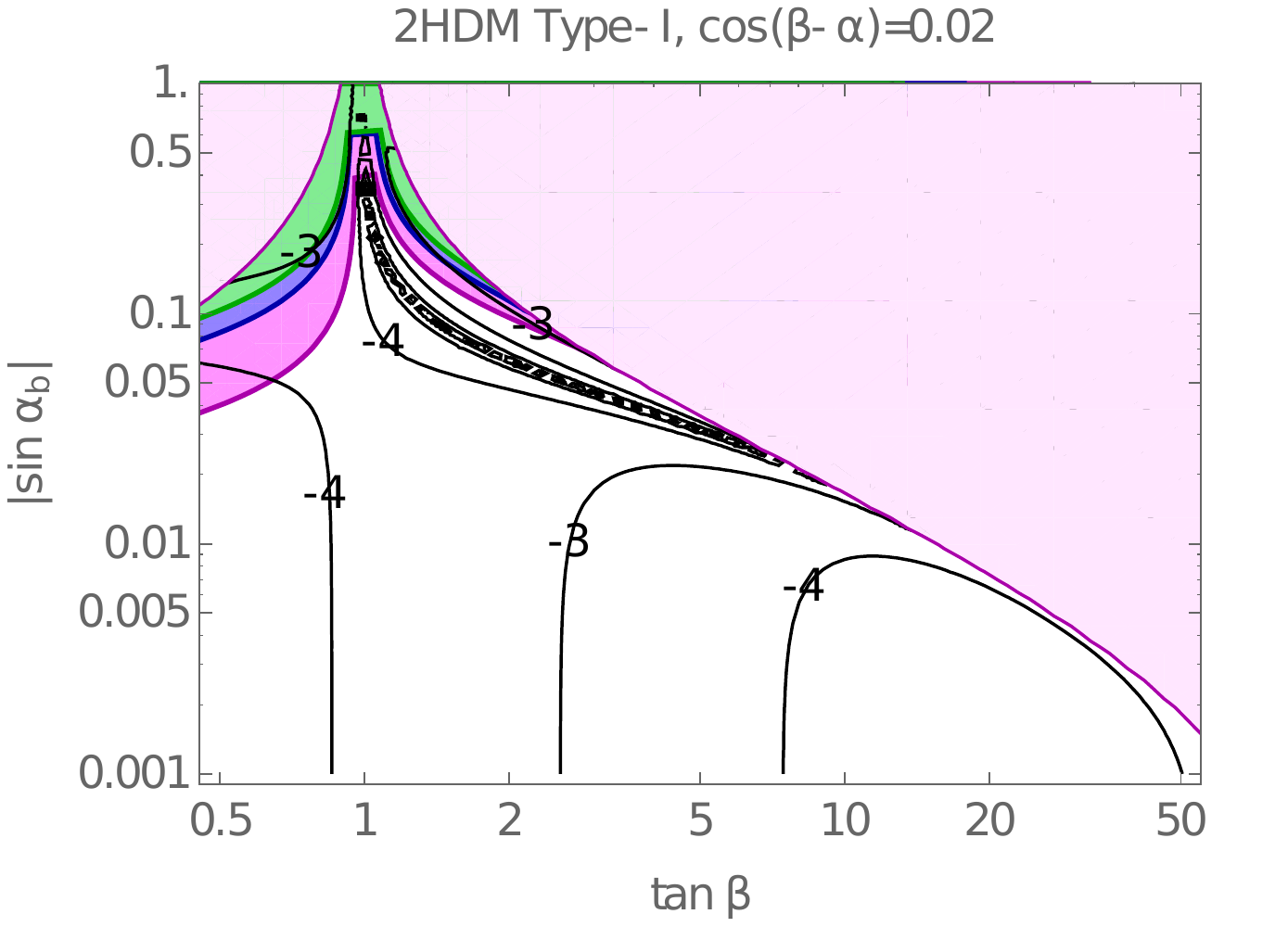}}     
\subfigure[\ Type-II Constraint on $h_2$]{\label{fig:T2cbma002h2}\includegraphics[width=70mm]{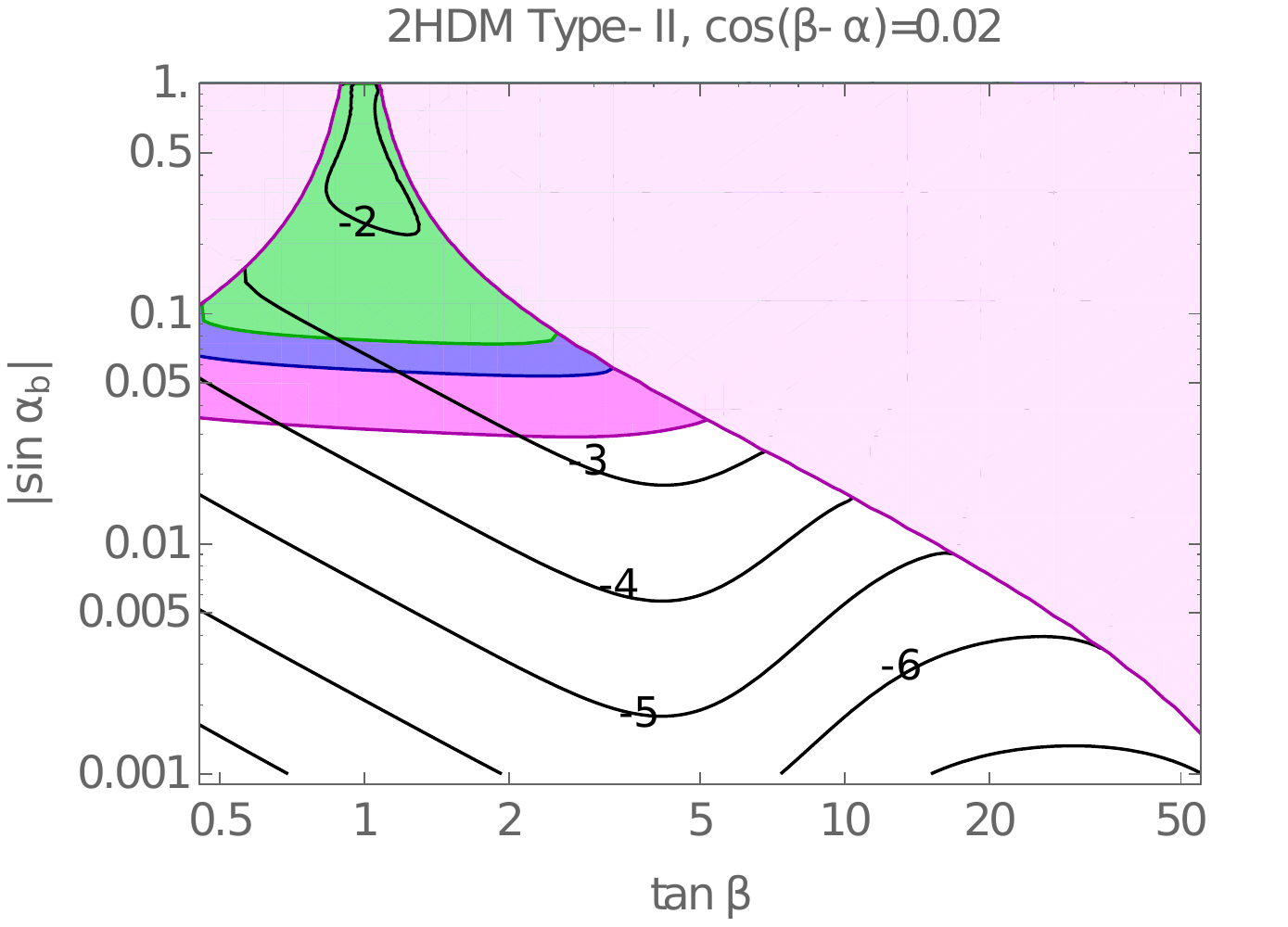}}
\subfigure[\ Type-II Constraint on $h_3$]{\label{fig:T2cbma002h3}\includegraphics[width=70mm]{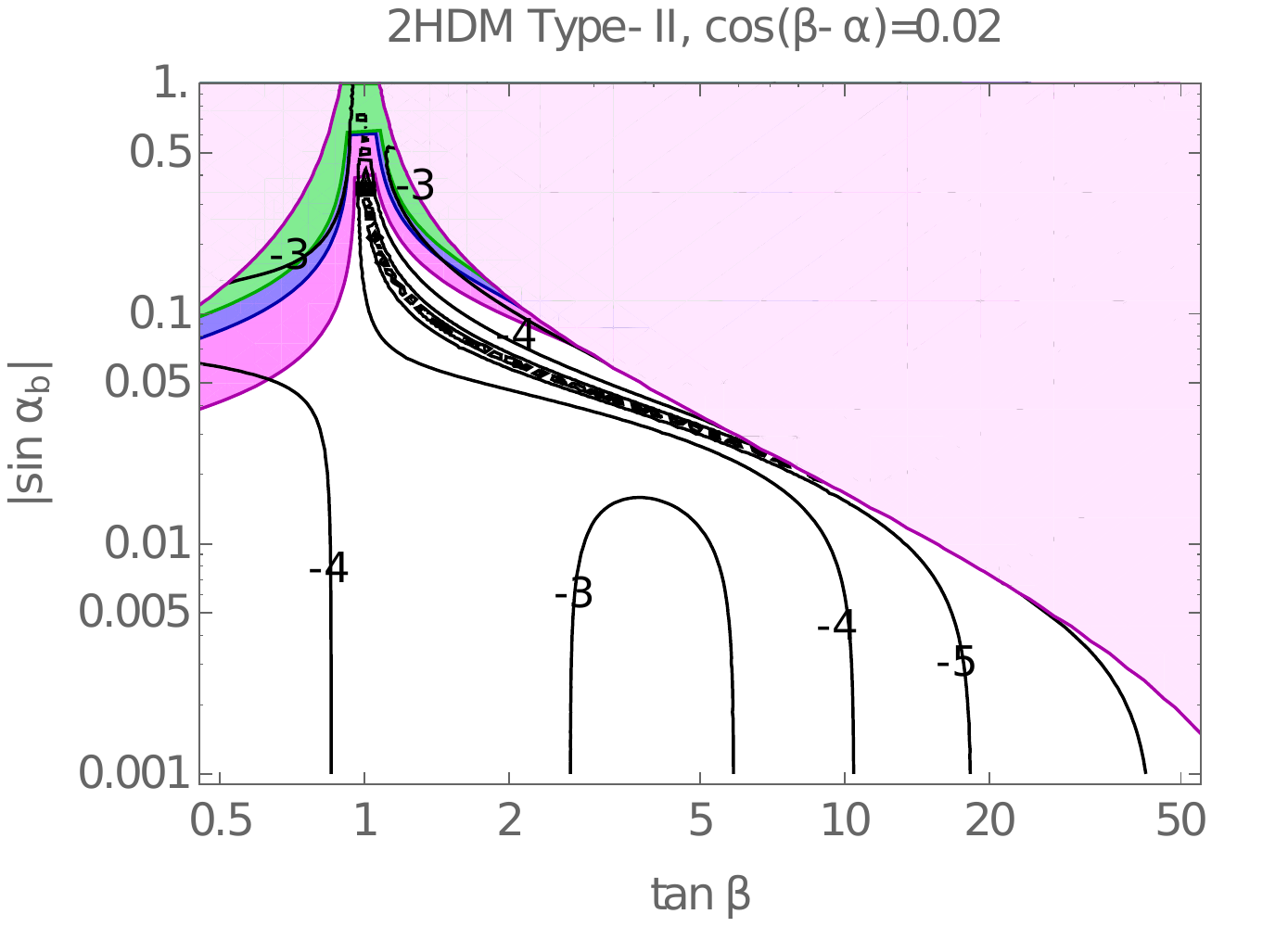}}
\caption{Exclusion limits for the heavy Higgs resonant productions in the Type-I and Type-II 2HDMs with $\cos(\beta-\alpha)=0.02$. The plots in the first and second row represent the constraints for the Type-I and Type-II models, respectively. The left (right) column shows the constraints from the resonance production of $h_2 (h_3)$. The pink region is theory-inaccessible. The green, blue, and magenta regions represent the exclusion limits for the LHC integrated luminosities equal to 100 fb$^{-1}$, 300 fb$^{-1}$ and 3000 fb$^{-1}$, respectively. The black contour represents the logarithm of $\log\overline{|M|^2_i}$ in Eq. \ref{amp2} with $s=m_{h_{2,3}}$ for $i=2,3$. }
\label{fig:T2cmba002h2h3}
\end{figure}

\begin{figure}
\centering     
\subfigure[\ Type-I constraint on $h_2$]{\label{fig:T1cbma01h2}\includegraphics[width=70mm]{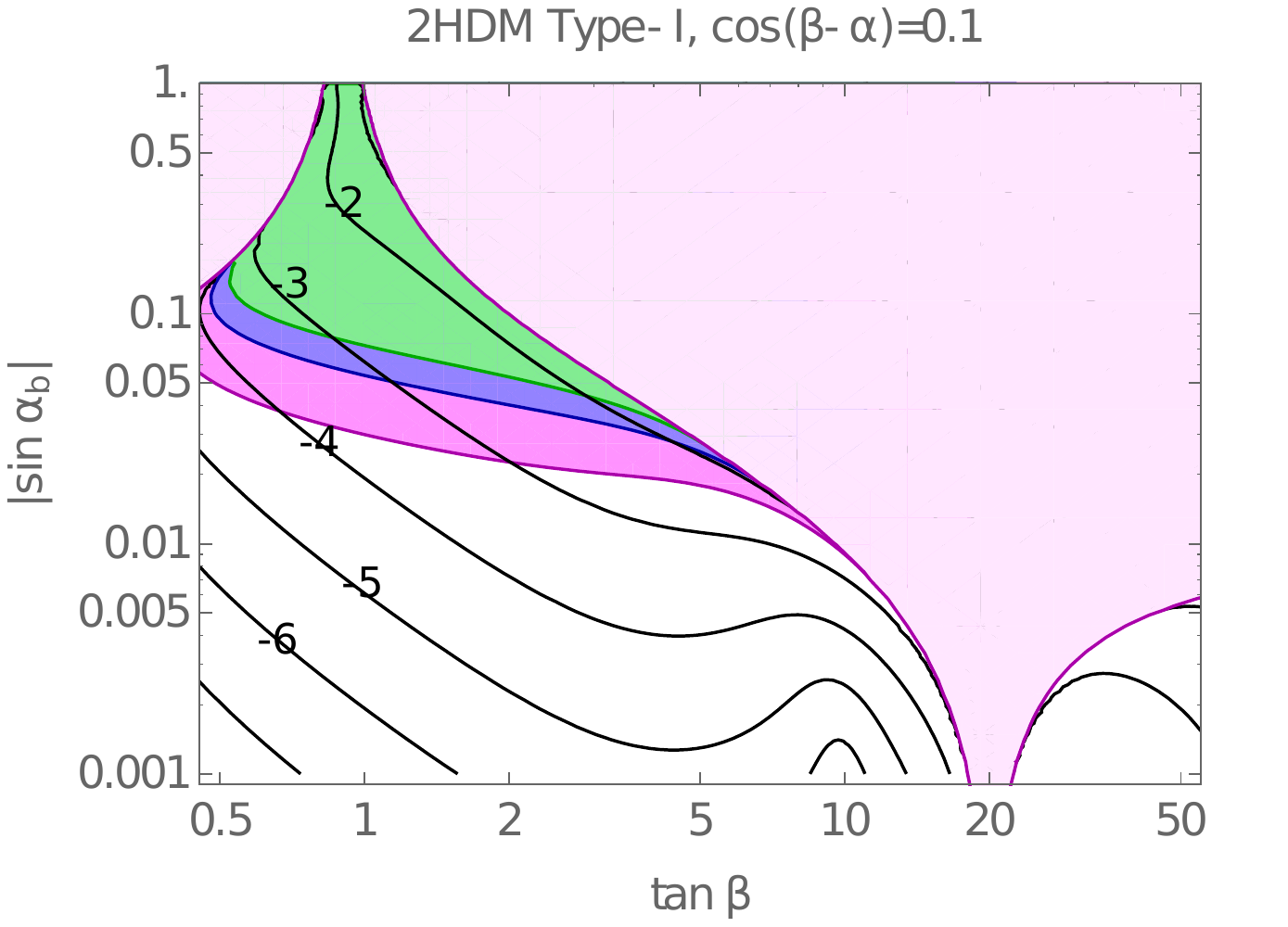}}
\subfigure[\ Type-I constraint on $h_3$]{\label{fig:T1cbma01h3}\includegraphics[width=70mm]{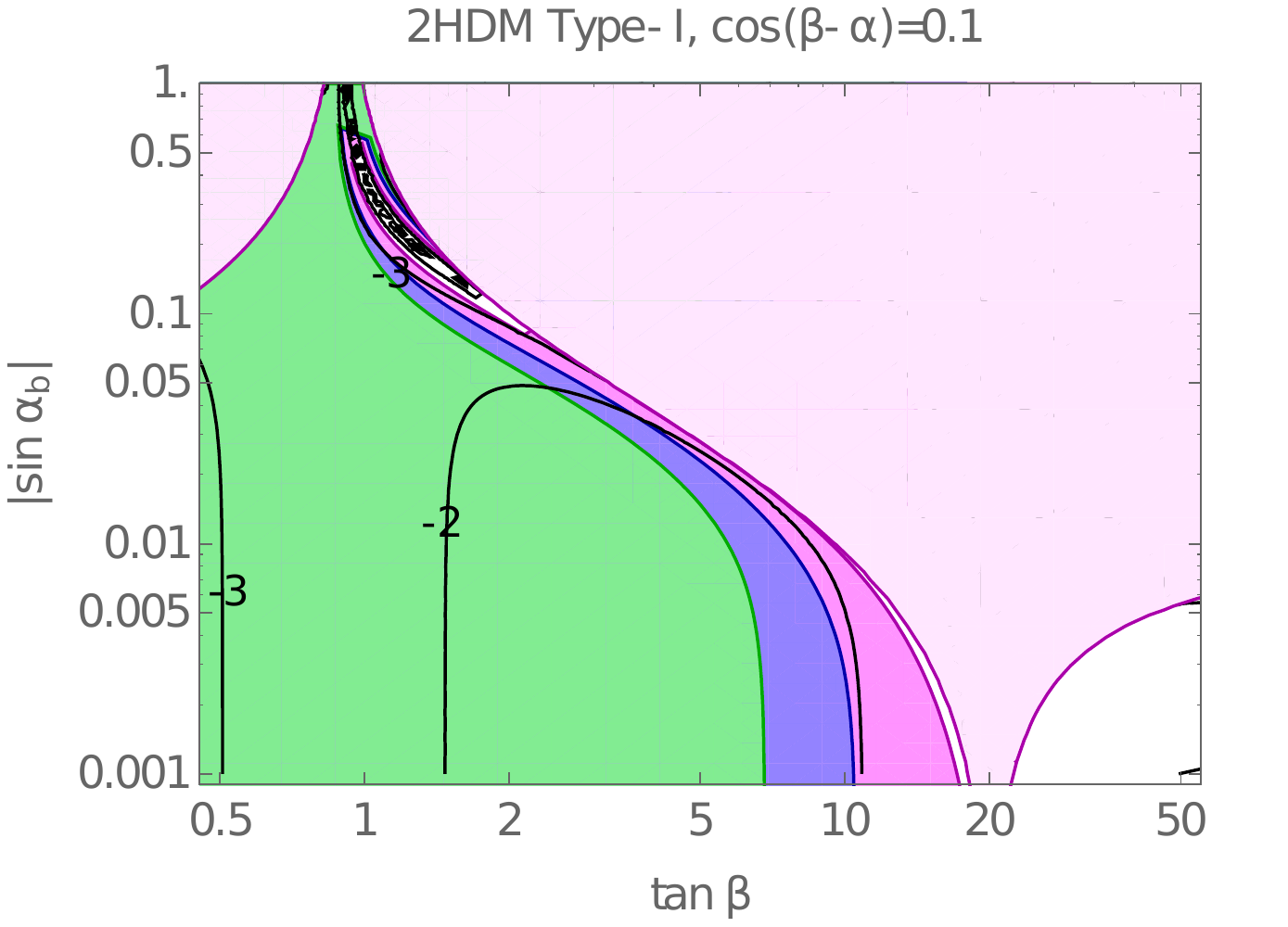}}\\
\subfigure[\ Type-II constraint on $h_2$]{\label{fig:T2cbma005h2}\includegraphics[width=70mm]{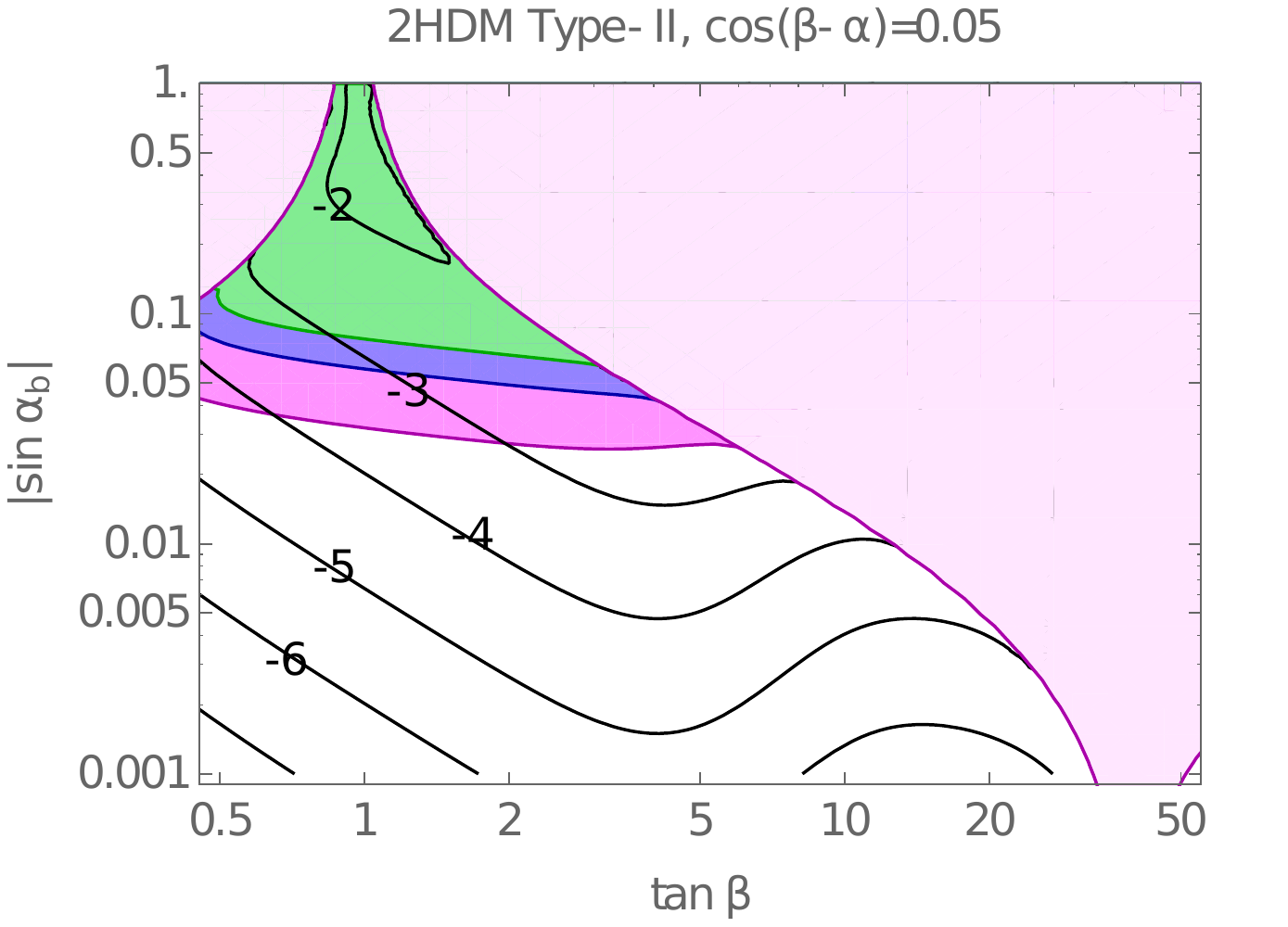}}
\subfigure[\ Type-II constraint on $h_3$]{\label{fig:T2cbma005h3}\includegraphics[width=70mm]{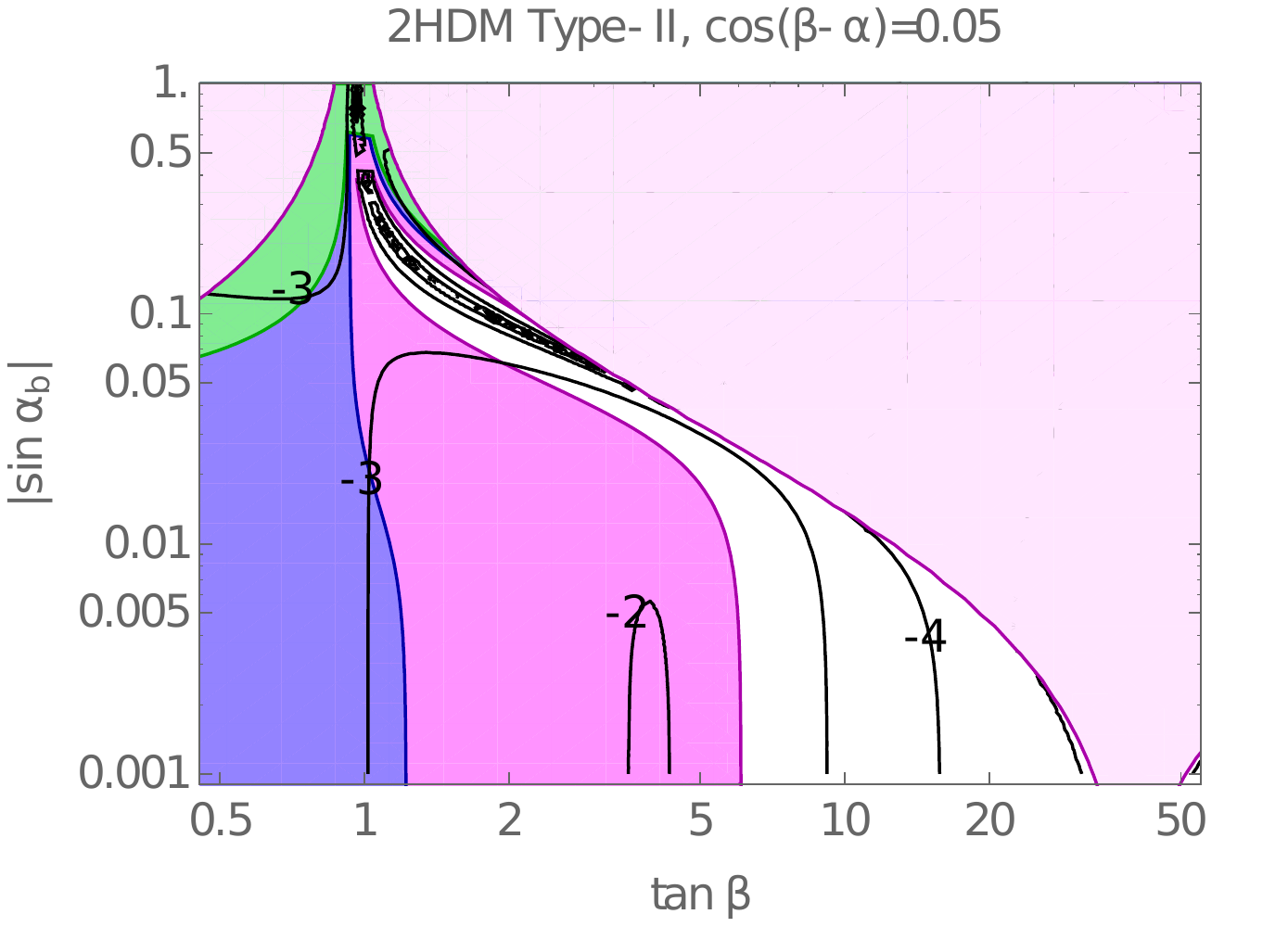}}
\caption{Exclusion limits for the heavy Higgs resonant productions with $\cos(\beta-\alpha)=0.1$ (Type-I) and $\cos(\beta-\alpha)=0.05$ (Type-II). The plots in the first and second row represent the constraints for the Type-I and Type-II models, respectively. The left (right) column shows the constraints from the resonance production of $h_2 (h_3)$. The pink region is theoretically inaccessible. The green, blue, and magenta regions represent the exclusion limits for the LHC integrated luminosities equal to 100 fb$^{-1}$, 300 fb$^{-1}$ and 3000 fb$^{-1}$, respectively. The black contour represents the logarithm of $\log \overline{|M|^2_i}$ in Eq. \ref{amp2} with $s=m_{h_{2,3}}$ for $i=2,3$. }
\label{fig:cbma01h2h3}
\end{figure}

We now translate our simulated exclusion limit into constraints on the parameter space of CPV 2HDMs. We use a benchmark point below which is consistent with the electroweak precision measurements and muon $g-2$ data as discussed in Ref.~\cite{Chen:2015gaa}:
\begin{equation}
m_{h_2}=550\ {\rm GeV}, \ m_{h_3}=600\ {\rm GeV}, \ m_{H^{\pm}}=620\ {\rm GeV}, \nu=1
\end{equation}
and show the constraints on $\sin \alpha_b$ vs $\tan\beta$ plane. The 95\% CLs exclusion limit is given by:
\begin{equation}
\sigma (gg\to h_{2,3})\times Br(h_{2,3}\to Zh_1)\times Br(h_1\to b\bar{b})<\sigma_L
\end{equation}
where $\sigma_L$ is the exclusion limit listed in  \Cref{table:BDT550limit,table:BDT600limit}. We assume that the resonance process is dominant when the invariant mass of two gluons is approaching the mass of the heavy Higgs. This is not always true in the parameter space we consider, especially in the limit of small $\theta$. The gluon fusion to $Zh_1$ box diagram may become important and interfere with the resonant triangle diagram. This may change the distribution of the invariant mass of $Zh_1$. Here, we will simply identify the region of parameter space that may suffer from this effect, leaving a detailed analysis for future study. To proceed, we will compare the relative scale of the amplitude squared of the resonant and non-resonant $gg\to Zh_1$ processes at the center of mass energy $\sqrt{s}=m_{h_2},$ and $m_{h_3}$. For the resonance contribution we use the following formula
\begin{eqnarray}\label{amp2}
\overline{|M_i|^2}&=&\frac{G_F \alpha_s^2|g_{iz1}|^2 s^2}{512 \pi^2 \sqrt{2}}[|\sum_{f=t,b} c_{f,i} F^H_{1/2}(\tau^i_f)|^2+|\sum_{f=t,b} \tilde{c}_{f,i} F^A_{1/2}(\tau^i_f)|^2]\nonumber \\ &&\times \frac{M_Z^2-2m^2_{h_1}-2s+(m^2_{h_i}-s)^2/M_Z^2}{(s-m^2_{h_i})^2+m^2_{h_i}\Gamma^2_{h_i}}
\end{eqnarray}
where $G_F$ and $\alpha_s$ are the Fermi constant and strong coupling constant, respectively, while $g_{iz1}$, $F^{A/H}_{1/2}$, $c_f$, $\tilde{c_f}$, and $\Gamma_{h_i}$ are defined in Sections \ref{sec:model} and \ref{prodec}.
For the non-resonant piece we obtain the scale of $\overline{M^2_\mathrm{box}}\simeq 10^{-5}$ from Ref.~\cite{Hespel:2015zea}. In presenting our results in Figs.~\ref{fig:Aligh2h3}-\ref{fig:cbma01h2h3}, we include contours of constant $\overline{|M_i|^2}$ in order to identify regions where the resonant and non-resonant contributions are commensurate in scale.

We now consider the prospective future reach at the LHC. In the alignment limit, the sensitivity comes primarily from the resonant production of $h_2$, as expected from Eqs.~\ref{g2z1} and \ref{g3z1}. We show the prospective exclusion regions associated with $h_{2,3}$ production separately in Fig.~\ref{fig:Aligh2h3}. The pink region is forbidden by the requirement of the electroweak symmetry breaking. The green, blue, and magenta regions represent the prospective exclusion limits for the LHC integrated luminosities equal to 100 fb$^{-1}$, 300 fb$^{-1}$ and 3000 fb$^{-1}$, respectively. The black contours correspond to $\log_{10}\overline{|M|^2_i}$ in Eq.~(\ref{amp2}) with $s=m_{h_{2,3}}$ for $i=2,3$. If we require $\overline{|M|^2_i} > 10^{-4}$ to guarantee the dominance of the resonant production, then there will be some parts of the prospective exclusion region for 3000 fb$^{-1}$ at low $\tan\beta$ that may not be valid for our analysis. One could observe that from Fig.~\ref{fig:T1Aligh3} and Fig.~\ref{fig:T2Aligh3} there is a loss of sensitivity for $h_3$ near $\tan\beta\approx1$ in the alignment limit for both Type-I and Type-II models. This is due to the cancellation effect in the coupling $g_{3z1}$ when $-\alpha=\beta=\pi /4$.

The situation is similar for non-vanishing but small  $\theta$. An illustration for $\cos(\beta-\alpha)=0.02$ is shown in Fig.~\ref{fig:T2cmba002h2h3}. However, for a large deviation, for example $\cos(\beta-\alpha)=0.05$ in the Type-II model and $\cos(\beta-\alpha)=0.1$ in the Type-I model, the constraints from the resonance production of $h_3$ become important, and can cover large part of the parameter space. In this situation, the effect of the non-resonant production is negligible due to a relatively large $\overline{|M|_3^2}$. This can be seen in Fig.~\ref{fig:cbma01h2h3}.

\begin{figure}
\centering     
\subfigure[\ Type-I Alignment limit current]{\label{fig:T1Alignp}\includegraphics[width=70mm]{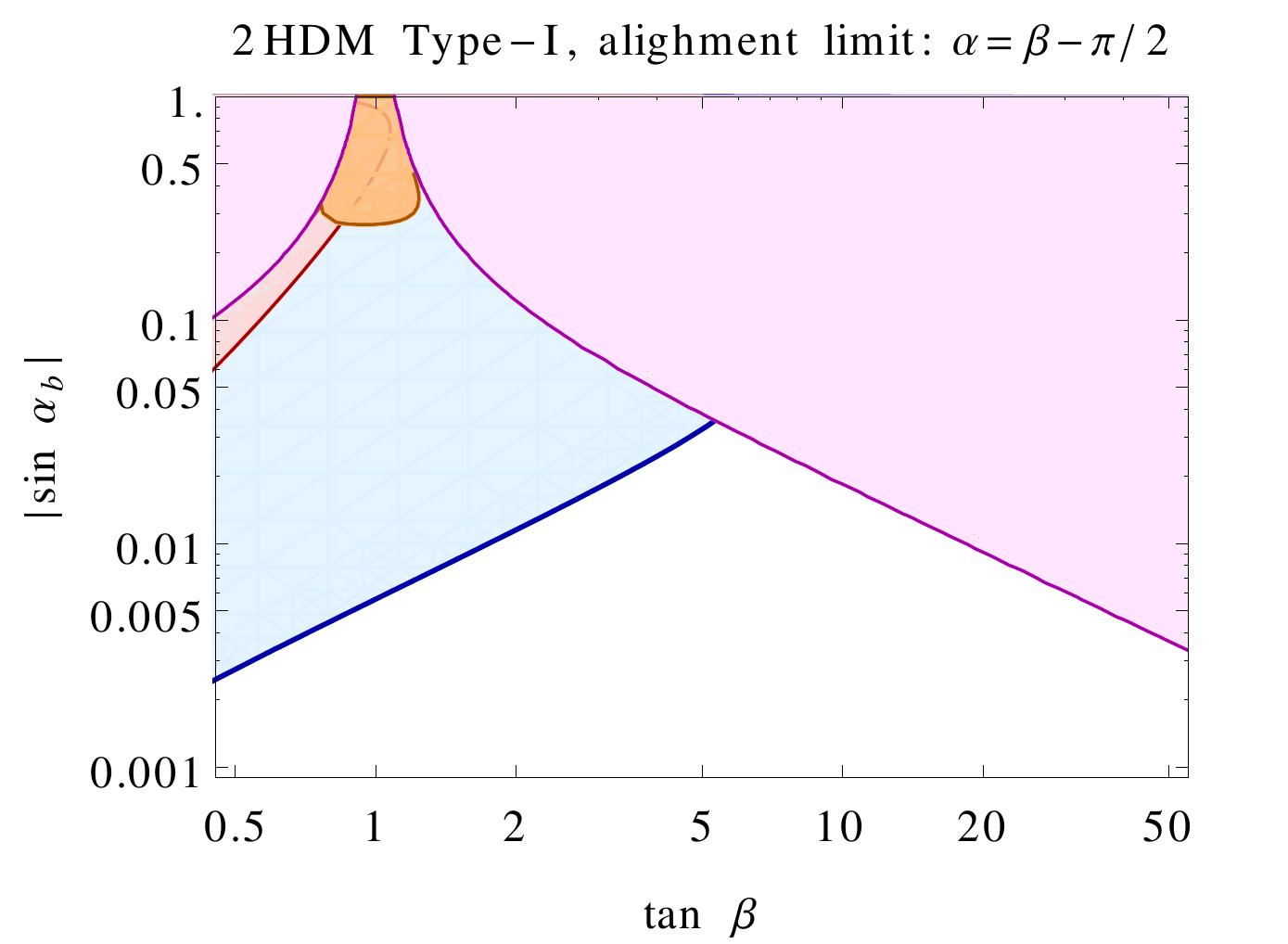}}
\subfigure[\ Type-I Alignment limit future]{\label{fig:T1Alignf}\includegraphics[width=70mm]{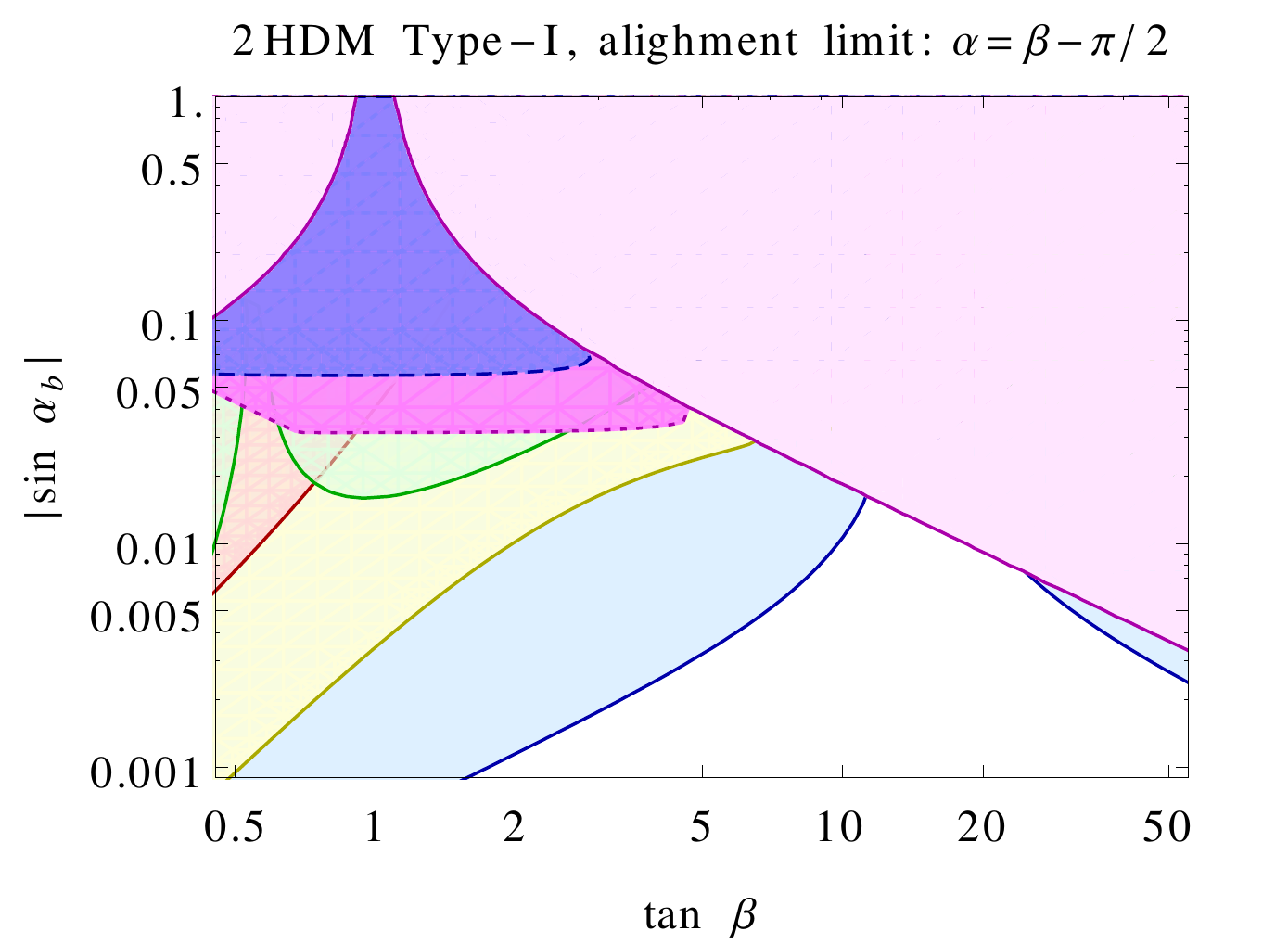}}\\
\subfigure[\ Type-I $\cos(\beta-\alpha)=0.02$ current]{\label{fig:T1cbma002p}\includegraphics[width=70mm]{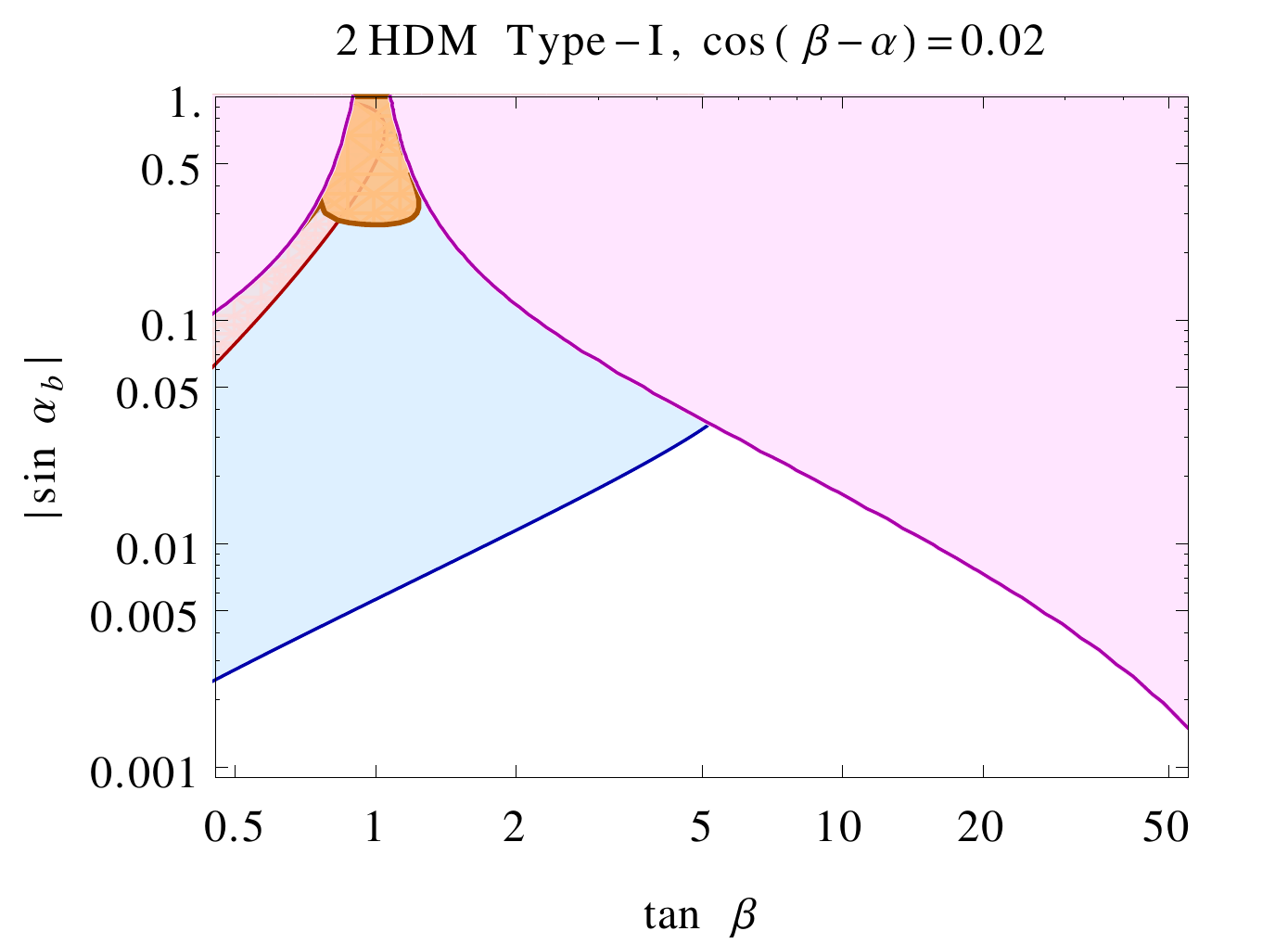}}
\subfigure[\ Type-I $\cos(\beta-\alpha)=0.02$ future]{\label{fig:T1cbma002f}\includegraphics[width=70mm]{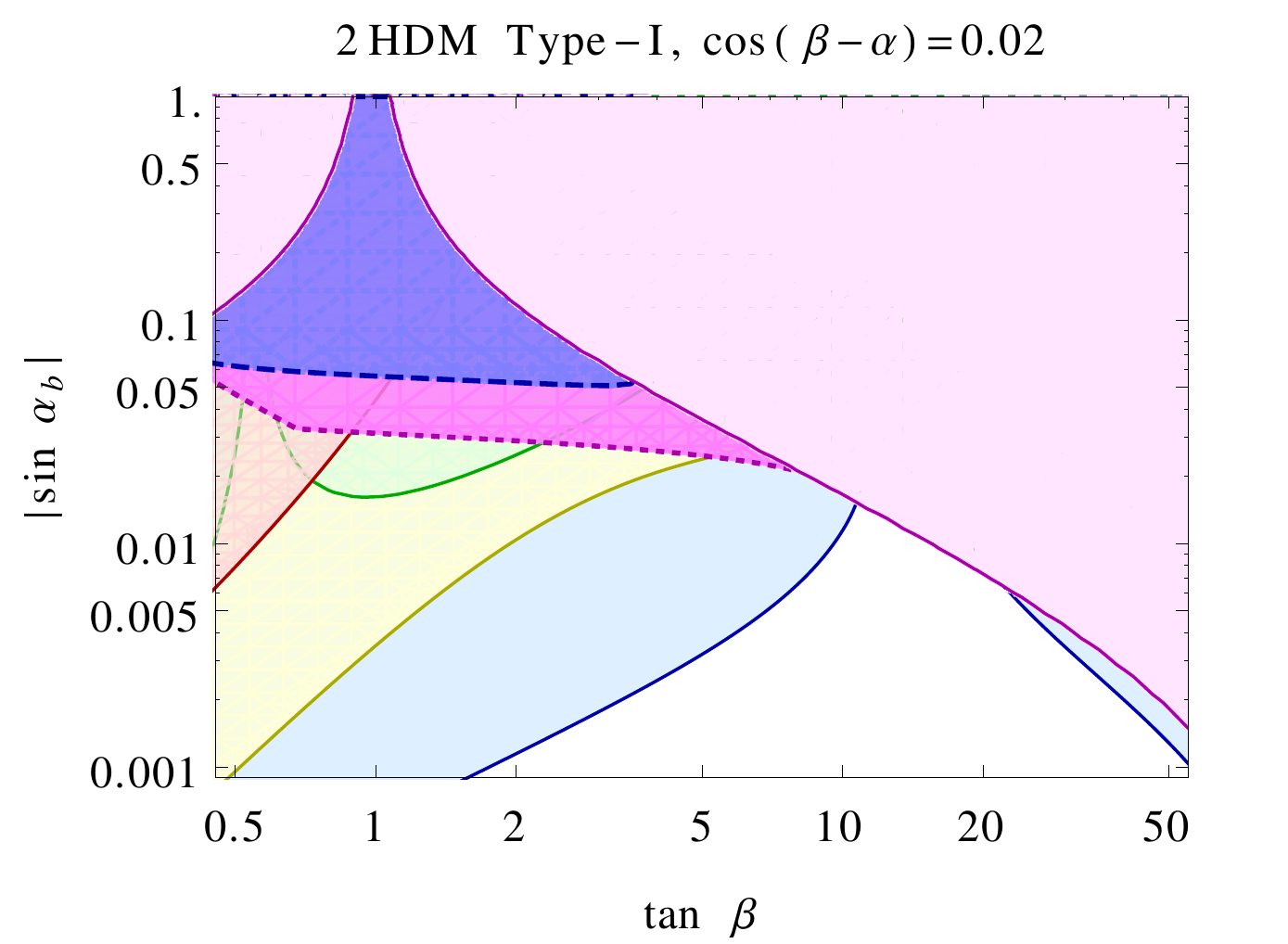}}\\
\subfigure[\ Type-I $\cos(\beta-\alpha)=0.1$ current]{\label{fig:T1cbma01p}\includegraphics[width=70mm]{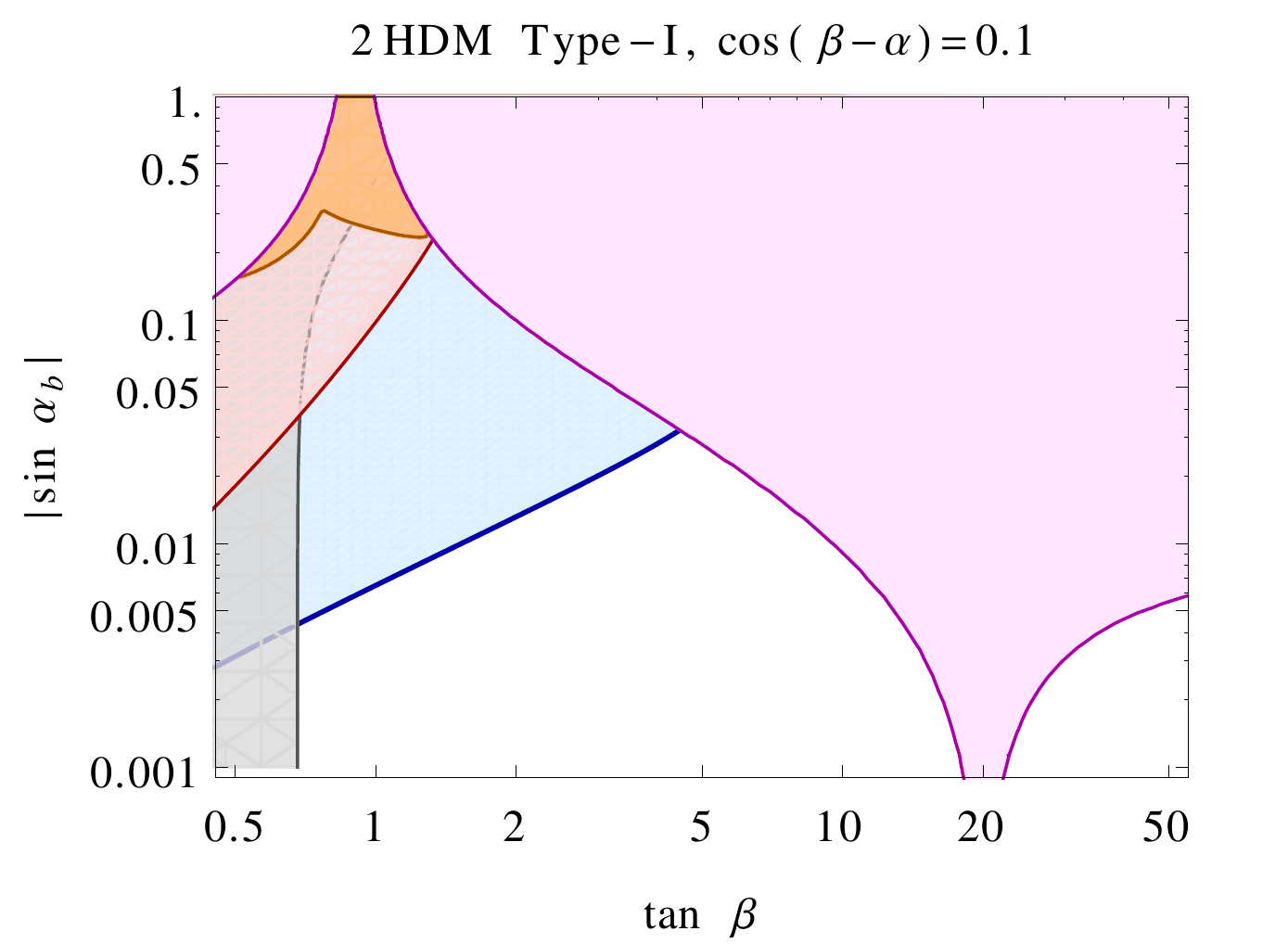}}
\subfigure[\ Type-I $\cos(\beta-\alpha)=0.1$ future]{\label{fig:T1cbma01f}\includegraphics[width=70mm]{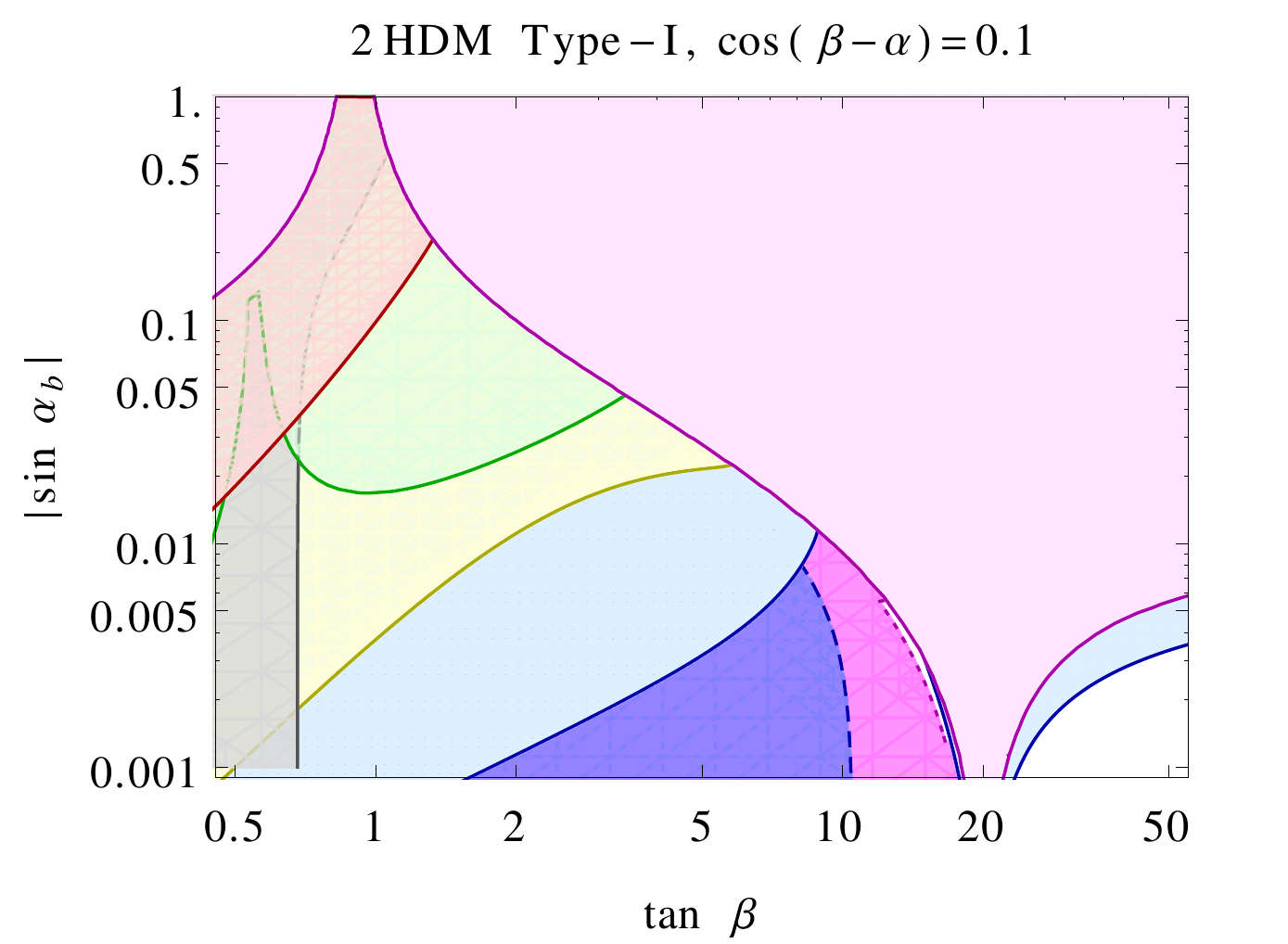}}
\caption{Exclusion regions for the collider and EDM experiments in the type-I 2HDM. The left (right) column is for the current (future) exclusion limit. The orange region is excluded by the current LHC data. The blue and magenta regions represent the future LHC limit with integrated luminosities equal to 300 fb$^{-1}$ and 3000 fb$^{-1}$, respectively. Light transparent red represents the constraint from mercury EDM, light blue denotes electron EDM, light transparent green stands for neutron EDM, and light yellow signifies future prospective radium EDM. The gray region is excluded by the coupling measurement of the SM-like Higgs and the pink region is theoretically inaccessible due to the absence of a real solution for $\alpha_c$. The benchmark point used here is $m_{h_2}=550\ {\rm GeV}, \ m_{h_3}=600\ {\rm GeV}, \ m_{H^{\pm}}=620\ {\rm GeV}, \nu=1$. }
\label{fig:T1}
\end{figure}

\begin{figure}
\centering     
\subfigure[\ Type-II Alignment limit current]{\label{fig:T2Alignp}\includegraphics[width=70mm]{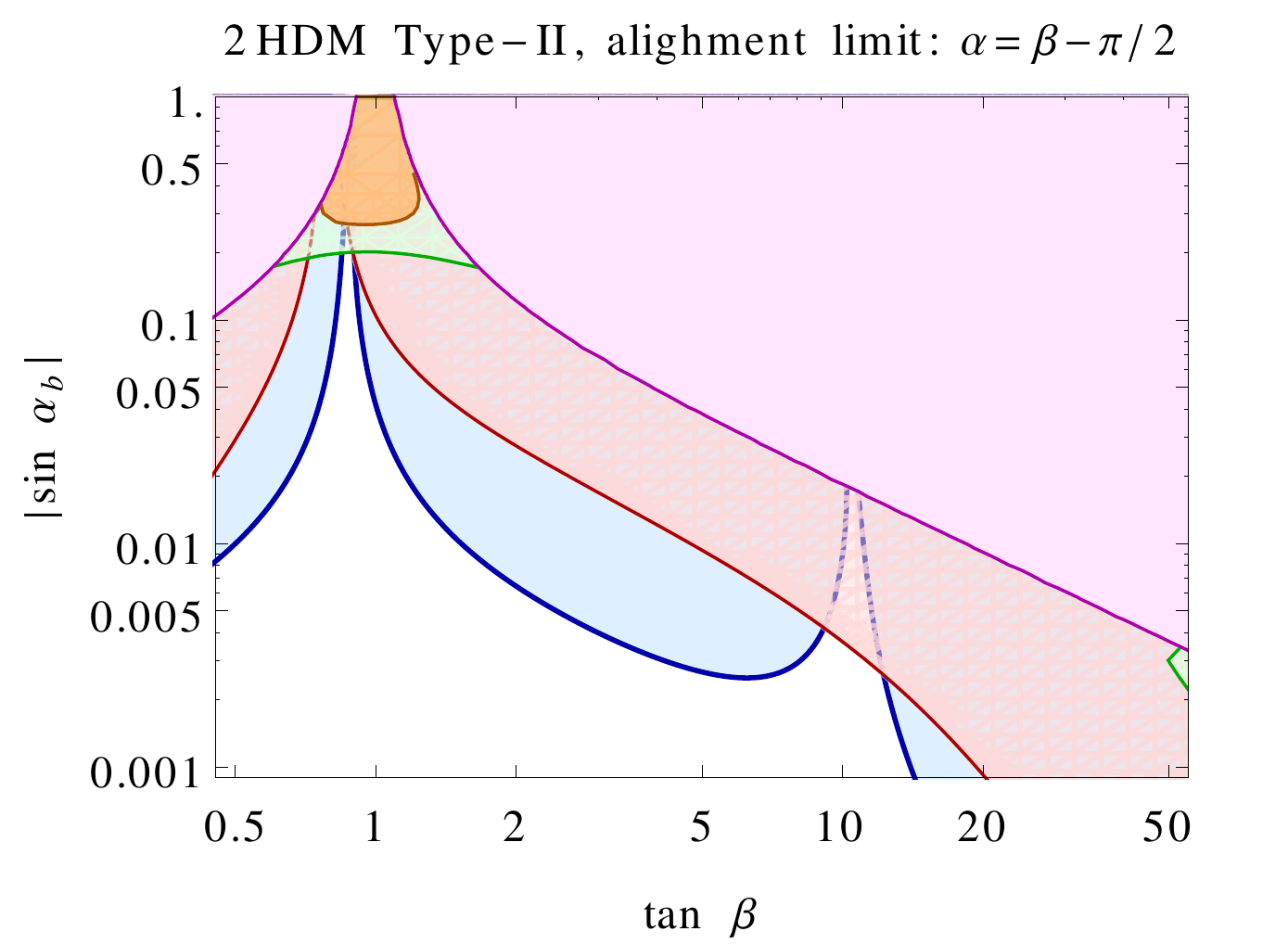}}
\subfigure[\ Type-II Alignment limit future]{\label{fig:T2Alignf}\includegraphics[width=70mm]{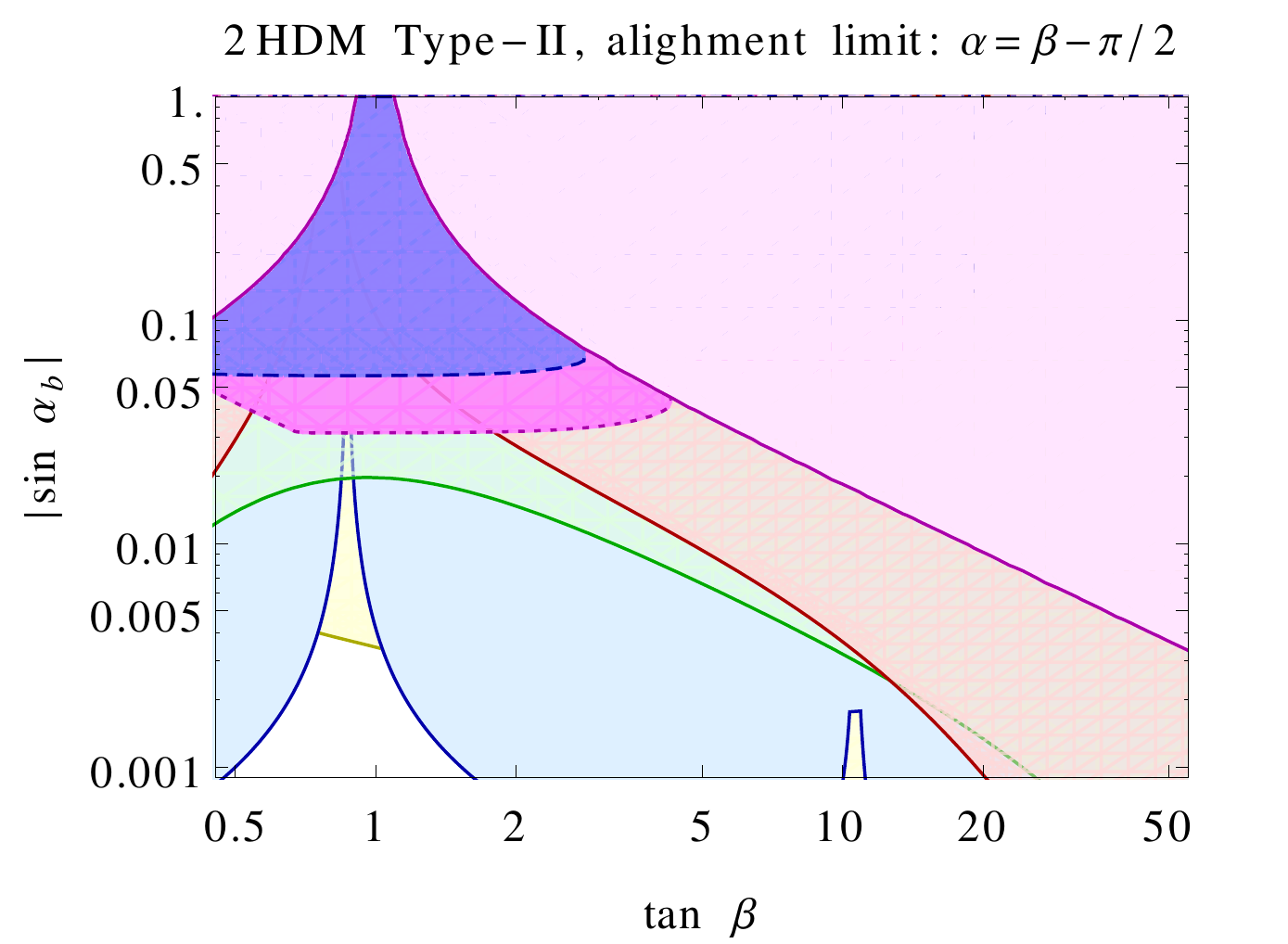}}\\
\subfigure[\ Type-II $\cos(\beta-\alpha)=0.02$ current]{\label{fig:T2cbma002p}\includegraphics[width=70mm]{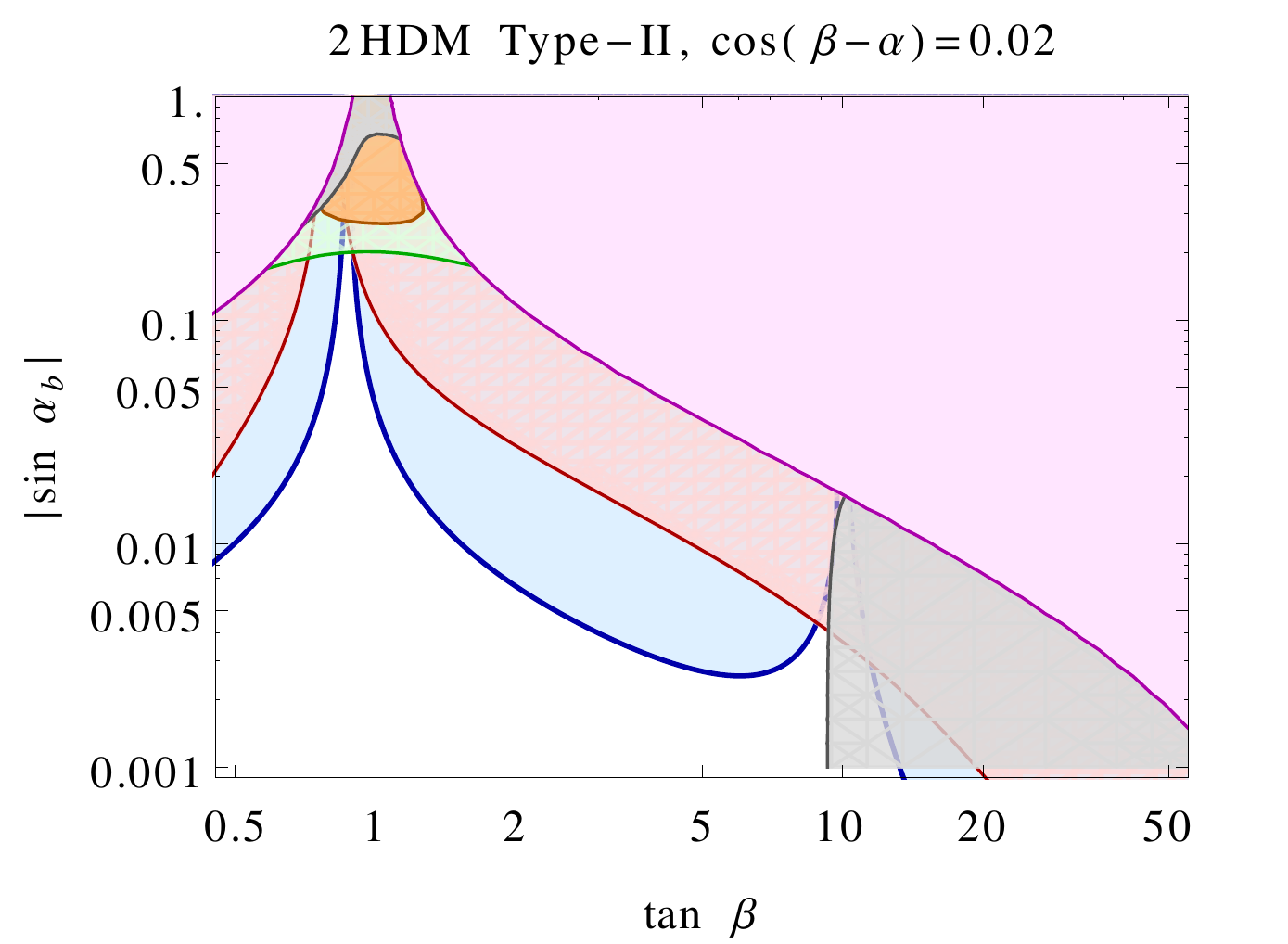}}
\subfigure[\ Type-II $\cos(\beta-\alpha)=0.02$ future]{\label{fig:T2cbma002f}\includegraphics[width=70mm]{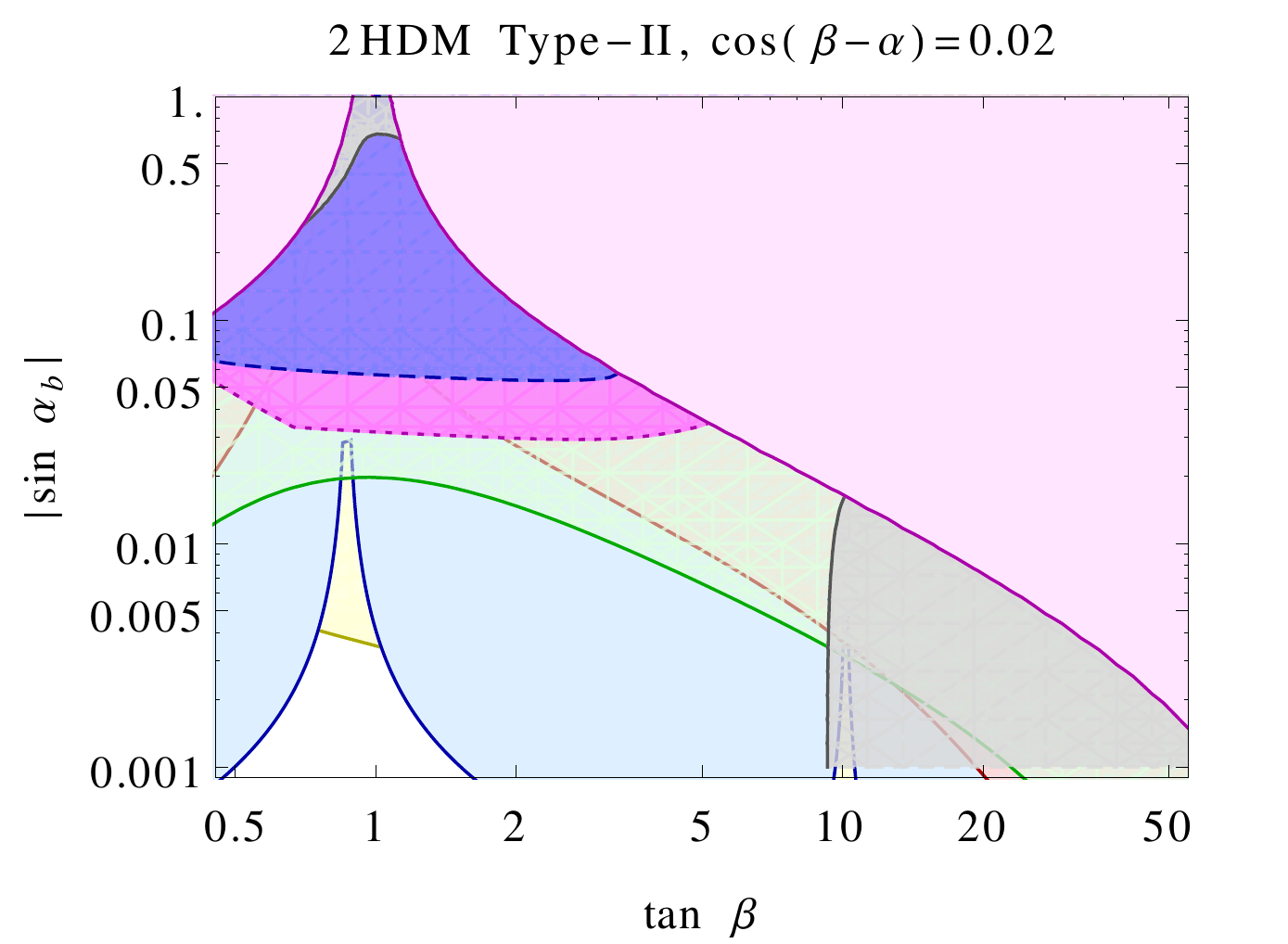}}\\
\subfigure[\ Type-II $\cos(\beta-\alpha)=0.05$ current]{\label{fig:T2cbma005p}\includegraphics[width=70mm]{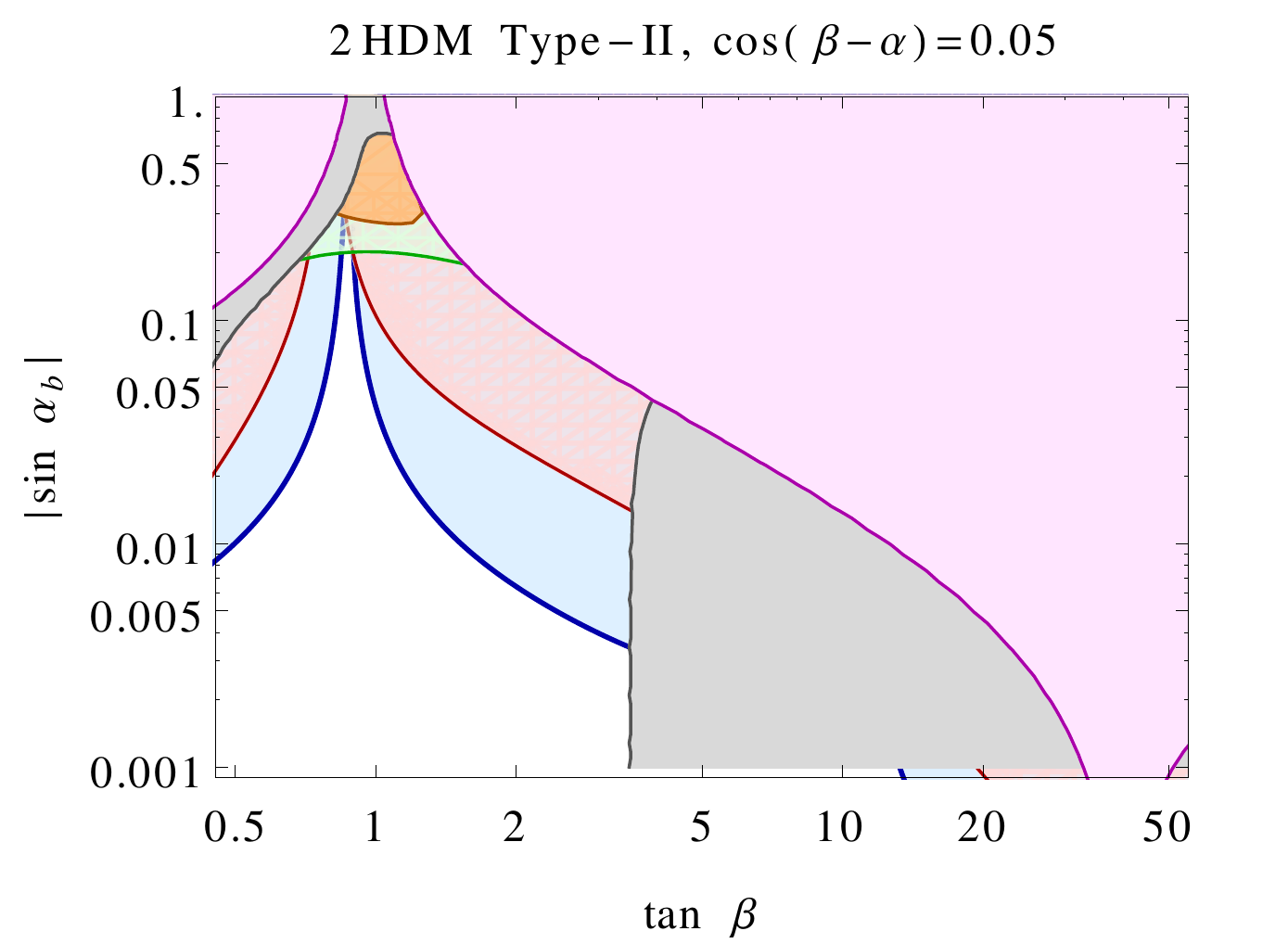}}
\subfigure[\ Type-II $\cos(\beta-\alpha)=0.05$ future]{\label{fig:T2cbma005f}\includegraphics[width=70mm]{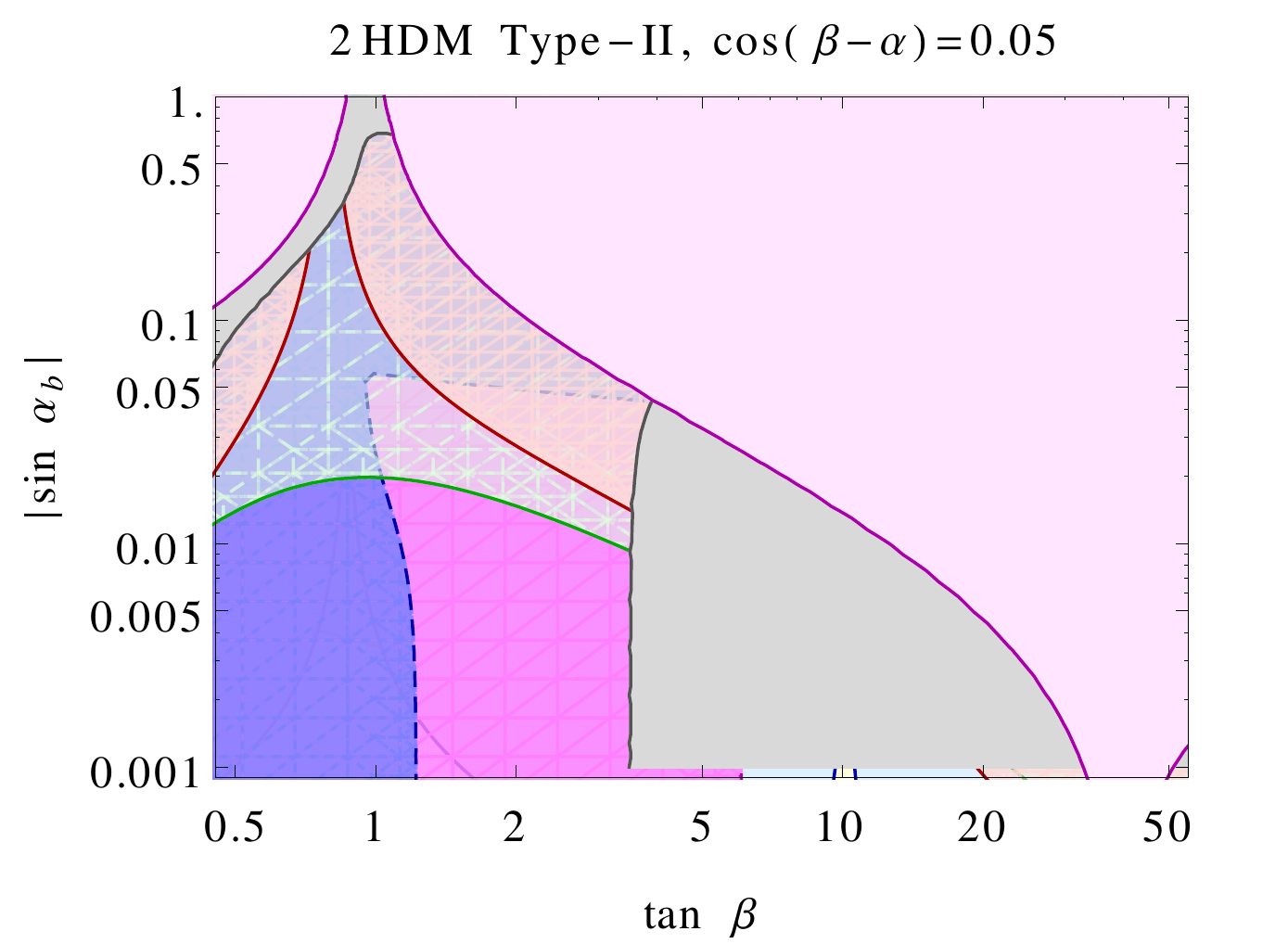}}
\caption{Exclusion regions for the collider and EDM experiments in the type-II 2HDM. The left (right) column is for the current (future) exclusion limit. The orange region is excluded by the current LHC data. The blue and magenta regions represent the future LHC limit with integrated luminosities equal to 300 fb$^{-1}$ and 3000 fb$^{-1}$, respectively. Light transparent red represents the constraint from mercury EDM, light blue denotes electron EDM, light transparent green stands for neutron EDM, and light yellow signifies future prospective radium EDM. The gray region is excluded by the coupling measurement of the SM-like Higgs and the pink region is theoretically inaccessible due to the absence of a real solution for $\alpha_c$. The benchmark point used here is $m_{h_2}=550\ {\rm GeV}, \ m_{h_3}=600\ {\rm GeV}, \ m_{H^{\pm}}=620\ {\rm GeV}, \nu=1$. }
\label{fig:T2}
\end{figure}

In addition, we take into account the constraint from the Higgs coupling measurements at 7 and 8 TeV LHC as what was done in Ref.~\cite{Chen:2015gaa}. The channels included in the $\chi^2$ analysis are: $h_1 \to$ {$WW$, $ZZ$, $\gamma\gamma$, $bb$, $\tau\tau$}. 
We also include the present and prospective constraints from EDM searches given in Ref.~\cite{Inoue:2014nva}, which are summarized in Table~\ref{EDMs}. We find that the constraints from LHC and low energy experiments are complementary.  \Cref{fig:T1,fig:T2} demonstrate the exclusion limits from both LHC and EDM searches. In each figure, the orange region gives the current LHC exclusion limit. The blue, and magenta regions represent prospective future LHC limits for integrated luminosities equal to 300 fb$^{-1}$ and 3000 fb$^{-1}$, respectively. The light green and light blue regions are excluded by the neutron EDM and electron EDM searches, respectively. The light red and light yellow represents current constraints from the mercury and prospective radium atomic EDM searches. The gray regions are excluded by the Higgs coupling measurements. The pink region is again as described above 
theory-inaccessible. There are also constraints that we do not show in Fig.~\ref{fig:T1} and Fig.~\ref{fig:T2} from heavy flavor physics~\cite{Enomoto:2015wbn}, which exclude the regions of parameter space with $\tan\beta$ less than 0.9 in both Type-I and Type-II models for the benchmark point we choose. These constraints can be relaxed if other new particles are introdueced in addition to the 2HDMs or some non-trivial flavor structure \cite{Campos:2017dgc,Celis:2016azn}. Figure 3 of Ref.~\cite{Enomoto:2015wbn} demonstrates the constraints on the $\tan\beta$ vs $m_{H^+}$ plane for the Type-I and Type-II 2HDMs. The most stringent bounds on $\tan\beta$ come from $B_s-\bar{B}_s$ mixing and $B^0_s\to\mu^+\mu^-$ for the Type-I model, and from $B_s-\bar{B}_s$ and $B_d-\bar{B}_d$ mixing for the Type-II model.  
\begin{table}[h]
\centering{\begin{tabular}{ c c c c c c }
\hline
\hline
Source & Current EDM (e cm)  & Projected EDM (e cm)   \\
\hline
Electron (e) & $d_{\rm e} < 8.7\times 10^{-29}$ at 90\% CL\cite{Baron:2013eja}& $d_{\rm e} < 8.7\times 10^{-30}$ \cite{Kumar:2013qya} \\

Neutron (n) & $d_{\rm n} < 2.9\times 10^{-26}$ at 90\% CL\cite{Baker:2006ts}& $d_{\rm n} < 2.9\times 10^{-28}$ \cite{Kumar:2013qya} \\

Mercury (Hg) & $d_{\rm Hg} < 7.4\times 10^{-30}$ at 95\% CL\cite{Graner:2016ses}&  - \\

Radium (Ra) &    -      & $d_{\rm Ra} < 10^{-27}$ \cite{Kumar:2013qya}  \\
\hline
\end{tabular} \ \ \ 
\caption{Current and projected EDM constraints in units of $e$-cm. For the projected limits we assume that the sensitivity of 
nEDM is improved by two orders of magnitude, and eEDM is improved by one order of magnitude.  The mercury EDM remains the same
while future projected sensitivity of the radium EDM is assumed to be $d_{\rm Ra} < 10^{-27}$ $e$-cm.
}\label{EDMs}}
\end{table}

In Figs.~\ref{fig:T1Alignp} and \ref{fig:T1Alignf}, we show the current and prospective exclusion regions for the Type-I model in the alignment limit. One can see that the reach of the collider search is not competitive with that of the electron EDM search even at the end of the HL-LHC phase, especially in the low $\tan\beta$ region. This is due to the fact that the collider search is sensitive to $\mathrm{Br}(h_{2,3}\to Zh_1)$ in addition to the $h_{2,3}$ production cross sections. In the alignment limit, the $Zh_1$ channel is fed mainly  from the decay of $h_2$, and the coupling $g_{2z1}$ is suppressed by the CPV angle $\alpha_b$ as shown in Eq.~(\ref{g2z1}). Moreover, for low $\tan\beta$, the couplings of $h_{2}$ to  quarks  are enhanced, which leads to a suppression on $\mathrm{Br}(h_2\to Zh_1)$ and an increasing gluon fusion $h_2$ production cross section. However,  the increase of the latter cannot compensate for the decreasing $\mathrm{Br}(h_2\to Zh_1)$. The net effect is a  reduced $Zh_1$ signal strength.  In contrast,  the electron EDM is sensitive to the pseudo-scalar couplings that are enhanced at low $\tan\beta$ in the Type-I model, so electron EDM searches exclude a large part of parameter space in the low $\tan\beta$ region in Fig.~\ref{fig:T1Alignf}. 

In Figs.~\ref{fig:T1cbma002p} to \ref{fig:T1cbma01f}, we present the current and prospective exclusion regions away from the alignment limit.  We can observe that the current collider constraints are not as strong as those from EDMs . However, the future LHC reach can be comparable to that of the EDMs searches, and even better at moderate $\tan\beta$. One can observe this feature from Eq.~(\ref{g3z1}), where $g_{3z1}$ is proportional to $\theta$ which describes the level of deviation from the alignment limit, in contrast to Eq.~(\ref{g2z1}) where $g_{2z1}$ is suppressed by the small CPV angle $\alpha_b$. One can also observe that in the large $\tan\beta$ region the LHC loses sensitivity. The reason is that at large $\tan\beta$, pseudo-scalar couplings of both $t$ and $b$ quarks to $h_3$ are suppressed, so the total partonic production cross-section $\hat{\sigma}(gg\to h_3)$ decreases as $\tan\beta$ increases. Despite the possible increase in $\mathrm{Br}(h_3\to Zh_1)$, the over all effect is a decreasing trend of signal rate towards large $\tan\beta$ resulting in an untestable region for the LHC search.

\begin{figure}
\centering     
\subfigure[\ Type-I production cross-section for $h_3$]{\label{fig:csh3t1}\includegraphics[width=70mm]{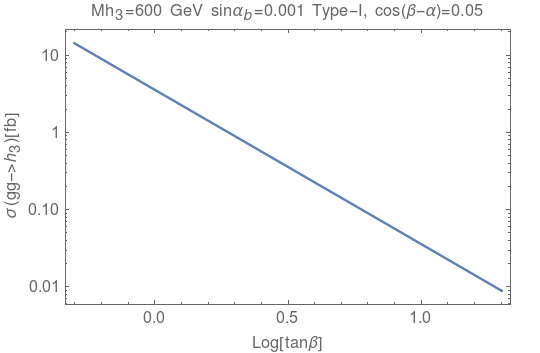}}
\subfigure[\ Type-II production cross-section for $h_3$]{\label{fig:csh3t2}\includegraphics[width=70mm]{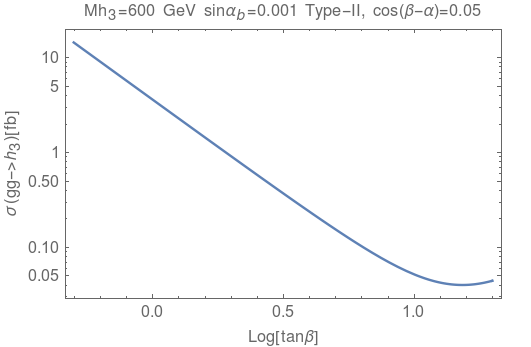}}
\caption{Production cross-section for $h_3$ for Type-I(left), Type-II(right) model.}
\end{figure}

In the Type-II model, the results are shown in Fig.~\ref{fig:T2}. In contrast to the Type-I model, the electron and mercury EDMs are not able to probe the parameter space when $\tan\beta$ is close to one due to the cancellation in Barr-Zee diagrams indicated in ~\cite{Inoue:2014nva} and ~\cite{Bian:2014zka}, whereas the neutron and radium EDMs retain sensitivity in this region. In the situation that is close to the alignment limit, as one can observe from Fig.~\ref{fig:T2Alignp} to ~\ref{fig:T2cbma002f}, the future LHC reach can help to test the region where $\tan\beta$ is close to one. However, the reach of future neutron and radium EDM constraints still exceeds that of the LHC. When the deviation from the alignment limit is as large as $\cos(\beta-\alpha)=0.05$, the future LHC may probe a large portion of the parameter space for reasons similar to those for the Type-I model:  $g_{3z1}$ is sensitive to this deviation and is not suppressed by the CPV angle. Moreover, some portions of the large $\tan\beta$ region cannot be accessed, but for reasons different from the Type-I case. In the Type-II model, the pseudo-scalar coupling of the $t$-quark to $h_3$ is suppressed at large $\tan\beta$, while the pseudo-scalar coupling of the $b$-quark to $h_3$ is enhanced. However, for the range of $\tan\beta$ we are interested in, the enhancement of the $b$-quark loop contribution to  $\hat{\sigma}(gg\to h_3)$ cannot compensate for the suppression of the $t$-quark loop effect as one can see from Fig. \ref{fig:csh3t2}. Thus, ${\hat\sigma} (gg\to h_3)$ decreases with increasing $\tan\beta$. As for $Br(h_3\to Zh_1)$, due to the increasing  $Br(h_3\to \bar{b}b)$  and decreasing $Br(h_3\to \bar{t}t)$, the overall effect leads to a decreasing $Br(h_3\to Zh_1)$. A decreasing production cross-section combined with a decreasing decay branching ratio makes the large $\tan\beta$ region relatively inaccessible for the LHC in the Type-II model.

Now we argue that the future result of EDM and LHC experiments are expected to be complementary to each other and combining information from two kinds of experiments would help us better determine if the 2HDMs are realized in the nature. Since the global fit of the Higgs coupling measurements constrains 2HDMs in the parameter space that are close to the alignment limit, we  summarize our results in two categories: 2HDMs are in the exact alignment limit ($\cos(\beta-\alpha)=0$), 2HDMs deviate from the alignment limit (i.e. $\cos(\beta-\alpha)\neq0$).

\begin{itemize}
\item 2HDMs in the alignment limit:

\begin{itemize}
\item \textbf{Future LHC makes a discovery}

As discussed above, in the alignment limit, the productions of the heavy Higgses $h_2$ and $h_3$ are purely determined by the size of CPV angle $\alpha_b$, so the reach of future LHC is merely sensitive to the CPV effect in the model. From Fig.~\ref{fig:T1Alignf}, one can observe that, in the Type-I model, the reach of future LHC is entirely inside the reach of future radium and electron EDM experiments. Thus, one can conclude that if Type-I model is true, with a discovery at the future LHC one should also observe non-zero radium and electron EDMs, otherwise the null results of radium and electron EDMs will veto the Type-I CPV 2HDM. A similar conclusion can be drawn for the Type-II model by observing Fig.~\ref{fig:T2Alignf}, where one can find that the LHC sensitive region is well within the reaches of radium and neutron EDMs. Hence, if the Type-II model is true, the discovery of the future LHC should lead to the observations of non-zero radium and neutron EDMs.

\item \textbf{Future LHC gives a null result}

For both Type-I and Type-II models, a null result from future LHC does not exclude the possibility of CPV in 2HDM as long as CPV angle $\alpha_b$ is sufficiently small. Meanwhile, any non-zero EDMs that correspond to the regions of parameter space within the reach of LHC would disfavor the CPV 2HDMs.
\end{itemize}

\item 2HDMs away from the alignment limit:

\begin{itemize}
\item \textbf{Future LHC makes a discovery}

In this case, the result from future LHC is not purely sensitive to the CPV effect since the coupling of $h_3$ to $Zh_1$ is proportional to the level of deviation from the alignment limit and is not suppressed by the CPV angle $\alpha_b$. A discovery at the future LHC may or may not imply a non-zero EDM result, a situation that depends largely on the magnitude of deviation from the alignment limit. 
Moreover, if the deviation is relatively small, such as $\cos(\beta-\alpha)=0.02$, the exclusion limit would mainly come from $h_2$, for which production is purely sensitive to the CPV angle $\alpha_b$, as shown in Fig.~\ref{fig:T1cbma002f} and Fig.~\ref{fig:T2cbma002f} for the Type-I and Type-II models, respectively. In addition, one would expect the non-zero radium and electron EDMs for the Type-I model and non-zero radium and neutron EDMs for the Type-II model. Thus, a null EDM result would disfavor both Type-I and Type-II CPV 2HDMs. 
On the other hand, if the deviation is relatively large, such as $\cos(\beta-\alpha)=0.1$ in the Type-I model (Fig.~\ref{fig:T1cbma01f}) and $\cos(\beta-\alpha)=0.05$ in the Type-II model (Fig.~\ref{fig:T2cbma005f}), the exclusion power would be dominated by the $h_3$ decay. As mentioned above, the discovery of $h_3$ does not necessarily lead to sizable EDMs, and therefore a more detailed study of the CP properties of the newly discovered particle would be needed.

\item \textbf{Future LHC gives a null result}

In this case, for sufficiently large deviations as shown in Fig.~\ref{fig:T1cbma01f} and Fig.~\ref{fig:T2cbma005f}, the discovery of any EDM results would indicate that the CPV source is not consistent with the CPV 2HDMs. On the other hand, for relatively a small deviation as shown in Fig.~\ref{fig:T1cbma002f} and Fig.~\ref{fig:T2cbma002f}, the CPV 2HDMs is still available if CPV angle $\alpha_b$ is sufficiently small. Moreover, any non-zero EDMs that correspond to the regions of parameter space within the reach of LHC would disfavor the CPV 2HDMs.

\end{itemize}

\end{itemize}

Finally, we comment on the potential constraints from the viability of successful EWBG in 2HDMs. One can potentially include the allowed regions where EWBG is viable in Figs.~\ref{fig:T1} and ~\ref{fig:T2}.
For example, the authors of Ref.~\cite{Dorsch:2016nrg} studied the CPV for EWBG and identified some regions of parameter space that seem favorable. They pointed out that the CP violating phase necessary for successful baryogenesis is sensitive to $\tan\beta$ and 
the masses of the heavy Higgses. That work also concentrated on parameter space region where the dominant decay channel of $h_3$ is to the $Z h_2$ final state ($A\to ZH$ in the CP-conserving limit), based on earlier studies of the electroweak phase transition\cite{Dorsch:2013wja,Dorsch:2014qja}. A strong first order electroweak phase transition favors -- but does not absolutely require -- regions of  parameter space leading to dominance of this decay mode. 
In this paper our main focus is on constraints of the CP violating phases from the LHC and EDMs in 2HDMs. It is possible that for the spectra considered here, the CPV 2HDMs can accommodate a strong first order electroweak phase transition and give rise to the CPV asymmetries needed for successful baryogenesis. A detailed and general analysis of this possibility is beyond the scope of the present paper and will be left for future study.

\section{Conclusions\label{conclusion}}
The CP properties of the SM-like Higgs boson provide a portal to investigate the possibility of CP violation beyond the Standard Model, an important ingredient for successful baryogenesis. A CP violating 2HDM with an approximate $Z_2$ symmetry may provide this source of CP-violation. Future searches for new scalars arising in the 2HDM, together with next generation EDM searches, may both discover the 2HDM and determine whether or not it contains the CPV interactions necessary for electroweak baryogenesis. In this paper, we have studied how well a future LHC search for a heavy Higgs with $Zh_1$ decay mode and $Z(\ell\ell)h_1(bb)$ final state  can probe the parameter space of CP violating 2HDMs.  We used the BDT method to estimate the future exclusion limits on the observable $\sigma(gg\to h_{2,3})\times Br(h_{2,3}\to Zh_1)\times Br(h_1\to b\bar{b})$. We then compared this reach with that of next generation searches for the EDMs of the electron, neutron and neutral atoms.

Our results in Figs.~\ref{fig:T1} and \ref{fig:T2}, lead to the following conclusions: (1) In the exact alignment limit, a discovery at the LHC would imply observable radium and electron EDMs for the Type-I model and observable radium and neutron EDMs for the Type-II model.  Null LHC results would still be consistent with CPV 2HDMs if the CPV angle is sufficiently small, but would not preclude observable EDMs. 
(2) Away from the alignment limit,  a discovery at the LHC may or may not imply non-zero EDM results, depending on the level of deviation from the alignment limit. If the  deviation is small, the LHC reach is mainly dominated by the production of the mostly CP-even Higgs ($h_2$). In this case, one can reach the same conclusion as  in the alignment limit for both Type-I and Type-II models. However, with a relatively larger deviation, an LHC discovery may not imply non-zero EDM results because the production of the mostly CP odd Higgs ($h_3$) is not purely sensitive to the variation of the CPV phases. In addition, the  LHC reach would cover most of the parameter space that can also be probed by the EDM searches. As a consequence, a null LHC result, together with observation of the EDMs, would disfavor the CPV 2HDMs. Finally, we also point out that our analysis may break down in the small $\tan\beta$ and $\alpha_b$ regions, due to the non-negligible interference effect between the resonant and non-resonant productions of $Zh_1$ through the gluon fusion channel. We leave the detailed study of this effect for future work.

\section*{Acknowledgement}
We thank Yue Zhang for helpful discussions and Satoru Inoue providing the codes generating the atomic EDM constraint. We also thank Huaike Guo for helping check the model file for CPV2HDM in the initial stage of this work. The work of C.-Y.C is supported by NSERC, Canada. Research at the Perimeter Institute is supported in part by the Government of Canada through NSERC and by the Province of Ontario through MEDT.  The work of H. Li and M.J.R.M was supported in part under U.S. Department of Energy Contract DE-SC0011095. M.J.R.M. is also grateful for the hospitality of the Department of Physics at the University of Arizona, where a portion of this work was completed.

\bigskip
\appendix

\section{Distributions of BDT Input Variables}\label{APPA}

We demonstrate the distributions of BDT input variables after our primary cuts described in Sec. ~\ref{sec:14Pre}
\begin{figure}[!htb]
\includegraphics[width=0.75\linewidth]{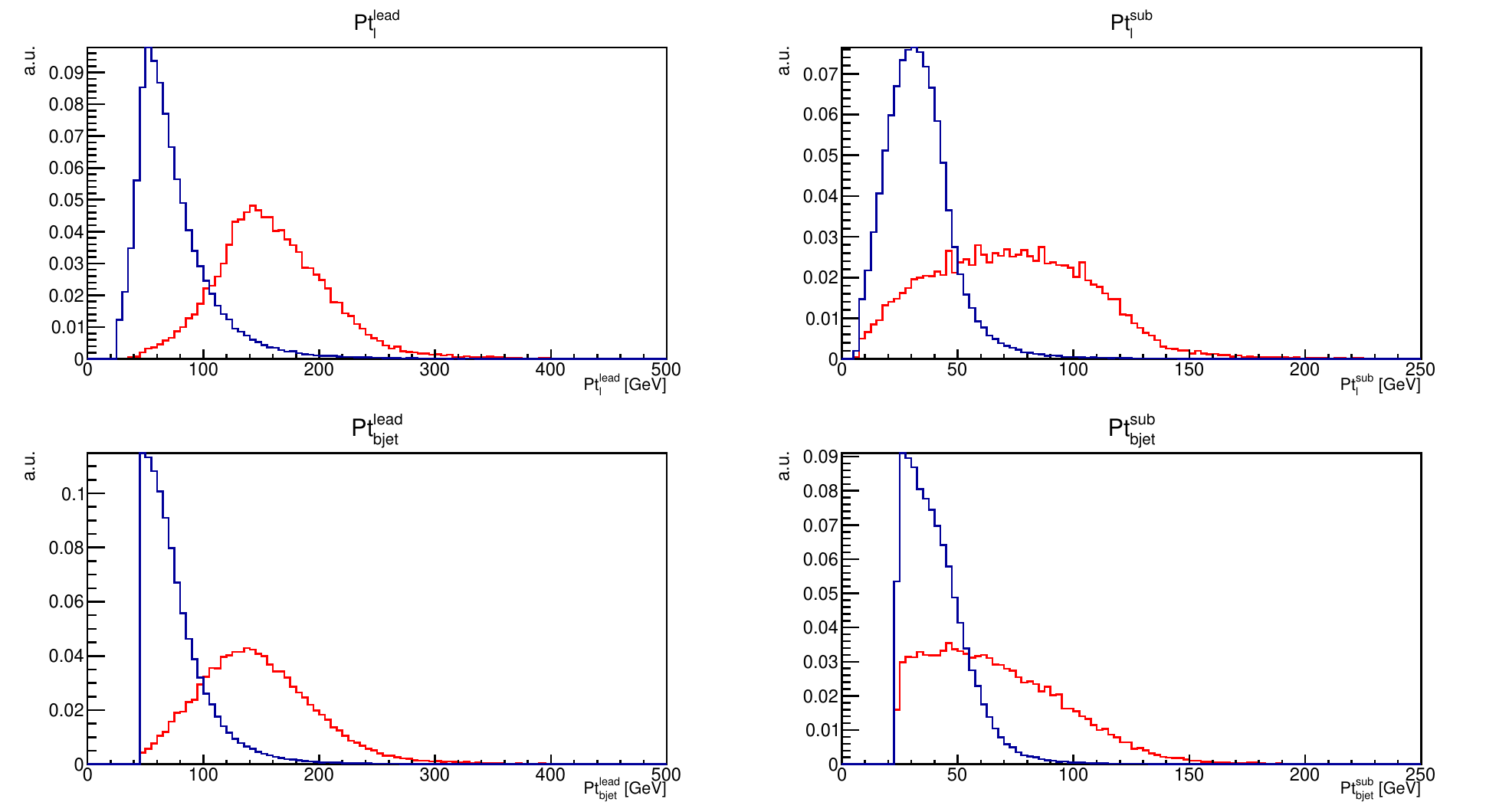}\caption{Plots indicated by their titles showing the distributions of the lepton leading $p_T$, lepton subleading $p_T$, b-jet leading $p_T$, and b-jet subleading $p_T$, respectively. The units of the horizontal axes are GeV. The red histogram is for signal with heavy Higgs mass 550 GeV, and the blue histogram is for the combined background.}\label{fig:MVA1}
\end{figure}
\begin{figure}[!htb]
\includegraphics[width=0.75\linewidth]{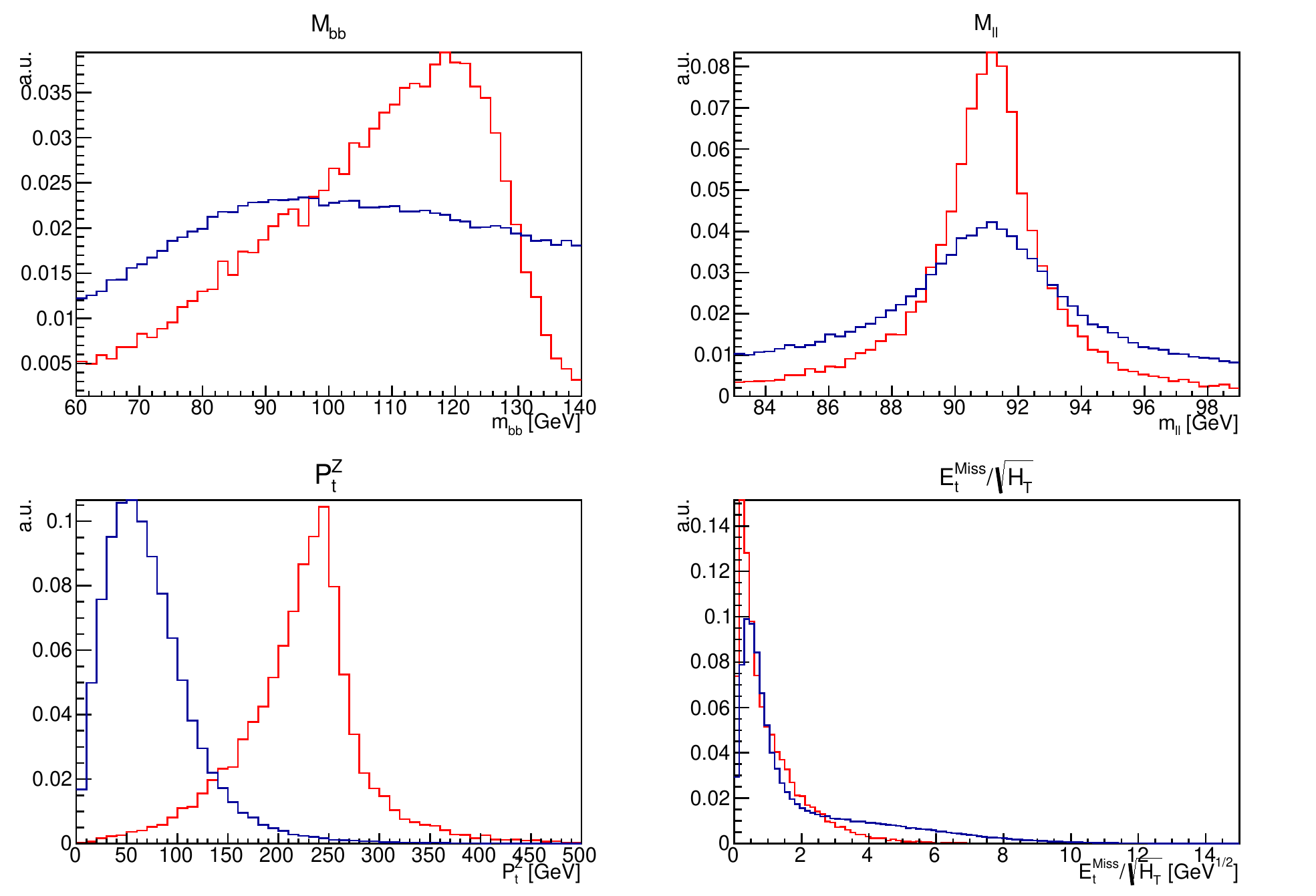}\caption{Plots indicated by their titles showing the distributions of the reconstructed invariant mass for dijet system $m_{bb}$, reconstructed invariant mass for dilepton system $m_{ll}$, $E^{miss}_T/\sqrt{H_T}$, and reconstructed transverse momentum for $Z$ boson $p_T^Z$, respectively. The units of the horizontal axes are GeV for $m_{bb}$, $m_{ll}$, $p^Z_T$, and GeV$^{1/2}$ for $E^{miss}_T/\sqrt{H_T}$. The red histogram is for the signal with heavy Higgs mass 550 GeV, and the blue histogram is for the combined background.}\label{fig:MVA2}
\end{figure}
\begin{figure}[!htb]
\includegraphics[width=0.75\linewidth]{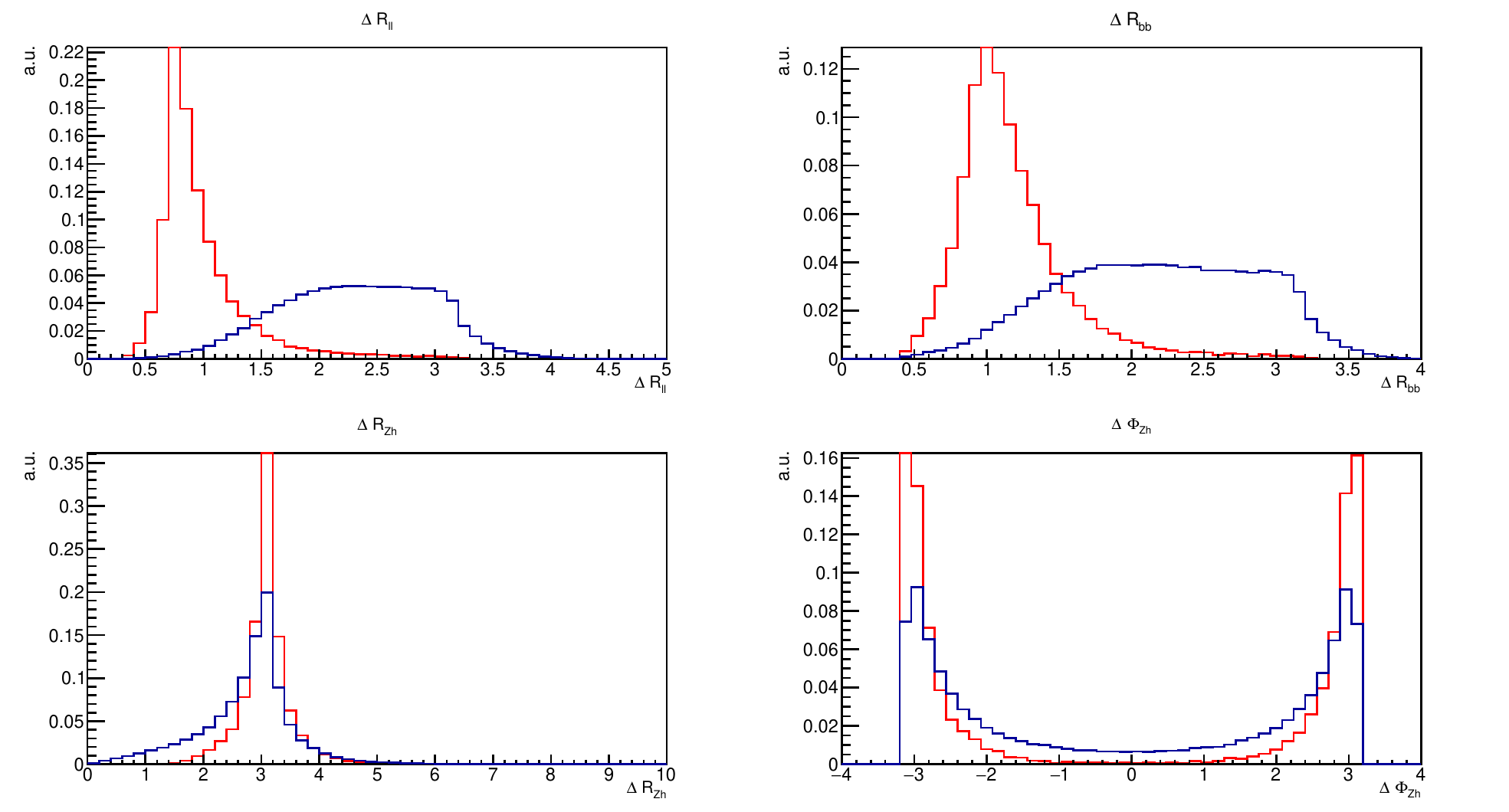}\caption{Plots indicated by their titles showing the distributions of $\Delta R_{ll}$, $\Delta R_{jj}$, $\Delta R_{Zh}$ and $\Delta \Phi_{Zh}$, respectively. The red histogram is for signal with heavy Higgs mass 550 GeV, and the blue histogram is for the combined background.}\label{fig:MVA3}
\end{figure}
\begin{figure}[!htb]
\includegraphics[width=0.6 \linewidth]{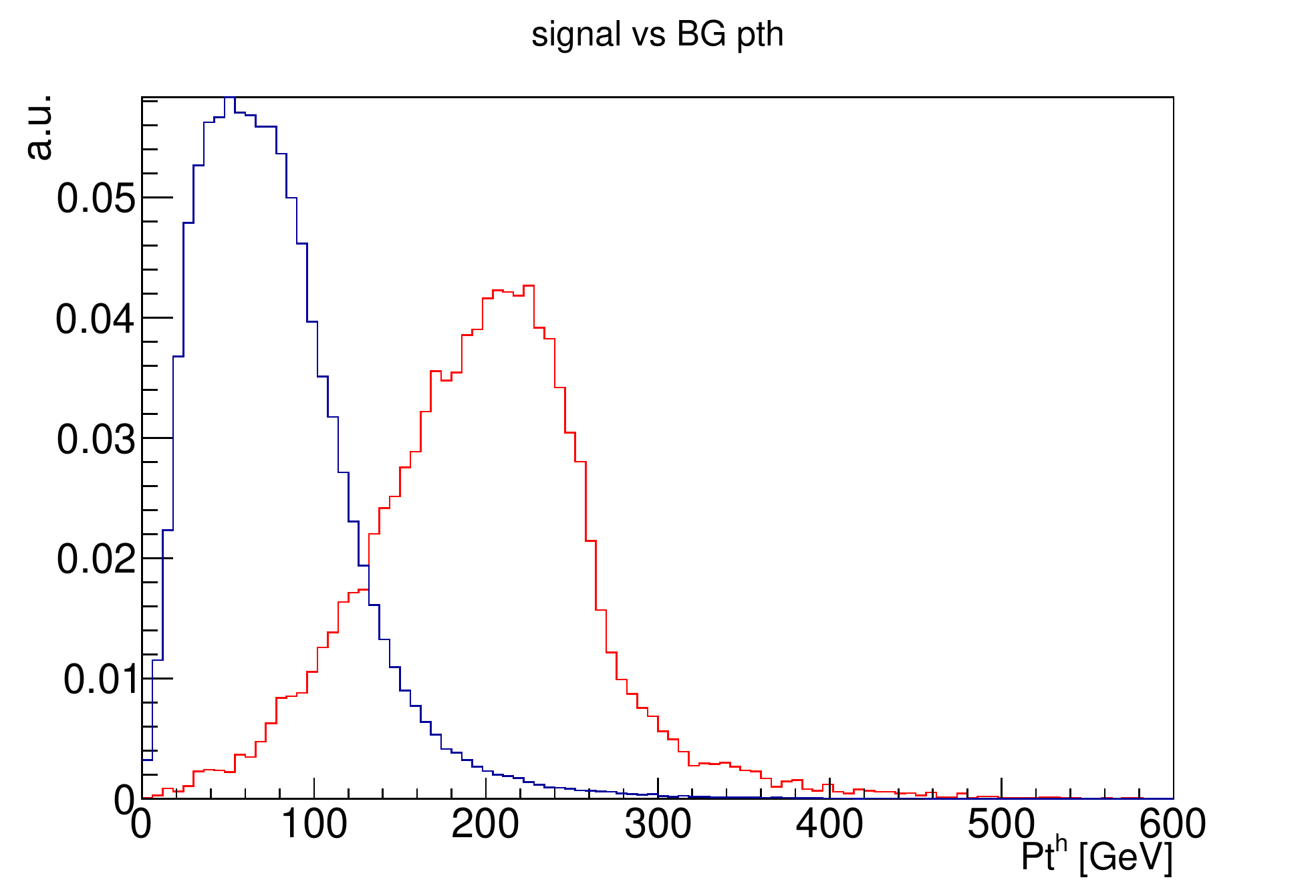}\caption{Reconstructed transverse momentum for Higgs $p_t^h$, the unit of the horizontal axis is GeV. The red histogram is for signal with heavy Higgs mass 550 GeV; the blue histogram is for the combined background.}\label{fig:MVA4}
\end{figure}

\section{Analytical Formulas for Higgs Tow Body Decays}\label{APPB}
Higgs two body decay rates are listed in the following,
\begin{itemize}
\item $h_i \to gg$, heavy Higgs decays to two gluons
\begin{eqnarray}
\Gamma(h_i \to gg) = \frac{\alpha_s^2 G_F m_{h_i}^3}{64\sqrt{2}\pi^3} \left[ \left|c_{t,i} F_{1/2}^H(\tau^i_{t}) +c_{b,i} F_{1/2}^H(\tau^i_{b})\right|^2 + \left|\tilde c_{t,i} F_{1/2}^A(\tau^i_{t}) + \tilde c_b^i F_{1/2}^A(\tau^i_{b})\right|^2 \right] . \ \ \ 
\end{eqnarray}
where the functions $F_{1/2}^H$ and $F_{1/2}^A$ and the variable $\tau^i_f$ are defined in Eqs.~\ref{fa} to \ref{ftau}.
\item $h_i\to Zh_1$, heavy Higgs decays to $Z$ boson and SM-like Higgs
\begin{eqnarray}\label{gAZH}
\Gamma(h_i \to Zh_1) &=& \frac{|g_{iz1}|^2}{16\pi m_{h_i}^3} \sqrt{\left( m_{h_i}^2 - (m_{h_1}+M_Z)^2 \right)\left( m_{h_i}^2 - (m_{h_1}-M_Z)^2 \right)}  \nonumber \\
&&\times \left[ - (2m_{h_i}^2+2m_{h_1}^2-M_Z^2) +\frac{1}{M_Z^2} (m_{h_i}^2-m_{h_1}^2)^2 \right]\, ,
\end{eqnarray}
where $g_{iz1}$ is defined in Eq.~(\ref{giz1}).
\item $h_i\to VV$, heavy Higgs decays to two vector bosons
\begin{eqnarray}
\Gamma(h_i \to VV) = \left( a_i \right)^2 \frac{G_F m_{h_i}^3}{16\sqrt{2}\pi} \delta_V \left( 1 - \frac{4M_V^2}{m_{h_i}^2} \right)^{1/2} 
\left[ 1 - \frac{4M_V^2}{m_{h_i}^2} + \frac{3}{4} \left( \frac{4M_V^2}{m_{h_i}^2} \right)^2 \right] \ ,
\end{eqnarray}
where $V=W,Z$ and $\delta_Z=1, \delta_W=2$, $i=2,3$.
\item $h_i \to f\bar{f}$, heavy Higgs decays to a fermion pair
\begin{eqnarray}
\Gamma(h_i \to \bar f f) = \left[ (c_{f,i})^2 + (\tilde c_{f,i})^2 \rule{0mm}{4mm}\right] \frac{N_cG_F m_f^2 m_{h_i}}{4\sqrt{2}\pi} \left( 1 - \frac{4m_f^2}{m_{h_i}^2} \right)^{3/2} \ ,
\end{eqnarray}
where $N_c=3$ for quarks, $N_c=1$ for leptons.
\item $h_i\to h_1 h_1$ heavy Higgs decays to a pair of SM Higgs
\begin{eqnarray}
\Gamma(h_i \to h_1 h_1) &=& \frac{g_{i11}^2 v^2}{8\pi m_{h_i}} \sqrt{1-\frac{4 m_{h_1}^2}{m_{h_i}^2}} \ ,
\label{trihiggs}
\end{eqnarray}
where $g_{i11}$ is defined by 
\beqa
g_{i11}=\frac{\partial^3 V}{\partial h_i\partial h_1\partial h_1}\mid_{H^\pm, h_i=0}\ \ \ .
\eeqa
\end{itemize}

\bibliographystyle{utphys}

\begin{thebibliography}{40}




\bibitem{Aad:2012tfa} 
  G.~Aad {\it et al.} [ATLAS Collaboration],
  Phys.\ Lett.\ B {\bf 716}, 1 (2012)
  doi:10.1016/j.physletb.2012.08.020
  [arXiv:1207.7214 [hep-ex]].
  
\bibitem{Chatrchyan:2012xdj} 
  S.~Chatrchyan {\it et al.} [CMS Collaboration],
  Phys.\ Lett.\ B {\bf 716}, 30 (2012)
  doi:10.1016/j.physletb.2012.08.021
  [arXiv:1207.7235 [hep-ex]].
  
\bibitem{Trodden:1998ym} 
  M.~Trodden,
  Rev.\ Mod.\ Phys.\  {\bf 71}, 1463 (1999)
  doi:10.1103/RevModPhys.71.1463
  [hep-ph/9803479].

\bibitem{Cline:2006ts} 
  J.~M.~Cline,
  hep-ph/0609145.

\bibitem{Morrissey:2012db} 
  D.~E.~Morrissey and M.~J.~Ramsey-Musolf,
  New J.\ Phys.\  {\bf 14}, 125003 (2012)
  doi:10.1088/1367-2630/14/12/125003
  [arXiv:1206.2942 [hep-ph]].




\bibitem{Sakharov:1967dj} 
  A.~D.~Sakharov,
  Pisma Zh.\ Eksp.\ Teor.\ Fiz.\  {\bf 5}, 32 (1967)
  [JETP Lett.\  {\bf 5}, 24 (1967)]
  [Sov.\ Phys.\ Usp.\  {\bf 34}, 392 (1991)]
  [Usp.\ Fiz.\ Nauk {\bf 161}, 61 (1991)].
  doi:10.1070/PU1991v034n05ABEH002497
  
\bibitem{Kajantie:1996mn} 
  K.~Kajantie, M.~Laine, K.~Rummukainen and M.~E.~Shaposhnikov,
  Phys.\ Rev.\ Lett.\  {\bf 77}, 2887 (1996)
  doi:10.1103/PhysRevLett.77.2887
  [hep-ph/9605288].

\bibitem{Csikor:1998eu} 
  F.~Csikor, Z.~Fodor and J.~Heitger,
  Phys.\ Rev.\ Lett.\  {\bf 82}, 21 (1999)
  doi:10.1103/PhysRevLett.82.21
  [hep-ph/9809291].
  
\bibitem{Rummukainen:1998as} 
  K.~Rummukainen, M.~Tsypin, K.~Kajantie, M.~Laine and M.~E.~Shaposhnikov,
  Nucl.\ Phys.\ B {\bf 532}, 283 (1998)
  doi:10.1016/S0550-3213(98)00494-5
  [hep-lat/9805013].
  
\bibitem{Dorsch:2013wja} 
  G.~C.~Dorsch, S.~J.~Huber and J.~M.~No,
  JHEP {\bf 1310}, 029 (2013)
  doi:10.1007/JHEP10(2013)029
  [arXiv:1305.6610 [hep-ph]].
  
\bibitem{Fromme:2006cm} 
  L.~Fromme, S.~J.~Huber and M.~Seniuch,
  JHEP {\bf 0611}, 038 (2006)
  doi:10.1088/1126-6708/2006/11/038
  [hep-ph/0605242].
  
\bibitem{Cline:1996mga} 
  J.~M.~Cline and P.~A.~Lemieux,
  Phys.\ Rev.\ D {\bf 55}, 3873 (1997)
  doi:10.1103/PhysRevD.55.3873
  [hep-ph/9609240].
  
\bibitem{Khachatryan:2014kca} 
  V.~Khachatryan {\it et al.} [CMS Collaboration],
  Phys.\ Rev.\ D {\bf 92}, no. 1, 012004 (2015)
  doi:10.1103/PhysRevD.92.012004
  [arXiv:1411.3441 [hep-ex]].

\bibitem{ATLAS:2013nma} 
  [ATLAS Collaboration],
  ATLAS-CONF-2013-013.
  
  
  \bibitem{Baron:2013eja} 
  J.~Baron {\it et al.} [ACME Collaboration],
  Science {\bf 343}, 269 (2014)
  doi:10.1126/science.1248213
  [arXiv:1310.7534 [physics.atom-ph]].

\bibitem{Baker:2006ts} 
  C.~A.~Baker {\it et al.},
  Phys.\ Rev.\ Lett.\  {\bf 97}, 131801 (2006)
  doi:10.1103/PhysRevLett.97.131801
  [hep-ex/0602020].

\bibitem{Griffith:2009zz} 
  W.~C.~Griffith, M.~D.~Swallows, T.~H.~Loftus, M.~V.~Romalis, B.~R.~Heckel and E.~N.~Fortson,
  Phys.\ Rev.\ Lett.\  {\bf 102}, 101601 (2009).
  doi:10.1103/PhysRevLett.102.101601

\bibitem{Kumar:2013qya} 
  K.~Kumar, Z.~T.~Lu and M.~J.~Ramsey-Musolf,
  arXiv:1312.5416 [hep-ph].


\bibitem{Inoue:2014nva} 
  S.~Inoue, M.~J.~Ramsey-Musolf and Y.~Zhang,
  Phys.\ Rev.\ D {\bf 89}, no. 11, 115023 (2014)
  doi:10.1103/PhysRevD.89.115023
  [arXiv:1403.4257 [hep-ph]].

\bibitem{Shu:2013uua} 
  J.~Shu and Y.~Zhang,
  Phys.\ Rev.\ Lett.\  {\bf 111}, no. 9, 091801 (2013)
  doi:10.1103/PhysRevLett.111.091801
  [arXiv:1304.0773 [hep-ph]].

\bibitem{Jung:2013hka} 
  M.~Jung and A.~Pich,
  JHEP {\bf 1404}, 076 (2014)
  doi:10.1007/JHEP04(2014)076
  [arXiv:1308.6283 [hep-ph]].


\bibitem{Yamanaka:2017mef} 
  N.~Yamanaka, B.~K.~Sahoo, N.~Yoshinaga, T.~Sato, K.~Asahi and B.~P.~Das,
  Eur.\ Phys.\ J.\ A {\bf 53}, 54 (2017)
  doi:10.1140/epja/i2017-12237-2
  [arXiv:1703.01570 [hep-ph]].

\bibitem{Bian:2014zka} 
  L.~Bian, T.~Liu and J.~Shu,
  Phys.\ Rev.\ Lett.\  {\bf 115}, 021801 (2015)
  doi:10.1103/PhysRevLett.115.021801
  [arXiv:1411.6695 [hep-ph]].

\bibitem{Chen:2015gaa} 
  C.~Y.~Chen, S.~Dawson and Y.~Zhang,
  JHEP {\bf 1506}, 056 (2015)
  doi:10.1007/JHEP06(2015)056
  [arXiv:1503.01114 [hep-ph]].
  
\bibitem{Bian:2017jpt} 
  L.~Bian, N.~Chen and Y.~Zhang,
  arXiv:1706.09425 [hep-ph].
  
\bibitem{Akeroyd:2016ymd} 
  A.~G.~Akeroyd {\it et al.},
  Eur.\ Phys.\ J.\ C {\bf 77}, no. 5, 276 (2017)
  doi:10.1140/epjc/s10052-017-4829-2
  [arXiv:1607.01320 [hep-ph]].
  
\bibitem{Dorsch:2016nrg} 
  G.~C.~Dorsch, S.~J.~Huber, T.~Konstandin and J.~M.~No,
  arXiv:1611.05874 [hep-ph].

  
\bibitem{Aad:2015wra} 
  G.~Aad {\it et al.} [ATLAS Collaboration],
  Phys.\ Lett.\ B {\bf 744}, 163 (2015)
  doi:10.1016/j.physletb.2015.03.054
  [arXiv:1502.04478 [hep-ex]].
  
\bibitem{adaboost} Y. Freund and R.E. Schapire (1996),
{\it Experiments with a new boosting algorithm}, Proc COLT, 209--217. ACM Press, New York (1996).  

  
  
  
\bibitem{Glashow:1976nt} 
  S.~L.~Glashow and S.~Weinberg,
  Phys.\ Rev.\ D {\bf 15}, 1958 (1977).
  doi:10.1103/PhysRevD.15.1958



  
\bibitem{ATLAS:2014kua} 
  The ATLAS collaboration [ATLAS Collaboration],
  ATLAS-CONF-2014-010.

\bibitem{CMS:2016qbe} 
  CMS Collaboration [CMS Collaboration],
  CMS-PAS-HIG-16-007.

\bibitem{HXsecpage}
\url{https://twiki.cern.ch/twiki/bin/view/LHCPhysics/CrossSections}

\bibitem{Alwall:2014hca} 
  J.~Alwall {\it et al.},
  JHEP {\bf 1407}, 079 (2014)
  doi:10.1007/JHEP07(2014)079
  [arXiv:1405.0301 [hep-ph]].
  
\bibitem{Sjostrand:2006za} 
  T.~Sjostrand, S.~Mrenna and P.~Z.~Skands,
  JHEP {\bf 0605}, 026 (2006)
  doi:10.1088/1126-6708/2006/05/026
  [hep-ph/0603175].
 
\bibitem{deFavereau:2013fsa} 
  J.~de Favereau {\it et al.} [DELPHES 3 Collaboration],
  JHEP {\bf 1402}, 057 (2014)
  doi:10.1007/JHEP02(2014)057
  [arXiv:1307.6346 [hep-ex]].



\bibitem{Cordero:2009kv} 
  F.~Febres Cordero, L.~Reina and D.~Wackeroth,
  Phys.\ Rev.\ D {\bf 80}, 034015 (2009)
  doi:10.1103/PhysRevD.80.034015
  [arXiv:0906.1923 [hep-ph]].

\bibitem{Czakon:2013goa} 
  M.~Czakon, P.~Fiedler and A.~Mitov,
  Phys.\ Rev.\ Lett.\  {\bf 110}, 252004 (2013)
  doi:10.1103/PhysRevLett.110.252004
  [arXiv:1303.6254 [hep-ph]].

\bibitem{ZhXsec8TeV}
   \url{https://twiki.cern.ch/twiki/bin/view/LHCPhysics/LHCHXSWGWHZH}\#ZhXsec8


\bibitem{ATLAS:2013gma} 
  [ATLAS Collaboration],
  ATLAS-CONF-2013-020.


\bibitem{Hocker:2007ht} 
  A.~Hocker {\it et al.},
  PoS ACAT {\bf }, 040 (2007)
  [physics/0703039 [PHYSICS]].
  
\bibitem{Hespel:2015zea} 
  B.~Hespel, F.~Maltoni and E.~Vryonidou,
  JHEP {\bf 1506}, 065 (2015)
  doi:10.1007/JHEP06(2015)065
  [arXiv:1503.01656 [hep-ph]].
  
\bibitem{Enomoto:2015wbn} 
  T.~Enomoto and R.~Watanabe,
  JHEP {\bf 1605}, 002 (2016)
  doi:10.1007/JHEP05(2016)002
  [arXiv:1511.05066 [hep-ph]].


\bibitem{Campos:2017dgc} 
  M.~D.~Campos, D.~Cogollo, M.~Lindner, T.~Melo, F.~S.~Queiroz and W.~Rodejohann,
  JHEP {\bf 1708}, 092 (2017)
  doi:10.1007/JHEP08(2017)092
  [arXiv:1705.05388 [hep-ph]].

\bibitem{Celis:2016azn} 
  A.~Celis, M.~Jung, X.~Q.~Li and A.~Pich,
  Phys.\ Lett.\ B {\bf 771}, 168 (2017)
  doi:10.1016/j.physletb.2017.05.037
  [arXiv:1612.07757 [hep-ph]].


  
\bibitem{Graner:2016ses} 
  B.~Graner, Y.~Chen, E.~G.~Lindahl and B.~R.~Heckel,
  Phys.\ Rev.\ Lett.\  {\bf 116}, no. 16, 161601 (2016)
  doi:10.1103/PhysRevLett.116.161601
  [arXiv:1601.04339 [physics.atom-ph]].

\bibitem{Barger:1989fj} 
  V.~D.~Barger, J.~L.~Hewett and R.~J.~N.~Phillips,
  Phys.\ Rev.\ D {\bf 41}, 3421 (1990).
  doi:10.1103/PhysRevD.41.3421

\bibitem{Branco:2011iw} 
  G.~C.~Branco, P.~M.~Ferreira, L.~Lavoura, M.~N.~Rebelo, M.~Sher and J.~P.~Silva,
  Phys.\ Rept.\  {\bf 516}, 1 (2012)
  doi:10.1016/j.physrep.2012.02.002
  [arXiv:1106.0034 [hep-ph]].
    
\bibitem{Dorsch:2014qja} 
  G.~C.~Dorsch, S.~J.~Huber, K.~Mimasu and J.~M.~No,
  Phys.\ Rev.\ Lett.\  {\bf 113}, no. 21, 211802 (2014)
  doi:10.1103/PhysRevLett.113.211802
  [arXiv:1405.5537 [hep-ph]].


  
\end{thebibliography}

\end{document}